\RequirePackage{silence}
\documentclass[fleqn,usenatbib,useAMS]{mnras} 
\usepackage{graphicx}
\usepackage{amsmath}
\DeclareMathOperator{\sech}{sech}
\usepackage{amsfonts}
\usepackage{float}
\usepackage{bm}
\usepackage[perpage,symbol*]{footmisc}
\usepackage{ae,aecompl}
\usepackage{array}
\usepackage{soul}
\usepackage{mathtools}
\usepackage{multirow}

\newcommand{\appropto}{\mathrel{\vcenter{
  \offinterlineskip\halign{\hfil$##$\cr 
    \propto\cr\noalign{\kern2pt}\sim\cr\noalign{\kern-2pt}}}}}

\newcommand{\ssim}{\,{\sim}\,} 

\DeclareRobustCommand{\perthousand}{%
  \ifmmode
    \text{\textperthousand}%
  \else
    \textperthousand
  \fi}

\title{Origin of the Local Group satellite planes} 

\author[Indranil Banik, David O'Ryan \& Hongsheng Zhao]{Indranil Banik$^{1}$\thanks{Email: \href{mailto:ib45@st-andrews.ac.uk}{ib45@st-andrews.ac.uk} (Indranil Banik)\newline $~~~~~~~~~~~~~~$ \href{mailto:hz4@st-andrews.ac.uk}{hz4@st-andrews.ac.uk} (Hongsheng Zhao)}, David O'Ryan$^{1}$ and Hongsheng Zhao$^{1}$\\
$^{1}$Scottish Universities Physics Alliance, University of St Andrews, North Haugh, St Andrews, Fife, KY16 9SS, UK}

\pubyear{2018}
\begin{document}
\label{firstpage}
\pagerange{\pageref{firstpage}--\pageref{lastpage}}

\maketitle

\begin{abstract} 

We attempt to understand the planes of satellite galaxies orbiting the Milky Way (MW) and M31 in the context of Modified Newtonian Dynamics (MOND), which implies a close MW-M31 flyby occurred ${\approx 8}$ Gyr ago. Using the timing argument, we obtain MW-M31 trajectories consistent with cosmological initial conditions and present observations. We adjust the present M31 proper motion within its uncertainty in order to simulate a range of orbital geometries and closest approach distances. Treating the MW and M31 as point masses, we follow the trajectories of surrounding test particle disks, thereby mapping out the tidal debris distribution.

Around each galaxy, the resulting tidal debris tends to cluster around a particular orbital pole. We find some models in which these preferred spin vectors align fairly well with those of the corresponding observed satellite planes. The radial distributions of material in the simulated satellite planes are similar to what we observe. Around the MW, our best-fitting model yields a significant fraction (0.22) of counter-rotating material, perhaps explaining why Sculptor counter-rotates within the MW satellite plane. In contrast, our model yields no counter-rotating material around M31. This is testable with proper motions of M31 satellites.

In our best model, the MW disk is thickened by the flyby 7.65 Gyr ago to a root mean square height of 0.75 kpc. This is similar to the observed age and thickness of the Galactic thick disk. Thus, the MW thick disk may have formed together with the MW and M31 satellite planes during a past MW-M31 flyby.

\end{abstract}

\begin{keywords}
galaxies: groups: individual: Local Group -- Galaxy: kinematics and dynamics -- Dark Matter -- methods: numerical -- methods: data analysis -- Galaxy: disc
\end{keywords}

\section{Introduction}
\label{Introduction}

The standard cosmological paradigm \citep[$\Lambda$CDM,][]{Ostriker_Steinhardt_1995} faces several challenges in the relatively well-observed Local Group \citep[e.g.][and references therein]{Kroupa_2015}. In particular, the satellite systems of its two major galaxies $-$ the Milky Way (MW) and Andromeda (M31) $-$ are both highly flattened. For the MW, this has been suspected for several decades \citep{Lynden_Bell_1976, Lynden_Bell_1982}. In light of subsequently discovered satellites, this led \citet{Kroupa_2005} to argue that its satellite system is inconsistent with $\Lambda$CDM as it is too anisotropic. Later proper motion measurements showed that most of its satellites co-rotate within a well-defined plane \citep{Metz_2008, Kroupa_2013}. Recently discovered ultra-faint satellites, globular clusters and tidal streams independently prefer a similarly oriented plane \citep{Pawlowski_2012, Pawlowski_Kroupa_2014}. Although some flattening is expected in $\Lambda$CDM \citep[e.g.][]{Butsky_2016}, it remains difficult to explain the very small thickness of the MW satellite system and its coherent rotation \citep{Ahmed_2017}.

An analogous situation was suspected around M31 \citep{Metz_2007, Metz_2009} and recently confirmed by \citet{Ibata_2013} using the homogeneous sky coverage of the Pan-Andromeda Archaeological Survey \citep{PANDAS}. Despite its greater distance, the detection of the M31 satellite plane is rather secure because we presently view it almost edge-on. It probably rotates coherently given the observed redshift gradient across it \citep{Ibata_2013}, though proper motions are needed to confirm this. As with the MW, it is difficult to understand the observed properties of the M31 satellite system in $\Lambda$CDM \citep{Ibata_2014}. The solution to these puzzling observations should also consider the two planes of non-satellite LG galaxies found by \citet{Pawlowski_2013}.


Just outside the LG, a satellite plane has recently been identified around Centaurus (Cen) A \citep{Muller_2018}. This structure is likely co-rotating because its members exhibit a radial velocity trend similar to M31. The satellites of M81 also appear to lie in a flattened distribution, though their kinematics are not currently known \citep{Chiboucas_2013}. These discoveries may be related to a vast plane of dwarf galaxies recently found near M101 \citep{Muller_2017}. This is not a satellite galaxy plane as it extends over 3 Mpc. Such a structure appears difficult to find in cosmological simulations of the $\Lambda$CDM paradigm \citep[][figure 8]{Gonzalez_2010}.

Although $\Lambda$CDM is widely considered to work well on large scales and at early times \citep[e.g.][]{Planck_2015}, it faces some tension regarding the assumptions of homogeneity and isotropy \citep{Kroupa_2015, Javanmardi_2015, Javanmardi_2017, Bengaly_2018}. The model also has difficulty explaining the strength of the 21 cm absorption feature corresponding to the hyperfine transition of cold neutral hydrogen at redshift ${z \approx 17}$ \citep{Bowman_2018}. While those authors attributed their unexpectedly strong detection to baryon-dark matter interactions, only a narrow range of parameters remains viable once laboratory and other constraints are considered \citep{Berlin_2018}.\footnote{The particle mass would have to be $10-80$ MeV/$c^2$ and it would need to carry a small but non-zero fraction of the electric charge. The cross-section for scattering with gas also needs to decline steeply with velocity.} \citet{Ewall_2018} considered a somewhat more conventional explanation based on a radio background from accreting black holes that formed out of stars or via direct collapse of gas clouds. They found that this scenario would have to be highly contrived to match other constraints, in particular the optical depth to reionisation \citep{Heinrich_2018}. For example, the black holes required by their model would need to stop growing at ${z \approx 16}$.

Given these difficulties and the many a priori predictions of MOND which were subsequently confirmed \citep[e.g.][]{Famaey_McGaugh_2012}, it is worth considering whether the matter content of the Universe might be purely baryonic \citep{McGaugh_1999}. Indeed, this seems to be one of the few viable scenarios for explaining the strong observed 21 cm absorption feature without significant fine-tuning \citep{McGaugh_2018}. Such a purely baryonic universe could have an expansion history very similar to $\Lambda$CDM \citep{Zhao_2007}.

Although high-redshift observations like this are interesting, the non-linear dynamics of gravitational systems are much easier to observe in the LG. Here, $\Lambda$CDM faces another recently discovered problem concerning the very high radial velocities of non-satellite LG dwarf galaxies at distances of ${1 - 3}$ Mpc \citep{Pawlowski_McGaugh_2014}. We investigated this problem based on a timing argument analysis of the LG \citep{Kahn_Woltjer_1959, Einasto_1982} extended to include test particles representing LG dwarfs. Following on from previous spherically symmetric dynamical models \citep{Sandage_1986, Jorge_2014}, we constructed an axisymmetric model of the LG consistent with the almost radial MW-M31 orbit \citep{Van_der_Marel_2012} and the close alignment of Cen A with this line \citep{Ma_1998}. Treating LG dwarfs as test particles in the gravitational field of these three massive moving objects, we investigated a wide range of model parameters using a full grid search \citep{Banik_Zhao_2016}.

None of the models provided a good fit, even when we made reasonable allowance for inaccuracies in our model as a representation of $\Lambda$CDM based on the scatter about the Hubble flow in detailed $N$-body simulations of it \citep{Aragon_Calvo_2011}. This is because several LG dwarfs have Galactocentric Radial Velocities (GRVs) much higher than expected in our best-fitting model, though the opposite is rarely the case \citep[][figure 9]{Banik_Zhao_2016}. We found that this high-velocity galaxy (HVG) problem should remain even when certain factors beyond our model are included, in particular the Large Magellanic Cloud and the Great Attractor.


To test whether the issue might be resolved using a three-dimensional (3D) model of the LG, we borrowed an algorithm described in \citet{Phelps_2013}. The typical mismatch between observed and predicted GRVs in the best-fitting model is actually slightly higher than in the 2D case, with a clear tendency persisting for faster outward motion than expected \citep[][figures 7 and 9]{Banik_Zhao_2017}. These results are similar to those obtained by \citet{Peebles_2017} using a similar algorithm. Despite using a very different method to \citet{Banik_Zhao_2016}, the conclusions remain broadly similar.


We recently revisited this analysis by performing a much more thorough search for the best-fitting model \citep[][section 4.1]{Banik_2017_anisotropy}. As our best-fitting model still yielded several HVGs that it could not account for, we considered possible $\Lambda$CDM-based explanations for them at some length (see their section 7.2). Not coming up with any plausible solutions, we looked for clues in the spatial distribution of the HVGs. This revealed that they lie very close to a plane extending over ${\ssim 3}$ Mpc, with an isotropic distribution ruled out at ${3 \sigma}$ (see their table 7).

$\Lambda$CDM has difficulty explaining the kinematics of even one HVG, let alone the 6 we identified. We therefore considered how these observations might be explained in the alternative theory called Modified Newtonian Dynamics \citep[MOND,][]{Milgrom_1983}. In MOND, galaxies are not surrounded by halos of dark matter. Instead, their dynamical effect is provided by an acceleration-dependent modification to gravity. The gravitational field strength $g$ at distance $r$ from an isolated point mass $M$ transitions from the usual $\frac{GM}{r^2}$ law at short range to
\begin{eqnarray}
	g ~=~ \frac{\sqrt{GMa_{_0}}}{r} ~~~\text{for } ~r \gg \sqrt{\frac{GM}{a_{_0}}}
	\label{Deep_MOND_limit}
\end{eqnarray}

MOND introduces Milgrom's constant $a_{_0}$ as a fundamental acceleration scale of nature below which the deviation from Newtonian dynamics becomes significant. Empirically, $a_{_0} \approx 1.2 \times {10}^{-10}$ m/s$^2$ to match galaxy rotation curves \citep{McGaugh_2011}. Remarkably, this is similar to the acceleration at which the classical energy density in a gravitational field \citep[][equation 9]{Peters_1981} becomes comparable to the dark energy density $u_{_\Lambda} = \rho_{_\Lambda} c^2$ implied by the accelerating expansion of the Universe \citep{Riess_1998}. Thus,
\begin{eqnarray}
	\frac{g^2}{8\rm{\pi}G} ~<~ u_{_\Lambda} ~~\Leftrightarrow~~ g ~\la~ 2\rm{\pi}a_{_0}
	\label{MOND_quantum_link}
\end{eqnarray}

This suggests that MOND may arise from quantum gravity effects \citep[e.g.][]{Milgrom_1999, Pazy_2013, Verlinde_2016, Smolin_2017}. Regardless of its underlying microphysical explanation, it can accurately match the rotation curves of a wide variety of both spiral and elliptical galaxies across a vast range in mass, surface brightness and gas fraction \citep[][and references therein]{Lelli_2017}. It is worth emphasising that MOND does all this based solely on the distribution of luminous matter. Given that most of these rotation curves were obtained in the decades after the MOND field equation was first published \citep{Bekenstein_Milgrom_1984}, it is clear that these achievements are successful a priori predictions. These predictions work due to underlying regularities in galaxy rotation curves that are difficult to reconcile with the collisionless dark matter halos of the $\Lambda$CDM paradigm \citep{Salucci_2017, Desmond_2016, Desmond_2017}.

In the LG, MOND implies a much stronger MW-M31 mutual attraction than does $\Lambda$CDM. Acting on their almost radial orbit \citep{Van_der_Marel_2012}, this leads to a close encounter ${9 \pm 2}$ Gyr ago \citep{Zhao_2013}. Due to the high MW-M31 relative velocity around the time of their flyby, they would likely have gravitationally slingshot several LG dwarfs out at high speed. The dwarfs flung out almost parallel to the motion of the perturber would have reached the fastest speeds. We used a MOND model of the LG to demonstrate this effect, showing that the HVGs ought to define the MW-M31 orbital plane \citep[][section 3]{Banik_2017_anisotropy}. Observationally, the HVGs do define a rather thin plane, with the MW-M31 line only ${16^\circ}$ out of this plane (see their table 4). Thus, we argued that the HVGs may preserve evidence of a past close MW-M31 flyby and their fast relative motion at that time.

During this flyby, a tidal tail would likely have formed and might later have condensed into satellite galaxies of the MW and M31. This phenomenon occurs in some observed galactic interactions \citep{Mirabel_1992} and in MOND simulations of them \citep{Tiret_2008, Renaud_2016}. Due to the way in which such tidal dwarf galaxies form out of a thin tidal tail, they would end up lying close to a plane and co-rotating within that plane \citep{Wetzstein_2007}, though a small fraction might well end up counter-rotating depending on the exact details \citep{Pawlowski_2011}. This might be how the MW and M31 satellite planes formed \citep{Zhao_2013}.

In principle, the LG satellite planes could form in some other galactic interaction(s) than a MW-M31 flyby \citep[e.g.][]{Hammer_2013}, an infeasible scenario in $\Lambda$CDM as dynamical friction between their overlapping dark matter halos would cause a rapid merger \citep{Privon_2013}. However, \emph{any} second-generation origin of the MW and M31 satellite planes runs into the issue that such satellites would be free of dark matter \citep{Barnes_1992, Wetzstein_2007}. This is due to the dissipationless nature of dark matter and its initial distribution in a dispersion-supported near-spherical halo. During a tidal interaction, dark matter of this form is clearly incapable of forming into a thin dense tidal tail out of which dwarf galaxies might condense. As a result, any tidal dwarf galaxies would be purely baryonic and thus have only a very small escape velocity, preventing them from subsequently accreting dark matter out of the MW halo. For this reason, $\Lambda$CDM struggles to explain the high observed internal velocity dispersions of the MW satellites coherently rotating in a thin plane \citep{McGaugh_Wolf_2010}. That work paid careful attention to the issue of whether tides were responsible for the high velocity dispersions, eventually ruling out this explanation in a $\Lambda$CDM context (see their figure 6).

A similar situation is evident around M31 $-$ its satellite galaxies also have rather high internal velocity dispersions \citep{McGaugh_Milgrom_2013} but mostly lie in a thin plane \citep{Ibata_2013}. The velocity dispersions could be explained if the M31 satellites formed primordially and thus retained their dark matter halos. However, this is difficult to reconcile with the anisotropy of the M31 satellite system as a whole \citep{Ibata_2014}, which is suggestive of a tidal origin.

After careful consideration of several proposed $\Lambda$CDM scenarios for how primordial satellites might end up in a thin plane, \citet{Pawlowski_2014} concluded that none of them agreed with observations for either the MW or M31. It was later shown that baryonic effects are unlikely to provide the necessary anisotropy if one sticks to a primordial origin for the satellites \citep{Pawlowski_2015}. This issue was revisited by performing a high-resolution $\Lambda$CDM hydrodynamical simulation of a MW analogue in a cosmological context \citep{Maji_2017}. Although this unpublished article claimed that the results were consistent with observations, it has recently been shown that this is not the case \citep{Pawlowski_2017}. Those authors showed that the satellite galaxy distribution of \citet{Maji_2017} was consistent with isotropy. However, the actual MW satellite system is inconsistent with isotropy at more than ${5 \sigma}$ once the survey footprint is taken into account \citep{Pawlowski_2016}.

More recent hydrodynamical $\Lambda$CDM simulations also fail to yield highly flattened satellite systems like those observed in the LG \citep{Ahmed_2017}. The mild amount of flattening in these simulations might not even be related to baryonic effects as similar results arise in dark matter-only simulations \citep{Garaldi_2017}. In any case, it is difficult to see how baryonic effects can help to form a 200 kpc-wide plane of primordial satellites \citep{Shao_2018}. On the other hand, tidally formed satellites would lack dark matter and thus have only a very small internal velocity dispersion, in disagreement with observations. For a recent review of the satellite plane problem, we refer the reader to \citet{Pawlowski_2018}. This considers both LG satellite planes and the recently discovered one around Cen A \citep{Muller_2018}.

Given how difficult it is for primordial satellites to explain the observed satellite planes, it is important to note that tidal dwarf galaxies ought to form in any viable cosmological model. In $\Lambda$CDM, these would have rather different properties to primordial dwarfs due to the former lacking significant quantities of dark matter \citep[][section 4]{Kroupa_2012}. However, observational evidence exists only for the latter \citep{Dabringhausen_2016}.

In MOND, this is to be expected because no galaxies have dark matter halos, whether formed primordially or as tidal dwarfs. MOND predicts that there will be a one-to-one correspondence between the baryonic distribution of a galaxy and its internal dynamics \emph{regardless of how it formed}. In the LG, this modification to gravity raises the expected internal velocity dispersions of \emph{purely baryonic} MW and M31 satellites enough to match observations \citep[][respectively]{McGaugh_Wolf_2010, McGaugh_Milgrom_2013}.

In addition to explaining the internal dynamics of LG satellites, MOND inevitably provides a past major tidal interaction in the LG out of which the MW and M31 satellite planes could have formed \citep{Zhao_2013}. Recent MOND-based $N$-body simulations of this interaction suggest that it could produce structures resembling the LG satellite planes \citep{Bilek_2017}. However, it is not clear if this scenario can explain the well-observed orientations of these planes \citep{Kroupa_2013}.

Here, we address this question by searching the parameter space much more thoroughly using restricted $N$-body models of the MW and M31 over a Hubble time. The galaxies are treated as point masses surrounded by test particle disks. We use a grid method to explore a range of MW-M31 relative tangential velocities consistent with the small observed proper motion of M31 \citep{Van_der_Marel_2012} and the unknown external gravitational field (EF) on the LG (Section \ref{g_ext_history}). The grid of models we try out thus contains three parameters $-$ the magnitude and direction of the MW-M31 relative tangential velocity and the EF from large scale structure in which the LG is embedded. In each case, we look at the orbital pole distribution of the tidal debris around each major LG galaxy. Ideally, we would find a clear clustering of orbital poles around the actually observed satellite plane angular momentum direction for that galaxy.

We also consider how much the MW disk is dynamically heated by its interaction with M31. This is to check the feasibility of our scenario for producing the MW thick disk \citep{Gilmore_1983}, a structure which seems to have formed fairly rapidly from its thin disk \citep{Hayden_2015}. The thick disk formed ${9 \pm 1}$ Gyr ago \citep{Quillen_2001}, a constraint we consider on the time of the MW-M31 flyby. More recent investigations suggest that there was a burst of star formation at that time \citep[][figure 2]{Snaith_2014}. An enhanced star formation rate could have further thickened the MW disk through the popping star cluster scenario \citep{Kroupa_2002}. The star formation rate of M31 also appears to have been much higher for lookback times ${\ga 9}$ Gyr \citep[][figure 12]{Dolphin_2017}. The disk heating which likely formed the Galactic thick disk appears to have been stronger in the outer parts of the MW, characteristic of a tidal effect \citep{Banik_2014}. This may explain why its thick disk has a longer scale length than its thin disk \citep{Juric_2008, Jayaraman_2013}.

Around each galaxy, there is usually a clear clustering of tidal debris orbital poles, though its direction generally differs for the MW and M31 as their disks are oriented differently. Sometimes, there is also an orbital pole peak in the opposite direction. Such a counter-rotating peak often contains a significant amount of material, especially for the MW. This is good to some extent because Sculptor counter-rotates within the plane preferred by most MW satellites \citep{Piatek_2006, Pawlowski_2011, Sohn_2017}. However, some models clearly have too high a counter-rotating fraction. Thus, we try to find models that have a reasonable fraction of counter-rotating material in addition to yielding orbital pole peaks that match the observed spin vectors of both LG satellite planes. We also expect the MW-M31 orbital plane to align with the HVG plane \citep[][co-ordinates in table 4]{Banik_2017_anisotropy}.

In this contribution, our objective is to test whether any single MOND model can match all these constraints and thereby provide a reasonable origin scenario for the MW and M31 satellite planes. Having introduced the problem in this section, we explain how we obtain the MW-M31 trajectory (Section \ref{MW_M31_trajectory}) and use it to advance test particles (Section \ref{MW_M31_disks}). We compare each model to observations using the method described in Section \ref{Comparison_with_observations}. Using a $\chi^2$ statistic (Equation \ref{chi_sq}), we identify the best-fitting model. The results of this model are described in Section \ref{Results}, where we argue that it matches observations as well as can be expected given the deficiencies of our model and observational uncertainties. Our conclusions are given in Section \ref{Conclusions}.

\section{Milky Way $-$ M31 trajectory}
\label{MW_M31_trajectory}

\subsection{The two-body force in an external field}
\label{Two_body_force}

\subsubsection{Deep-MOND limit}
\label{Two_body_force_DML}

We begin by determining the past MW-M31 trajectory in MOND. We use the quasilinear formulation of MOND \citep[QUMOND,][]{QUMOND} because it is easier to handle numerically. In spherical symmetry, this yields identical results to the more traditional aquadratic Lagrangian version of MOND \citep[AQUAL,][]{Bekenstein_Milgrom_1984} as long as the same interpolating function is used to transition from the Newtonian to the low-acceleration (deep-MOND) regimes. However, even in more complicated situations, AQUAL and QUMOND yield forces that only differ by $\la 5\%$ \citep{Banik_2015}. $N$-body simulations of both MOND formulations yield quite similar results \citep{Candlish_2016}.

QUMOND uses the Newtonian gravitational field $\bm{g}_{_N}$ to determine the true gravitational field $\bm{g}$. Thus, the first step is to find $\bm{g}_{_N}$ by solving the usual Poisson equation sourced by the baryonic density $\rho_{_b}$.
\begin{eqnarray}
	\nabla \cdot \bm{g}_{_N} ~=~ -4\pi G \rho_{_b}
	\label{Poisson_equation}
\end{eqnarray}

To determine $\bm{g}$, this Poisson equation must be solved again with a forcing analytically dependent on $\bm{g}_{_N}$. Thus, QUMOND does not require a non-linear grid relaxation stage, a major improvement on AQUAL.
\begin{eqnarray}
	\label{QUMOND_equation}
	\overbrace{\nabla \cdot \bm{g}}^{\propto \rho_{_{PDM}} + \rho_{_b}} ~&=&~ \nabla \cdot \left[ \bm{g}_{_N} \nu \left( g_{_N} \right) \right] ~~\text{ where} \\
	\nu \left( g_{_N} \right) &=& \frac{1}{2} ~+~ \sqrt{\frac{1}{4} + \frac{a_{_0}}{g_{_N}}}	
\end{eqnarray}

$\nu \left( g_{_N} \right)$ is the interpolating function used to transition between the Newtonian and deep-MOND regimes. Here, we use the `simple' form of this function \citep{Famaey_Binney_2005}. This works very well at explaining the rotation curve of the MW \citep[][figure 3]{Iocco_Bertone_2015} and the kinematics of ${\approx 6000}$ elliptical galaxies probing accelerations up to ${\approx 30 a_{_0}}$ \citep{Chae_2017}. The source term for the gravitational field is $\nabla \cdot \left( \nu \bm{g}_{_N} \right)$, which can be thought of as an `effective' density $\rho$ composed of the actual density $\rho_{_b}$ and an extra term which we define to be the phantom dark matter density $\rho_{_{PDM}}$. This is the distribution of dark matter that would be necessary in Newtonian gravity to generate the same total gravitational field as QUMOND yields from the baryons alone.

To get the gravity between the MW and M31 at some separation $d_{_{M31}}$, we assume that they are point masses in the deep-MOND regime, though we will later relax this assumption when their relative acceleration $\ddot{d}_{_{M31}}$ is a significant fraction of $a_{_0}$ (Section \ref{Newtonian_limit}). If the MW and M31 were the only objects in the Universe, then \citet{Zhao_2010} showed that we could approximate their mutual gravitational acceleration as
\begin{eqnarray}
	\label{g_iso}
	g_{iso} ~&=&~ \frac{Q\sqrt{GM_{_{LG}}a_0}}{d_{_{M31}}} \\
	Q ~&=&~ \frac{2 \left( 1 - {q_{_1}}^\frac{3}{2} - {q_{_2}}^\frac{3}{2}\right)}{3 q_{_1} q_{_2}}
\end{eqnarray}

Here, the total MW and M31 mass is $M_{_{LG}}$, of which a fraction $q_{_1}$ is in the MW and the remaining $q_{_2}$ resides in M31. Equation \ref{g_iso} suggests that $\ddot{d}_{_{M31}} {\approx 0.017~a_{_0}}$. However, the MOND field equations imply that $\ddot{d}_{_{M31}}$ is affected by the EF in which the LG is embedded \citep{Milgrom_1986}. This is true even if the EF is due to objects so far outside the LG that they affect the MW and M31 equally (i.e. there are no tides). Including this effect, the deep-MOND limit gravity $g_{_M}$ between the MW and M31 can be written as
\begin{eqnarray}
	\label{g_M}
	g_{_M} &=& \frac{\sqrt{GMa_0}}{r} f\left( \tilde{r} \right) \\
	\tilde{r} &\equiv & \frac{d_{_{M31}}}{r_t} \\
	r_t &=& \frac{\sqrt{GMa_0}\left( 3 + \cos^2 \phi \right)}{4Qg_{ext}}
\end{eqnarray}

We have assumed that the form of $f\left( \tilde{r} \right)$ is independent of the angle $\phi$ between the MW-M31 axis and $\widehat{\bm{g}}_{_{ext}}$, with the angular dependence introduced by the EF accounted for by making $r_t$ angle-dependent. The function $f\left( \tilde{r} \right)$ should be $Q$ when $\tilde{r} \ll 1$ or equivalently when $g_{_{ext}} \ll g_{iso}$, representing the isolated regime \citep{Zhao_2010}. In the opposite EF-dominated case, $f \to \frac{Q}{\tilde{r}}$ but the dependence on $Q$ is cancelled by how this also affects $r_t$ \citep[][equation 37]{Banik_2015}. It is in this limit that $f$ becomes angle-dependent as the MW and M31 start to feel the preferred direction introduced by the EF.

The EF on the whole LG is $g_{ext} \approx 0.03~a_{_0}$ \citep{Famaey_2007}, suggesting that it may have a significant effect. This is especially true as the MW and M31 spend a significant amount of time near apocentre, where the EF is expected to be particularly important. However, the EF is clearly not dominant i.e. much stronger than $g_{iso}$ (Equation \ref{g_iso}). To include the EF and determine $f\left( \tilde{r} \right)$, we thus need a numerical scheme.

For this purpose, we use a method similar to that outlined in section 2.2 of \citet{Banik_2017_escape}. We first obtain $\bm{g}_{_N}$ using the superposition principle (i.e. the force from a combination of masses is the sum of the forces due to each one on its own). This is done by adding the Newtonian-equivalent external field $\bm{g}_{_{N, ext}}$ to the forces from the two point masses. $\bm{g}_{_{N, ext}}$ is what the EF on the LG would have been if the Universe had the same matter distribution as at present but was governed by Newtonian gravity rather than QUMOND. Assuming the EF is caused by a distant point mass (something we justify in Section \ref{Tides}), we can use the spherically symmetric relation between $\bm{g}_{_{N, ext}}$ and the true EF of $\bm{g}_{_{ext}}$.
\begin{eqnarray}
	\overbrace{\nu \left(\frac{\left| \bm{g}_{_{N, ext}} \right|}{a_{_0}} \right)}^{\nu_{_{ext}}} \bm{g}_{_{N, ext}} &=& \bm{g}_{ext}
	\label{g_N_ext}
\end{eqnarray}

In the deep-MOND limit, this reduces to
\begin{eqnarray}
	g_{N,ext} ~=~ \frac{{g_{ext}}^2}{a_{_0}}
	\label{g_N_ext_DML}
\end{eqnarray}

Having obtained $\bm{g}_{_N}$, we can use Equation \ref{QUMOND_equation} to obtain $\nabla \cdot \bm{g}$, the effective matter distribution. We apply a direct summation procedure to this in order to obtain $\bm{g}$.
\begin{eqnarray}
	\bm{g \left( \bm{r} \right)} ~=~ \int \nabla \cdot \bm{g} \left( \bm{r'}\right) \frac{\left( \bm{r} - \bm{r'} \right)}{4 \pi |\bm{r} - \bm{r'}|^3} ~d^3\bm{r'}
	\label{g_direct_sum}
\end{eqnarray}

Due to the axisymmetry of the problem, the phantom dark matter can be thought of as a large number of uniform density rings. To find $\bm{g}$ on the MW and M31 and thus their relative acceleration, we sum the Newtonian gravity contributed by each of these rings. This is simple to do as the MW and M31 both lie on the symmetry axis.


We use Equation \ref{M_dyn_MOND} to estimate that the mass ratio between the MW and M31 is such that ${q_{_1} = 0.3}$ based on the flatline level of the MW rotation curve being ${180}$ km/s \citep{Kafle_2012} and that of M31 being ${225}$ km/s \citep{Carignan_2006}. With ${q_{_1}}$ fixed as we vary $g_{_{ext}}$, our numerical results can be summarised as
\begin{eqnarray}
	\label{u}
	f\left( \tilde{r} \right) = \frac{Q}{\sqrt{1 + \tilde{r}^2}}\left[ 1 + \overbrace{0.032~\rm{exp} \left({-\frac{\left( Ln~ \tilde{r} - 0.42\right)^2}{2\times 0.65^2}}\right)}^u\right]
\end{eqnarray}

\begin{figure}
	\centering
		\includegraphics [width = 8.5cm] {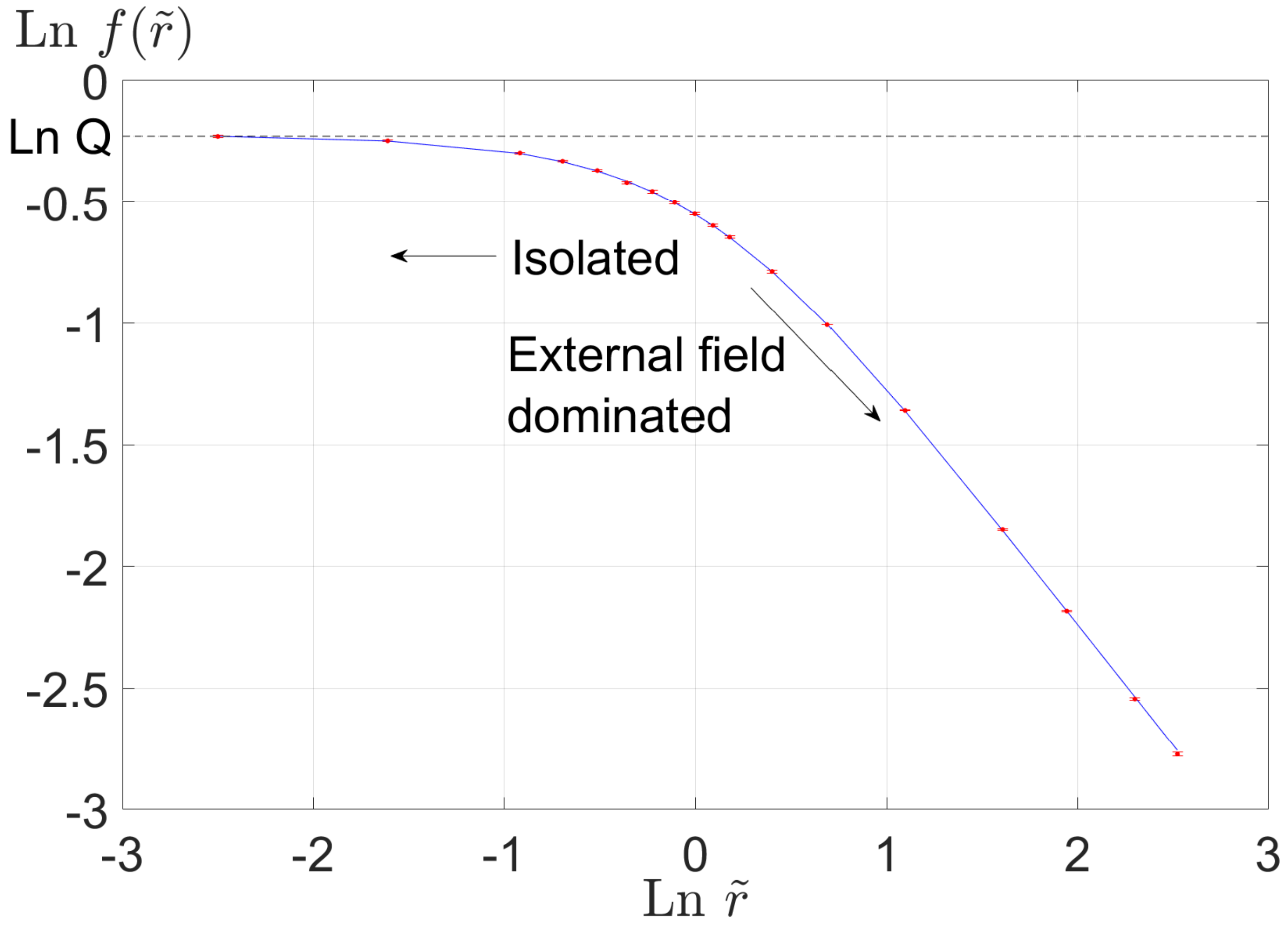}		
\caption{Natural logarithm (Ln) of $f\left( \tilde{r} \right)$ (Equation \ref{g_M}) against $Ln~\tilde{r}$, for two point masses with a 3:7 mass ratio and a total mass $M$ embedded in an external field aligned with the masses. Equivalently, the graph shows the deep-MOND limit force between the masses relative to the naive expectation of Equation \ref{Deep_MOND_limit}, which is off by a constant factor $Q = 0.794$ for ${\tilde{r} \ll 1}$ (weak external field) but becomes much less accurate in a strong external field. Our numerical results (red dots with uncertainties) reveal the precise form of the transition between these analytic limits, with the case of no external field $\left( \tilde{r} = 0 \right)$ arbitrarily plotted at $Ln~\tilde{r} = -2.5$ instead of the correct value of $-\infty$. All forces shown here are determined to within 1\%. Indistinguishable results are obtained if we set $\bm{g}_{_{ext}} \to -\bm{g}_{_{ext}}$. We fit these results using Equation \ref{u} (blue line). The total $\chi^2$ is 16.1 for 18 points and 3 fit parameters, indicating an acceptable fit.}
	\label{QUMOND_residuals}
\end{figure}

We determine $f\left( \tilde{r} \right)$ by performing 18 force calculations which vary $\tilde{r}$ over the range $\left(0 - 12.5 \right)$, thus probing a similar range in $g_{ext}/g_{iso}$. We fit $f\left( \tilde{r} \right)$ to these numerical results using three degrees of freedom corresponding to the maximum fractional deviation of $f$ from our initial guess of $\frac{Q}{\sqrt{1 + \tilde{r}^2}}$, the value of $\tilde{r}$ where this occurs and how quickly $f$ approaches our initial guess on either side. It is evident that we accurately recover the expected behaviour in the asymptotic limits (Figure \ref{QUMOND_residuals}). With 18 force calculations and 3 degrees of freedom in our fitting function, we obtain a total $\chi^2$ of 16.1. Therefore, Equation \ref{u} provides a good fit to our numerical results.

\subsubsection{Going beyond the deep-MOND limit}
\label{Newtonian_limit}

The mutual acceleration between the MW and M31 can be estimated using Equation \ref{u}, a fit to numerical results in the deep-MOND limit. Although this is valid most of the time, it becomes inaccurate when the MW and M31 are near pericentre. At this time, their larger mutual attraction simplifies the problem in another way $-$ it becomes more accurate to neglect the EF on the LG. Thus, we first estimate the Newtonian attraction between the MW and M31 with total mass $M_{_{LG}}$.
\begin{eqnarray}
	g_{_N} ~=~ \frac{GM_{_{LG}}}{{d_{_{M31}}}^2}
\end{eqnarray}

We need to determine whether it is more important to consider the EF or the violation of the deep-MOND limit. We do this by comparing $d_{_{M31}}$ to the geometric means of the MW-M31 separations at which $g_{_N} = a_{_0}$ and where $g_{_N} = g_{_{N,ext}}$, the Newtonian-equivalent EF on the LG (Equation \ref{g_N_ext}). For a weak EF, this distinction of cases can be achieved simply by determining whether $g_{_N} < g_{_{ext}}$. If it is, then the EF is more important. As $g_{_{ext}} \ll a_{_0}$, it is not as important to consider violation of the deep-MOND limit. In this case, we therefore use our previously derived result for $g_{_M}$ including the EF but assuming the deep-MOND limit (Equation \ref{g_M}).

However, if $g_{_N} > g_{_{ext}}$, it is not as important to consider the EF because $g_{_{ext}} \gg g_{_{N,ext}}$ for a weak EF (Equation \ref{g_N_ext_DML}). In this case, we neglect the EF but do not assume the deep-MOND limit. This requires a way to transition between that limit and the high-acceleration Newtonian limit. We do this using the `simple' interpolating function \citep{Famaey_Binney_2005} in another sequence of numerical MW-M31 mutual gravity calculations, each of which involves solving Equation \ref{QUMOND_equation}. This extra step is necessary to better determine the MW-M31 trajectory in the period around the MW-M31 encounter as it is not reasonable to assume the deep-MOND limit then.

As the EF is time-dependent (Section \ref{g_ext_history}), it is difficult to obtain $\ddot{d}_{_{M31}}$ including both the EF and a variable MW-M31 separation. This can only be done by having a two-dimensional grid of numerical simulations as there are two physical scales in the problem $-$ the MOND radius (Equation \ref{Deep_MOND_limit}) and the EF radius governing the transition from isolated to EF-dominated (where $g_{_N} = g_{_{N,ext}}$). However, we feel that it is unnecessary to consider both scales simultaneously because they are well separated, a consequence of the EF being much weaker than $a_{_0}$. Thus, if the EF is important, then by definition we must have that $g \la g_{_{ext}}$. This implies that $g \ll a_{_0}$, making it acceptable to assume the deep-MOND limit. Conversely, a high-acceleration situation will only mildly be affected by a weak EF.

\subsection{The external field on the Local Group}
\label{g_ext_history}

A viable EF history $g_{ext} \left( t \right)$ must explain the present peculiar velocity $\bm{v}_{pec}$ of the LG i.e. its deviation from the Hubble flow.
\begin{eqnarray}
	\bm{v}_{pec} ~\equiv ~ \dot{\bm{r}}_{_{LG}} - H\bm{r}_{_{LG}}
\end{eqnarray}

Here, $\bm{r}_{_{LG}}$ is the position of the LG relative to a co-moving observer. We use an overlying dot to indicate a time derivative (e.g. $\dot{q} \equiv \frac{\partial q}{\partial t}$ for any quantity $q$). The Hubble parameter $H \equiv \frac{\dot{a}}{a}$ where $a \left( t \right)$ is the cosmic scale-factor relative to its present value (i.e. $a \equiv 1$ at the current time $t = t_f$).

For the LG, $\bm{v}_{pec}$ has a value close to 630 km/s towards Galactic co-ordinates $\left(276^\circ, 30^\circ \right)$ when measured with respect to the rest frame of the cosmic microwave background \citep[CMB,][]{Kogut_1993}. For reasons discussed in Section \ref{Section_perturber_trajectories}, we assume that this has always been the EF direction $\bm{\widehat{g}}_{ext} \equiv \bm{{g}}_{ext} \div \left| \bm{{g}}_{ext} \right|$.

Peculiar velocities must be caused by the time-integrated effect of $\bm{g}_{ext} \left( t \right)$. Assuming that the EF arises due to very distant objects that have always been in the same direction as viewed from the LG, we get that
\begin{eqnarray}
	\int_{0}^{t_f} a~g_{ext}~dt ~=~ v_{pec}
	\label{v_pec_constraint}
\end{eqnarray}

The integrating factor $a \left(t \right)$ accounts for the effect of Hubble drag, which tends to reduce peculiar velocities in the absence of further forcing \citep[as derived in e.g.][section 2.1]{Banik_Zhao_2016}.\footnote{$v_{pec} \propto \frac{1}{a \left( t \right)}$ for a test particle in an otherwise homogeneous universe, making $av_{pec}$ constant.} There is no unique way to solve for $g_{ext} \left( t \right)$, but neither is it wholly unconstrained. To make further progress, we assume that $g_{ext}$ is due to a very distant point mass. Thus,
\begin{eqnarray}
	g_{ext} \propto
\left\{
	\begin{array}{ll}
		\frac{1}{a}  & \mbox{if } a \geq a_{peak} \\
		\left( a - a_c \right) & \mbox{if } a_c < a < a_{peak} \\
		0 & \mbox{if } a \leq a_c
	\end{array}
\right.
\end{eqnarray}

It is not possible that $g_{ext} \propto a^{-1}$ indefinitely into the past because the mass must correspond to an inhomogeneity in the Universe which was much smaller in the past. Thus, we make the further assumption that at very early times ($a < a_c = 0.1$), $g_{ext} = 0$. To avoid a discontinuous jump in $g_{ext}$, we assume that $g_{ext} ~\propto ~ \left( a - a_c \right)$ when $a_c < a < a_{peak}$.

To transform $g_{ext} \left( a \right)$ into $g_{ext} \left( t \right)$ and thus solve Equation \ref{v_pec_constraint}, we need a specific form for the background cosmology $a \left( t \right)$. We use a standard flat\footnote{$\Omega_{m,0} + \Omega_{\Lambda, 0} = 1$} dark energy-dominated cosmology with parameters given in Table \ref{Cosmology}.
\begin{eqnarray}
	\frac{\overset{..}{a}}{a} &=& -\frac{4 \rm{\pi} G}{3} \left( \rho_m - 2 \rho_\Lambda \right) \\
	&=& {H_{_0}}^2 \left(- \frac{1}{2} \Omega_{m,0}~a^{-3} + \Omega_{\Lambda, 0}  \right)
	\label{Friedmann_equation}
\end{eqnarray}

Here, $\rho_m$ and $\rho_\Lambda$ are the cosmic mean densities of matter and dark energy, respectively. Currently, these components comprise fractions $\Omega_{m,0}$ and $\Omega_{\Lambda, 0}$ of the cosmic critical density $\frac{3{H_{_0}}^2}{8 \pi G}$. Defining time $t$ to start when ${a = 0}$ and requiring that ${\overset{.}{a} = H_{_0}}$ when ${a = 1}$, we obtain the unique solution
\begin{eqnarray}
	a \left( t \right) &=& {{\left( \frac{{{\Omega }_{m,0}}}{{{\Omega }_{\Lambda ,0}}} \right)}^{\frac{1}{3}}}{{\sinh }^{\frac{2}{3}}}\left( \frac{3}{2}\sqrt{{{\Omega }_{\Lambda ,0}}}{{H}_{0}}t \right)
	\label{Expansion_history}
\end{eqnarray}

\begin{table}
  \centering
		\begin{tabular}{lll}
			\hline
			Name & Meaning and units & Value\\
			\hline
			$\Omega_{m,0}$ & Fraction of critical density in matter & 0.315 \\
			$\Omega_{\Lambda,0}$ & Fraction of critical density in dark energy & 0.685 \\
			$H_{_0}$ & Hubble constant, km/s/Mpc & 67.3 \\
			\hline
		\end{tabular}
	\caption{Cosmological parameters used in our work \citep{Planck_2015}. The critical density of the Universe at any epoch is $\frac{3H^2}{8 \pi G}$. Present values are indicated with $_0$ subscripts. Our MW-M31 trajectory calculations start when $a = 0.05$, which occurs 193 Myr after the Big Bang (Equation \ref{Expansion_history}).}
  \label{Cosmology}
\end{table}

The present values of $H_{_0}$ and $\Omega_{m,0}$ uniquely determine the present age of the Universe $t_f$ via inversion of this equation to solve for when $a = 1$. We also use it to determine when $a = 0.05$, thereby fixing the start time of our simulations.

Equation \ref{Expansion_history} is valid for $\Lambda$CDM but may be inaccurate for a MOND model. However, we expect only small deviations from our adopted cosmology because this works fairly well observationally \citep[e.g.][]{Joudaki_2017}. Moreover, even a purely baryonic universe could have an expansion history very similar to $\Lambda$CDM \citep{Zhao_2007}.

In our model, $g_{ext}$ peaks when $a = a_{peak}$ at a value higher than the present value ($g_{ext, 0}$) by the factor $\frac{1}{a_{peak}}$. We consider models to be realistic if $a_{peak} \approx 0.77$, when the Universe stopped being matter-dominated. This requires $g_{ext, 0} \approx 0.023~a_0$, though we try a range of values to see which works best (Section \ref{Results}).

\subsection{Tides on the Local Group}
\label{Tides}

\subsubsection{Perturber trajectories}
\label{Section_perturber_trajectories}

Any method for determining tides exerted by massive perturbers on the LG requires us to estimate where these perturbers were at earlier times. To do this, we treat the LG as a single point mass at the MW-M31 barycentre. We then use Equation \ref{g_M} to get the relative gravitational acceleration between the LG particle at position $\bm{r}_{_{LG}}$ and a perturber at position $\bm{r}_{_{p}}$, neglecting the other perturbers (i.e. applying the superposition principle to the effects of the perturbers, something we will justify shortly). As the perturbers we consider are all very distant (Table \ref{Perturbers}), we assume that each perturber has always been in the same direction as perceived from the LG barycentre $\bm{r}_{_{LG}}$. The distance $d_{p}$ of the perturber from $\bm{r}_{_{LG}}$ is thus the only dynamical quantity. We vary its initial value at some early time $t_{_i}$ in order to match its presently observed value. This is a basic form of timing argument analysis \citep{Kahn_Woltjer_1959}, a technique we will use more extensively in Section \ref{Timing_argument_NR}.
\begin{eqnarray}
	\label{Cosmic_acceleration_term}
	\ddot{d}_{p} ~&=&~ \frac{\ddot{a}}{a} d_{p} ~+~ g_{_{M}} \\
	\dot{d}_{p} ~&=&~ H d_{p} ~~\text{initially}
\end{eqnarray}

Here, $g_{_{M}}$ is the mutual gravitational acceleration between the perturber and the LG according to Equation \ref{g_M}, which assumes the deep-MOND limit. The relevant mass $M$ is now the total of the LG and the perturber. Their vector separation and mass ratio enter the problem in the same way as we used these quantities for the MW and M31 in Section \ref{Two_body_force_DML}. Because the Universe was expanding nearly homogeneously at very early times \citep{Planck_2015}, we assume that the LG and the perturber both started exactly on the Hubble flow when our simulations start (redshift 19 or ${a = 0.05}$).

In this problem, the deep-MOND limit is a very good approximation due to the large distance between the LG and perturbers like Cen A. It should also be reasonable to neglect the angular momentum of the LG as a whole with respect to these rather distant (${\ga 3}$ Mpc) perturbers.

\begin{table}
 \begin{tabular}{lllll}
	\hline
  Name & $b$ & $l$ & $d_{MW}$ & $M$ \\
  \hline
  Centaurus A & $19.4173^\circ$ & $309.5159^\circ$ & 3.8 & 6 \\
  M81 & $40.9001^\circ$ & $142.0918^\circ$ & 3.6 & 2.2 \\
  IC 342 & $10.5799^\circ$ & $138.1726^\circ$ & 3.45 & 2.2 \\
  \hline
 \end{tabular} 
 \caption{Properties of mass concentrations outside the Local Group which we consider. Sky positions are shown using Galactic co-ordinates. The Galactocentric distances $d_{MW}$ are given in Mpc for Cen A \citep{Harris_2010}, M81 \citep{Gerke_2011} and IC 342 \citep{Wu_2014}. Baryonic masses $M$ in units of $10^{11} M_\odot$ are from \citet{Karachentsev_2005} for Cen A and IC 342 while \citet{Karachentsev_2006} is used for M81.}
 \label{Perturbers}
\end{table}

Although we numerically calibrate the form of $g_{_{M}}$ at a particular mass ratio between the objects, we assume it is valid for the mass ratios between the LG and the perturbers listed in Table \ref{Perturbers}. This is partly because the mass ratios are similar to the 3:7 case we investigated in detail. Moreover, the greater distances imply that the EF is more significant. In the limit where the EF dominates, we do not need a numerical scheme to determine the force between the LG and a perturber. In this limit, the superposition principle also becomes valid as the phantom dark matter density is linear in the matter distribution if it is embedded in a dominant EF \citep[][equation 25]{Banik_2015}. This is because a perturbation theory approach becomes valid, suppressing the non-linearity usually inherent to MOND.

When determining $g_{_M}$, we need to know the EF acting on the LG-perturber system. In principle, some of the EF on the LG could arise due to the perturber itself such that it affects the MW-M31 mutual attraction but should not be considered when determining the LG-perturber attraction. For simplicity, we assume that the EF on the LG mostly arises due to very distant objects rather than the relatively nearby (within 5 Mpc) perturbers we consider (Table \ref{Perturbers}). This can be justified by considering the effects of Cen A, the perturber likely to impose the largest EF on the LG due to its high mass and relative proximity. Supposing that most of the EF on the LG arose due to Cen A, we can consider the LG-Cen A problem in isolation to put an upper limit on their mutual acceleration. In this case, the gravity on the LG is $g_{_{iso}} \approx \frac{\sqrt{GMa_{_0}}}{d_{p}}$ where $M$ is the total (LG $+$ Cen A) mass. For a distance to Cen A of $d_{p} = 3.8$ Mpc, this yields $0.0070 a_{_0}$. As the actual EF on the LG is very likely larger (Section \ref{g_ext_history}), the LG-Cen A problem can't be treated in isolation.

If instead we assume that the EF from more distant objects dominates and set $g_{_{ext}} = 0.02 a_{_0}$, then we get that Cen A pulls on the LG by $g_{_{EFE}} \approx \frac{GMa_{_0}}{{d_{p}}^2 g_{_{ext}}}$ i.e. by only $0.0024 a_{_0}$. As this is much less than $g_{_{ext}}$, it is self-consistent to assume that the EF on the LG mostly arises from objects much further away than the perturbers we consider. However, it is not self-consistent to suppose that most of this EF is provided by nearby objects. As Cen A is the most massive perturber we consider, this justifies our approximation that each LG-perturber problem can be considered embedded in the dominant EF on the LG found in Section \ref{g_ext_history}. As a result, it becomes valid to superpose the forces due to each perturber. For simplicity, we neglect how perturbers affect each other as they are all rather distant and in different sky directions (Table \ref{Perturbers}).


When running our algorithm for the first time, we treat the LG as a single point mass because we do not yet know $d_{_{M31}} \left( t \right)$. The perturber trajectories thus found allow us to obtain the MW-M31 trajectory. We take advantage of this by running the algorithm on a second pass, this time determining what $g_{_M}$ would be if the LG was at the position of the MW. After repeating this calculation for the M31 position, we take a weighted average of these determinations according to the MW:M31 mass ratio.
\begin{samepage}
\begin{equation*}
g_{_{M}} = 
\end{equation*}
\begin{eqnarray}
	\frac{\sqrt{GMa_0}}{d_{p}} f\left( \tilde{r} \right) \left( 1 + \frac{q_{_{MW}}q_{_{M31}}n\left(n\cos^2 \theta + 1\right){d_{_{M31}}}^2}{2{d_{p}}^2}\right)
	\label{Analytic_correction_g_M}
\end{eqnarray}
\begin{eqnarray}
	n &\equiv & 1 ~-~ \frac{\partial Ln~f \left( \tilde{r} \right)}{\partial \tilde{r}} ~~\left( g_{_{M}} \appropto {\tilde{r}}^{-n} \right) \\
	&=& 1 ~+~ \frac{\tilde{r}^2}{\tilde{r}^2 + 1} ~+~ \frac{\left( Ln~ \tilde{r} - 0.42 \right)u}{0.65^2\left(1+u\right)} \\
	\tilde{r} &=& \frac{4d_{p}Qg_{ext}}{\sqrt{GMa_0}\left( 3 + \left( \widehat{\bm{r}} \cdot \bm{\widehat{g}}_{ext}\right)^2 \right)}
\end{eqnarray}
\end{samepage}

Here, $M$ is the total mass of the LG and the perturber while $Q$ is calculated for the ratio between their masses. $q_{_{MW}}$ and $q_{_{M31}}$ retain their previous meanings and values. $u$ is found according to Equation \ref{u} based on the scaled distance $\tilde{r}$. The second order correction to $g_{_{M}}$ found here depends on the angle $\theta$ between the MW$-$M31 line and the direction from the LG barycentre to the perturber.

\begin{figure}
	\centering
		\includegraphics [width = 8.5cm] {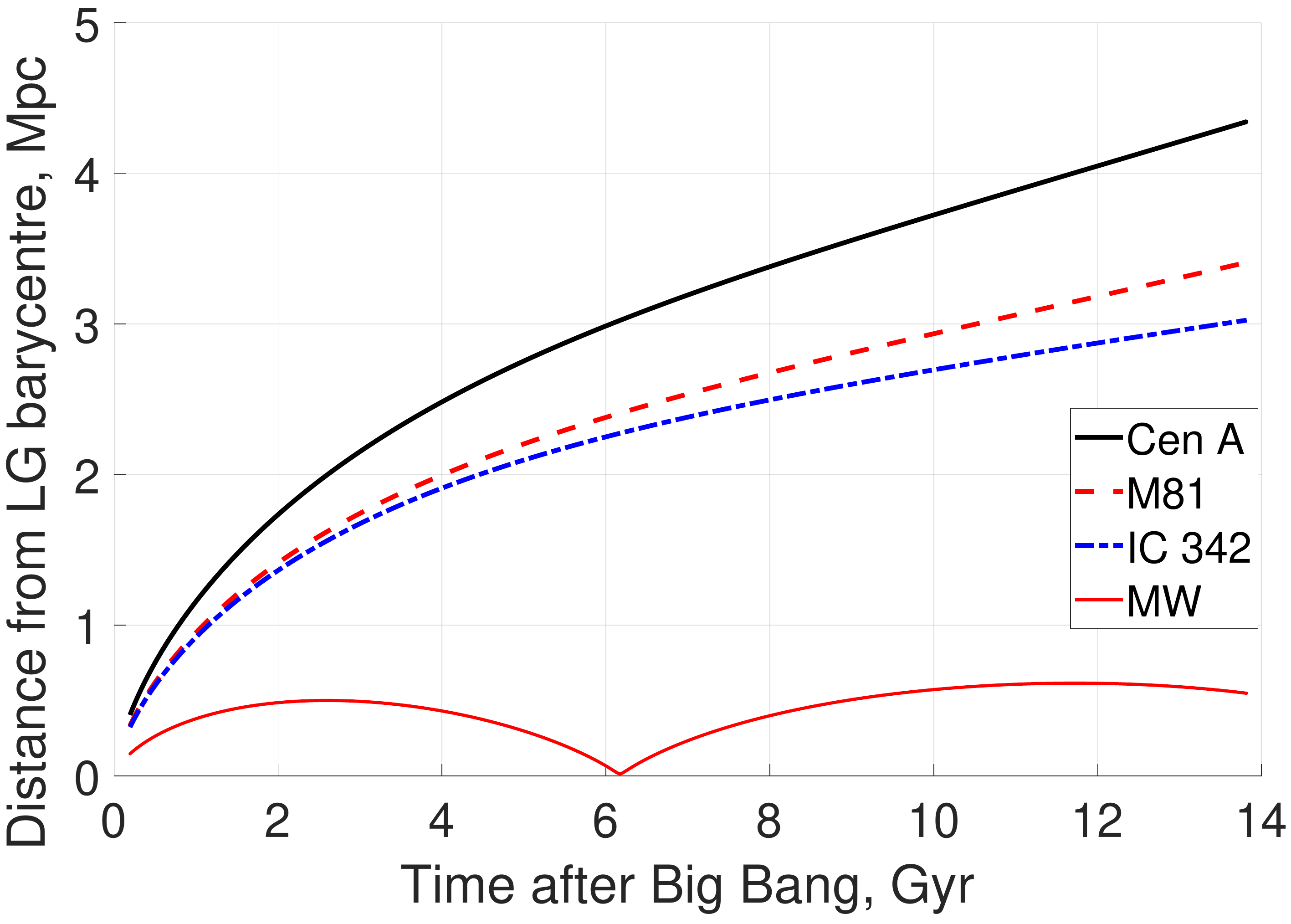}
	\caption{Trajectories of the perturbers in Table \ref{Perturbers}, i.e. their distance from the LG barycentre $\bm{r}_{_{LG}}$ over time. The MW is shown here for comparison (M31 is much closer to $\bm{r}_{_{LG}}$ due to its higher mass).}
	\label{Perturber_trajectories}
\end{figure}

Using the long bracketed term in Equation \ref{Analytic_correction_g_M} improves the accuracy of $\ddot{d}_{_p}$ and allows us to refine their trajectory. We only do this once as it leads to fairly small corrections. The final perturber trajectories $d_{p} \left( t \right)$ are shown in Figure \ref{Perturber_trajectories} for our best-fitting model (Section \ref{Results}).

\subsubsection{Effect of perturbers on the MW-M31 trajectory}
\label{Perturber_effect}

We estimate the effect of tides raised by each perturber on the LG based on how much their mutual attraction alters if the LG particle is placed at the location of the MW instead of M31. This difference in forces is projected onto the MW-M31 line to estimate how each perturber affects the MW-M31 mutual acceleration. For simplicity, we neglect the fact that there would also be a tidal torque on the MW-M31 orbit which would change its angular momentum.

In general, the gravitational force between two objects is not along the line connecting them \citep[e.g.][figure 1]{Banik_2015}. When handling the trajectory of the LG with respect to a perturber, we neglect this because we expect distant perturbers to be on nearly radial Hubble flow orbits. However, when determining the effect of each perturber on the LG, the important consideration is how this differs depending on whether $\bm{r}_{_{LG}} = \bm{r}_{_{MW}}$ or $\bm{r}_{_{LG}} = \bm{r}_{_{M31}}$. Thus, it may be important to consider the tangential force i.e. the force on the LG orthogonal to the LG$-$perturber line.

We find this by determining the gravity due to a point mass embedded in a constant EF. At long range, the EF dominates and so the gravitational field becomes analytic \citep[][equation 37]{Banik_2015}. To get the tangential force at smaller distances, we construct a force library for a point mass embedded in an EF in the deep-MOND limit. In this way, we can find the tangential force due to the mass, something without a Newtonian analogue.

Our results are only valid if one of the objects is much less massive than the other. At present, we have no way of handling intermediate mass ratios as we can only rigorously solve problems that are at least axisymmetric. Thus, we guess how the relative tangential gravity $\bm{g}_{tan}$ between two objects varies with their mass ratio (Appendix \ref{Tangential_force}).

In this way, we estimate the gravity $\bm{g}_{_{LG}}$ exerted by each perturber on the MW and M31, both directly towards the perturber and in the orthogonal direction (within the plane defined by the EF and the LG-perturber line). The difference between the gravity on the MW and M31 is then projected onto the line connecting them so as to estimate $g_{_{tide}}$, our symbol for how much each perturber tidally alters the MW-M31 mutual attraction $\ddot{d}_{_{M31}}$.
\begin{eqnarray}
	\label{MW_M31_line}
	g_{_{tide}} &=& \left( \left. \bm{g}_{_{LG}} \right|_{\bm{r}_{_{LG}} = \bm{r}_{_{M31}}} - \left. \bm{g}_{_{LG}} \right|_{\bm{r}_{_{LG}} = \bm{r}_{_{MW}}} \right) \cdot \widehat{\bm{d}}_{_{M31}} \nonumber \\
	\widehat{\bm{d}}_{_{M31}} &=& \frac{\bm{r}_{_{M31}} - \bm{r}_{_{MW}}}{\left| \bm{r}_{_{M31}} - \bm{r}_{_{MW}} \right|}
\end{eqnarray}

We assume that the EF on the LG arises due to a mass concentration similar to the Great Attractor \citep[GA,][]{Mieske_2005}. Thus, we take the EF to be entirely caused by a point mass currently 84 Mpc away and always in the same direction as viewed from the LG barycentre (Section \ref{g_ext_history}). Because of the large distance to the GA, we assume that its distance from the LG scaled directly with $a \left( t \right)$ i.e. it was always on the Hubble flow. Although this may not have been true at very early times, forces then have only a very small effect on present peculiar velocities due to Hubble drag (Equation \ref{v_pec_constraint}). The actual effect is even more pronounced \citep[][figure 4]{Banik_Zhao_2016}.

At 84 Mpc, the GA is expected to raise only a small tide on the LG. However, for completeness, we include this using the distant tide approximation. In this scheme, we approximate the tidal effect of the EF as an extra contribution of
\begin{eqnarray}
	g_{_{tide,GA}} ~=~ \frac{\left( 2\cos^2 \theta_{ext} - 1 \right)d_{_{M31}}g_{_{ext}}}{d_{_{GA}}}
	\label{Tide_GA}
\end{eqnarray}

Here, $\theta_{ext}$ is the angle between the MW-M31 line and $\bm{\widehat{g}}_{ext}$. The factor of $\left( 2\cos^2 \theta_{ext} - 1 \right)$ is appropriate for an inverse distance gravity law (Equation \ref{Deep_MOND_limit}). This applies to the LG-GA problem due to the very low accelerations involved and our assumption that it can be considered in isolation.


\subsection{The angular momentum barrier}
\label{Angular_momentum_barrier}

The MW-M31 trajectory can't be approximated as radial if they undergo a past close encounter. For simplicity, we assume that their mutual orbital angular momentum ${\bm{h}}_{_{LG}}$ has remained constant since their first turnaround when the Universe was ${\ssim 3}$ Gyr old. This is because ${\bm{h}}_{_{LG}}$ must have been gained by tidal torques, which are most significant when the MW and M31 are furthest apart. Although we expect the MW and M31 to have gone through two apocentres \citep[][figure 4]{Banik_2017_anisotropy}, any large scale structures are much further away in the latter epoch, suggesting that tidal torques then are less significant than at the first MW-M31 turnaround.

Moreover, we will see later that the main constraint on our simulations arises from ${\bm{h}}_{_{LG}}$ around the time of the MW-M31 flyby. Their present proper motion is not constrained all that well and is in any case dependent on other uncertain quantities like the masses of the Large Magellanic Cloud (LMC) and M33 relative to the MW and M31, respectively (Section \ref{Discussion_timing_argument}). Thus, our analysis would not be much affected if a dwarf galaxy recently merged with M31, totally changing $\widehat{\bm{h}}_{_{LG}}$ as ${\bm{h}}_{_{LG}}$ is so small.\footnote{Mergers are possible in MOND as dynamical friction still exists when the baryonic parts of two galaxies overlap. As these are much smaller than galactic dark matter halos in $\Lambda$CDM, mergers would be much rarer in MOND, though its stronger long-range gravity would lead to more gravitational focusing.} In this case, our best-fitting model would have the same ${\bm{h}}_{_{LG}}$, the only difference being that this would be slightly less consistent with the observed proper motion of M31. Thus, it is beyond the scope of our analysis to consider whether ${\bm{h}}_{_{LG}}$ has changed since the MW-M31 flyby. Instead, we consider all possible orientations for $\widehat{\bm{h}}_{_{LG}}$ and a wide range of values for $\left| {\bm{h}}_{_{LG}} \right|$. This ensures that the plausible ranges are covered, thus giving MOND a fair chance of explaining the LG satellite planes.

\subsection{Satisfying the timing argument}
\label{Timing_argument_NR}

Combining the contributions to the MW-M31 mutual attraction $\ddot{d}_{_{M31}}$ from cosmological acceleration ($\frac{\ddot{a}}{a}d_{_{M31}}$), angular momentum (${{{h_{_{LG}}}^2}/{{d_{_{M31}}}^3}}$), self-gravity $g$ (Section \ref{Two_body_force}) and the tidal contribution $g_{_{tide}}$ due to each perturber (Equations \ref{MW_M31_line} and \ref{Tide_GA}), we get that
\begin{eqnarray}
	\label{Combined_MW_M31_attraction}
	\ddot{d}_{_{M31}} ~=~ \frac{\ddot{a}}{a}d_{_{M31}} ~+~ \frac{{h_{_{LG}}}^2}{{d_{_{M31}}}^3} ~+~ g ~+~ \sum_{Perturbers} g_{_{tide}}
\end{eqnarray}

We use this to integrate the MW-M31 trajectory back from present conditions, assuming a particular MW-M31 specific angular momentum $\bm{h}_{_{LG}}$ in the period after their first turnaround but a purely radial orbit before then. As the MW-M31 line $\widehat{\bm{d}}_{_{M31}}$ rotates, it changes orientation relative to the EF and the directions towards nearby perturbers. Thus, the results of our simulations depend on $\widehat{{\bm{h}}}_{_{LG}}$.
\begin{eqnarray}
	\label{MW_M31_line_rotation}
	\dot{\widehat{\bm{d}}}_{_{M31}} ~=~ \frac{\bm{h}_{_{LG}} \times \widehat{\bm{d}}_{_{M31}}}{{{d}_{_{M31}}}^2}
\end{eqnarray}

$\widehat{\bm{d}}_{_{M31}}$ rotates substantially during our simulation, but almost all of this rotation occurs close to the time of the MW-M31 flyby as ${d}_{_{M31}}$ is rather small then. Thus, it is not too important to accurately know $\bm{h}_{_{LG}}$ at other times.

A critical constraint on our problem is provided by the \emph{timing argument} \citep{Kahn_Woltjer_1959}. The near-uniformity of the CMB indicates that the Universe was expanding almost homogeneously at early times \citep{Planck_2015}. Thus, our backwards integration should leave the MW-M31 separation satisfying
\begin{eqnarray}
	\dot{d}_{_{M31}} ~=~ H_{_i}d_{_{M31}} ~~\text{when } t = t_i
	\label{Timing_argument_condition_MW_M31}
\end{eqnarray}

\begin{figure}
	\centering
		\includegraphics [width = 8.5cm] {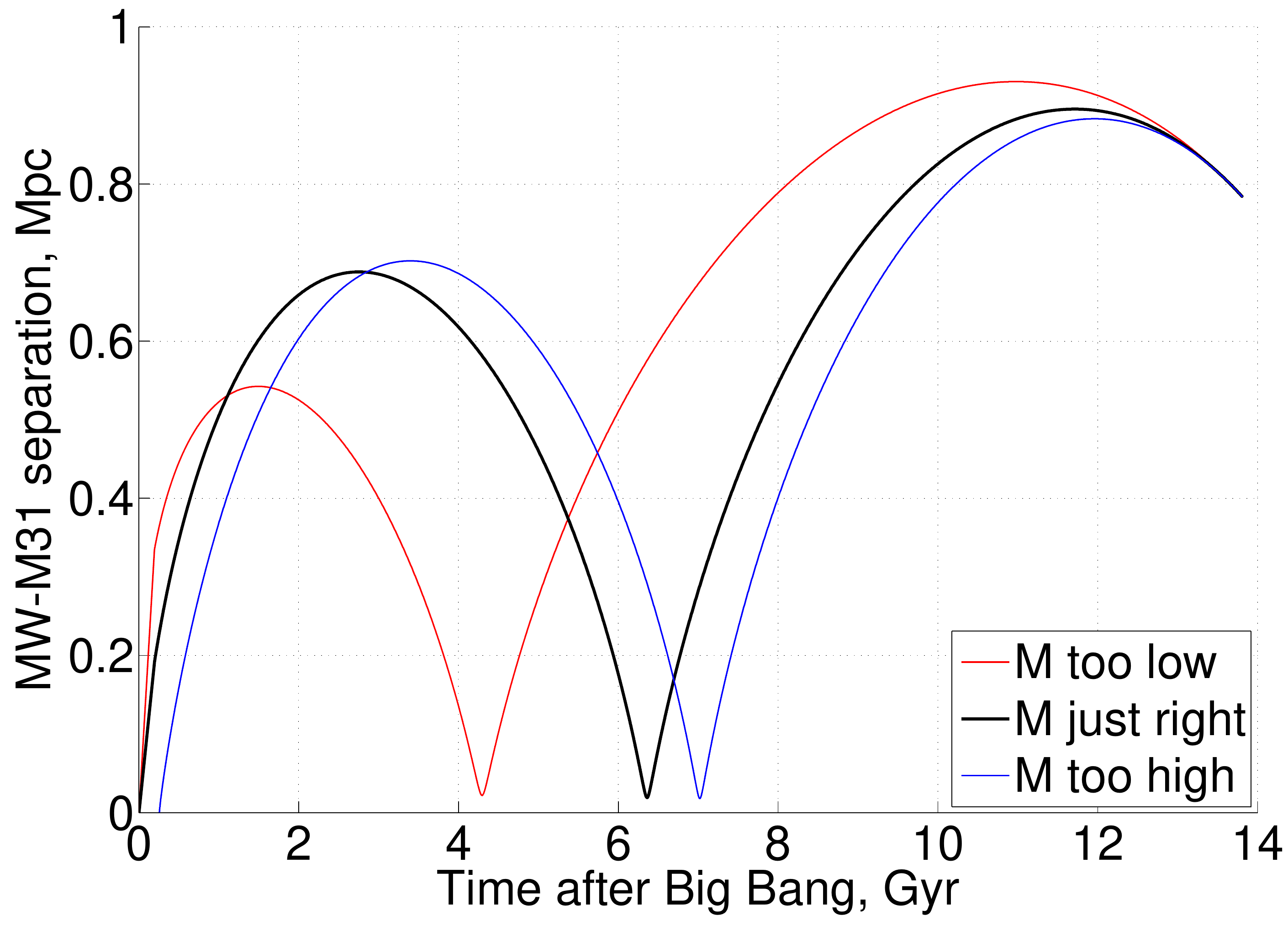}		
	\caption{Three solutions for the MW-M31 separation $d_{_{M31}} \left( t \right)$, all with the same present separation and relative velocity but using different total masses $M$. Only one of the trajectories (thick black) satisfies the timing argument (Equation \ref{Timing_argument_condition_MW_M31}). If $M$ is too low (red curve), $v_{pec} \propto \frac{1}{a} \to +\infty$ while if $M$ is too high (blue curve), $v_{pec} \to -\infty$ as $a \to 0$. In neither case does $d_{_{M31}}$ smoothly approach 0 at early times.}
	\label{Timing_argument_invalid_trajectories}
\end{figure}

Our simulations cover the period from $t = t_{_f}$ (the present epoch) back to $t = t_{_i}$, the limit of our backwards integration. We refer to this as the start time of our simulations, even though in a numerical sense they start at $t_{_f}$. We use Equation \ref{Expansion_history} to determine $t_{_i}$ by solving it for when $a = 0.05$ (redshift 19). This allows us to determine the Hubble parameter $H_{_i}$ at that epoch.

In general, a particular set of model parameters will not satisfy Equation \ref{Timing_argument_condition_MW_M31}. We vary the total LG mass using a Newton-Raphson procedure to ensure this boundary condition is satisfied. This occurs at a particular mass called the timing argument mass. In Figure \ref{Timing_argument_invalid_trajectories}, we show how $d_{_{M31}} \left( t \right)$ is affected by using a lower or higher LG mass. Visually, it is possible to see why these other trajectories are incorrect $-$ loosely speaking, they do not `naturally' (i.e. without the cosmic acceleration term) lead to a MW-M31 pericentre at the time of the Big Bang.

\section{The MW and M31 disks}
\label{MW_M31_disks}

To enable investigation of a wide range of model parameters governing a past flyby encounter between the MW and M31, we model them as point masses surrounded by test particle disks. For simplicity, we neglect the EF but do not assume the deep-MOND limit as this is not accurate close to the MW and M31, regions critical for our analysis.

At any instant in time, the gravitational field $\bm{g}$ governing our test particles is axisymmetric, though the symmetry axis $\widehat{\bm{d}}_{_{M31}}$ changes with time according to Equation \ref{MW_M31_line_rotation}. We can thus obtain $\bm{g}$ by numerically solving Equation \ref{QUMOND_equation} using the direct summation method (Equation \ref{g_direct_sum}). This is straightforward for points on the symmetry axis of the problem. At other points, we reduce the computational burden by utilising a `ring library' which stores $\bm{g}_{_N}$ due to a unit radius ring. Thus, instead of having to further split each ring into a finite number of elements, we can simply interpolate within our densely allocated ring library to find the gravity exerted by the ring at the point where we wish to know its contribution to $\bm{g}$.


As our numerical scheme is inaccurate very close to the MW and M31, we terminate the trajectories of test particles that get within a disk scale length of either galaxy (Table \ref{Disk_parameters}, stellar disk used for MW). This should not affect our results too much because we are mostly interested in their satellites. These lie several disk scale lengths from their hosts, making a point mass approximation reasonably accurate for a theory where galaxies lack extended dark matter halos.

\begin{table}
  \centering
    \begin{tabular}{ll}
		\hline
		Disk scale length of.. & Value (kpc)\\ \hline
		MW stars & 2.15 \\
		MW gas & 7 \\
		M31 & 5.3 \\
		\hline
 \end{tabular}
  \caption{Parameters of the MW and M31 disks. We use kpc for the exponential scale lengths of the MW stellar disk \citep{Bovy_2013}, its gas disk \citep{McMillan_2017} and the M31 disk \citep{Courteau_2011}. We model the MW disk surface density as a double exponential with 0.1764 of its mass in its gas disk and the rest in its stellar disk \citep[][table 1]{Banik_2017_escape}. M31 is also treated as an exponential disk, though its mass is always $\frac{7}{3} \times$ the total MW mass. Disk orientations are discussed in Table \ref{Disk_spin_initial}.}
  \label{Disk_parameters}
\end{table}

Each test particle is initially set up at position $\bm{r}$ on a circular orbit around galaxy $i$ of mass $M_i$ and position $\bm{r}_{_i}$. The circular speed is based on a gravitational field directly towards galaxy $i$ of strength
\begin{eqnarray}
    \label{g}
	g &=& \frac{g_{_N}}{2} ~+~ \sqrt{\left( \frac{g_{_N}}{2} \right)^2 + g_{_N}a_{_0}} \\
	g_{_N} &=& \frac{GM_i}{\left| \bm{r} - \bm{r}_{_i} \right|^2}
\end{eqnarray}

We add the systemic velocity of galaxy $i$ to the circular velocity $v_{_c} = \sqrt{\left| \bm{r} - \bm{r}_{_i} \right|g}$. Our estimates of $v_{_c}$ are valid if we neglect the EF and the other galaxy. For a particle very close to one galaxy, this should be a reasonable assumption. However, starting when $a = 0.05$ leads to a rather small MW-M31 separation, causing some MW particles to actually start closer to M31.

Despite starting quite close, the MW and M31 subsequently spend several Gyr at rather large separations. During this time, we expect their disks to dissipatively settle down as they should have been gas-rich at early times. However, our simulations lack dissipative processes. To help address this issue, we start with completely dynamically cold MW and M31 disks at a fairly large separation and a relative velocity consistent with the MW-M31 trajectory from Section \ref{MW_M31_trajectory}. Thus, we start the test particle integrations 2 Gyr after the Big Bang. This increases the MW-M31 separation from 200 kpc to 700 kpc, thereby preventing any issues due to M31 significantly perturbing MW particles before the flyby. Our results are not much different if we start integrating the test particle trajectories when $a = 0.05$, though we consider such an early start less realistic.

Our grids of initial test particle positions are uniform in distance from the nearest galaxy and azimuthal angle with respect to it (Table \ref{Disk_resolution}). Although our particles do not have any mass, they nonetheless represent a certain amount of mass in the real world. To account for this, we integrate the assumed surface density law (Table \ref{Disk_parameters}) to obtain the mass represented by each particle, later using these masses as statistical weights when analysing our simulations (Section \ref{Comparison_with_observations}). We treat M31 as an exponential disk but use a double exponential model for the MW, whose adopted mass distribution is the same as that used in \citet{Banik_2017_escape} to calculate its rotation and escape velocity curves in MOND.

\begin{table}
  \centering
    \begin{tabular}{lll}
		\hline
		Parameter & Value for MW & Value for M31 \\ \hline
		Radial resolution (kpc) & 0.215 & 0.53 \\
		Azimuthal resolution & $2.8125^\circ$ & $2.8125^\circ$ \\
		Inner radial limit (kpc) & 2.15 & 5.3 \\
		Outer radial limit (kpc) & 107.5 & 106 \\
		\hline
 \end{tabular}
  \caption{Resolutions and extents of the test particle grids used to simulate the MW and M31. We terminate trajectories of particles that get within a disk scale length of each galaxy (using the stellar disk for the MW).}
  \label{Disk_resolution}
\end{table}

Using our numerically determined gravitational field for the MW-M31 trajectory found in Section \ref{MW_M31_trajectory}, we advance test particles using a fourth-order Runge-Kutta method. The timestep is quantised and varied by integer powers of 2 so that there are at least 70 steps per dynamical time, which we define as the ratio of distance to velocity relative to each major LG galaxy. Naturally, we use whichever galaxy returns a smaller dynamical time, regardless of whether this is the galaxy in which the particle started. At early times, cosmic expansion becomes important. Thus, we also require our timestep to be below $\frac{1}{200}$ of the Universe's age. If necessary, our algorithm can reduce the timestep to 213 kyr. This should be sufficient to handle very early times and/or particles quite close to the centre of the MW or M31. For more distant particles, a longer timestep is generally used, though this never exceeds 6.8 Myr.

\subsection{Disk orientations}
\label{Disk_orientations}

Due to the MW-M31 flyby, their disks end up in a slightly different orientation to what they start with. To account for this, we adjust the initial disk spin vectors of the MW and M31 so as to match their presently observed orientations. Because our simulations use test particle disks, it is possible to adjust the disks independently. We use a 2D Newton-Raphson procedure to match the presently observed orientation of each disk.\footnote{We accelerate this process by assuming a change to the initial disk spin of either galaxy causes an equal change to its final disk spin (e.g. tilting the MW $5^\circ$ further north initially causes its present spin to also tilt $5^\circ$ further north). This greatly increases the reliability of our algorithm as well as removing the burden of estimating derivatives through finite differencing, reducing the computational cost ${3\times}$.}

We determine the final MW disk spin vector in our simulations by fitting a plane to the particles within a cylindrical radius of 10.75 kpc and within 21.5 kpc of the disk plane (using the present MW disk orientation $-$ we are only interested in broadly selecting the relevant region). For M31, we use 21.2 kpc radially and 15.9 kpc vertically. Our plane-fitting procedure is similar to that used in section 5.1 of \citet{Banik_2017_anisotropy} but also considers the unequal statistical weights of the particles and applies outlier rejection with a threshold of 2.58$\sigma$, corresponding to the 99\% confidence interval of a Gaussian distribution. As rejecting outliers reduces the root mean square (rms) plane thickness, we repeat the procedure iteratively until the algorithm no longer finds any outliers. Our results do not depend much on the details of the outlier rejection system.

To match the presently observed orientations of the MW and M31 disks, we need their initial orientations to be as shown in Table \ref{Disk_spin_initial}. Fortunately, both disks precess by only a few degrees. Thus, we are always sure which hemisphere contains the total angular momentum of each disk. This allows us to get by with a plane-fitting algorithm, though we do also check the angular momenta of the particles to which we are fitting a plane.

We assume that the MW and M31 have a constant mass throughout the entire simulation. We will see later that only a small fraction of their mass ends up in their satellite regions, so the galaxies lose rather little mass in our best-fitting model (Section \ref{SP_mass}). In general, including a small amount of mass loss at the time of the MW-M31 flyby further reduces their total timing argument mass while pushing their flyby further into the past. This makes the results slightly more consistent with observations. However, we neglect this effect to avoid having too many model parameters and because there would also be some accretion onto the MW and M31.


\begin{table}
 \begin{tabular}{lll}
	\hline
	Initial disk & Initial & Final \\
  spin vector of.. & direction & direction \\ \hline
	Milky Way (MW) & \multirow{2}{*}{$\begin{bmatrix} 86.61^\circ \\ -87.48^\circ \end{bmatrix}$} & \multirow{2}{*}{$\begin{bmatrix} 0^\circ \\ -90^\circ \end{bmatrix}$} \\
	& & \\ [5pt]
	Andromeda (M31) & \multirow{2}{*}{$\begin{bmatrix} 243.35^\circ \\ -23.61^\circ \end{bmatrix}$} & \multirow{2}{*}{$\begin{bmatrix} 238.65^\circ \\ -26.89^\circ \end{bmatrix}$} \\
	& & \\
  \hline
 \end{tabular}
 \caption{Initial spin vectors of the MW and M31 disks in Galactic co-ordinates. These differ slightly from their presently observed orientations (last column) because the disks precess during their close encounter. The present M31 disk orientation is obtained from \citet[][table 1]{Banik_2017_anisotropy} and changes by $5.37^\circ$ while that of the MW changes by only $2.52^\circ$.}
 \label{Disk_spin_initial}
\end{table}

\section{Comparison with observations}
\label{Comparison_with_observations}


We compare each model with observations using a $\chi^2$ statistic that includes contributions for the satellite plane orientations $\chi^2_{_{SP,i}}$ (Section \ref{Satellite_plane_angular_momenta}) and counter-rotating fractions $\chi^2_{_{CR,i}}$ (Section \ref{Counter_rotating_fraction}) as well as $\chi^2_{_{TD}}$ for how the MW disk is heated by the flyby (Section \ref{MW_thick_disk_z_rms}) and $\chi^2_{_{HVG}}$ (Section \ref{HVG_alignment}) for the alignment of the MW-M31 orbital plane with the plane of HVGs identified by \citet{Banik_2017_anisotropy}.
\begin{eqnarray}
	\chi^2 = \sum_{i = \text{MW, M31}} \left( \chi^2_{_{SP,i}} + \chi^2_{_{CR,i}} \right) ~+~ \chi^2_{_{TD}} ~+~ \chi^2_{_{HVG}}
	\label{chi_sq}
\end{eqnarray}

Although the flyby must have affected both the MW and M31 disks, this is much more difficult to test for M31 than for the MW because M31 is an external galaxy viewed nearly face-on \citep{Chemin_2009}. The accretion history of M31 is also more complicated and includes recent interactions with other galaxies, significantly complicating any analysis \citep{Sadoun_2014}. For example, the M31 disk is likely being perturbed by the close proximity to M32 \citep[][table 4]{Weisz_2014}. In future, accurate ages and kinematics of M31 stars could help distinguish the effects of more recent interactions. The work of \citet{Dolphin_2017} may be an important step in this regard.

\subsection{Satellite plane angular momenta}
\label{Satellite_plane_angular_momenta}

The main objective of this investigation is to compare our simulations with observations of the MW and M31 satellite planes. To facilitate this comparison, we first need to select which simulated particles correspond to satellites of e.g. the MW. We define a `satellite region' around it by considering MW particles within an outer cut-off radius ${R_{_{max}}}$ of the MW as this encompasses the region usually considered in the satellite plane problem. In addition, we discard particles too close to its disk by excluding the region with $r_{plane} < r_{{min}}$ and $\left| z \right| < z_{{min}}$ in cylindrical polar co-ordinates $\left(r_{plane}, \phi, z \right)$ centred on the MW with axis along the observed MW disk spin vector. Our adopted values for these parameters are given in Table \ref{Satellite_plane_filter}. We exclude everything within 15 disk scale lengths of its disk plane out to a large distance, completely removing the MW disk remnant. However, most of its simulated satellite plane should still remain outside our excluded region if it is similar to the corresponding observed structure. This is because the MW satellite plane is much larger than $z_{_{min}}$ and almost polar relative to the Galactic disk \citep[][figure 1]{Pawlowski_2018}. For M31, we use a similar procedure to the MW, though with a slightly higher $z_{_{min}}$ to compensate for its larger disk.

\begin{table}
 \begin{tabular}{lll}
	\hline
	Quantity & Value for MW & Value for M31 \\ \hline
  $R_{_{max}}$ & 250 kpc & 250 kpc \\
  $r_{_{min}}$ & 172 kpc & 250 kpc \\
  $z_{_{min}}$ & 32.25 kpc & 53 kpc \\
  \hline
 \end{tabular}
 \caption{Our adopted filters for selecting simulation particles corresponding to the satellite galaxy regions around the MW and M31. Variables are defined in Section \ref{Satellite_plane_angular_momenta}. Although we could have used $r_{_{min}} = R_{_{max}}$ for both galaxies, in practice the shorter scale length of the MW meant a lower $r_{_{min}}$ was sufficient.}
 \label{Satellite_plane_filter}
\end{table}



For simplicity, we neglect particles starting in one galaxy and ending up within the satellite region of the other galaxy. There is little tendency for particles to end up very far from their original galaxy (Figures \ref{MW_radial_distribution} and \ref{M31_radial_distribution}).

After selecting `satellite' particles in this way, we create a 2D histogram of the orbital poles $\widehat{\bm{h}} \equiv \frac{\bm{h}}{\left| \bm{h} \right|}$ of these particles relative to their nearest galaxy, where
\begin{eqnarray}
    \bm{h} ~=~ \left(\bm{r} - \bm{r}_{_i} \right) \times \left(\bm{v} - \bm{v}_{_i} \right)
    \label{h_rel}
\end{eqnarray}

Here, the nearest galaxy to the particle is galaxy $i$ with position $\bm{r}_{_i}$ and velocity $\bm{v}_{_i}$. We assign the statistical weight of the particle to the bin in Galactic latitude and longitude defined by its orbital pole $\widehat{\bm{h}}$. This results in a 2D histogram for the MW and another one for M31. Each of these histograms almost always has a clear peak indicating a preferred orbital pole, though its direction varies with model parameters and between the galaxies.

Based on the bin that is assigned the most mass, we identify the peak $\widehat{\bm{h}}_{_{peak}}$ in each orbital pole histogram. We then quantify the angular dispersion of this peak, considering only particles with $\widehat{\bm{h}}$ in the same hemisphere as the peak i.e. with $\widehat{\bm{h}} \cdot \widehat{\bm{h}}_{_{peak}} > 0$. We also quantify the angular dispersion of the remaining particles' orbital poles about the direction $-\widehat{\bm{h}}_{_{peak}}$. This allows for the possibility of a counter-rotating peak i.e. a secondary peak in the orbital pole distribution at $-\widehat{\bm{h}}_{_{peak}}$.

To make our results more robust, we apply a $\sigma$-clipping procedure to reject outliers from the primary ($\equiv$ co-rotating) and counter-rotating peaks. The threshold we use is 2.58$\sigma$, corresponding to the 99\% confidence interval of a Gaussian distribution.\footnote{If this threshold angle exceeds ${90^\circ}$, we do not perform outlier rejection $-$ this often happens for the MW.} The angular dispersion of each peak is defined as the weighted rms angle to $\pm \widehat{\bm{h}}_{_{peak}}$ of particles whose orbital pole lies in the appropriate hemisphere. Considering the statistical weight $W_j$ of each particle $j$, the angular dispersion of the counter-rotating peak is thus
\begin{eqnarray}
	\sigma_{_-} ~=~ \sqrt{\frac{\sum_j W_j \left[ \overbrace{\cos^{-1} \left( -\widehat{\bm{h}}_{_j} \cdot \widehat{\bm{h}}_{_{peak}} \right)}^{<90^\circ}\right]^2}{\sum_j W_j}}
\end{eqnarray}

After this outlier rejection stage, we obtain the total angular momentum $\bm{h}_{_{tot, \pm}}$ of the remaining particles in each peak, taking care of their statistical weights and using $+$ to denote the co-rotating side ($-$ for the counter-rotating side). We use this to refine our estimate of $\widehat{\bm{h}}_{_{peak}}$, considering particles corresponding to both orbital pole peaks.
\begin{eqnarray}
    \widehat{\bm{h}}_{_{peak}} ~\propto ~ \bm{h}_{_{tot, +}} - \bm{h}_{_{tot, -}}
\end{eqnarray}


This leads to a revised measure of each peak's angular dispersion. Using our newly estimated ${\widehat{\bm{h}}_{_{peak}}}$ and ${\sigma_{_\pm}}$, we repeat the $\sigma$-clipping procedure and thus update the peak locations and dispersions. This is continued iteratively until the algorithm converges (finds no more outliers), which usually takes ${\ssim 10}$ steps. In this way, we obtain the directions of the co-rotating and counter-rotating peaks as well as their angular dispersions. We use ${\widehat{\bm{h}}_{_{peak}}}$ to increment the $\chi^2$ of our model by
\begin{eqnarray}
    \Delta \chi^2 ~&=&~ \sum_{i = {MW,M31}} \chi^2_{_{SP,i}} \\
		\chi^2_{_{SP,i}} ~&=&~ \left[ \frac{\cos^{-1} \left( \widehat{\bm{h}}_{_{peak}} \cdot \widehat{\bm{h}}_{_{SP,i}} \right)}{\sigma_{_{SP,i}}} \right]^2
\end{eqnarray}

\begin{table}
 \begin{tabular}{lll}
	\hline
	Satellite plane spin vector of.. & Direction & Error \\ \hline
  Milky Way (MW) & $\left( 176.4^\circ, -15.0^\circ \right)$ & $25^\circ$ \\
  Andromeda (M31) & $\left( 206.2^\circ, 7.8^\circ \right)$ & $15^\circ$ \\ [5pt]
  \hline
 \end{tabular}
 \caption{Observed directions of the MW and M31 satellite plane angular momenta, shown in Galactic co-ordinates. Both point roughly in the direction of the Galactic anti-centre. The MW satellite plane orientation is from \citet[][section 3]{Kroupa_2013} while that of M31 is from their section 4. The error budgets used in our analysis (last column) allow for both observational and modelling uncertainties.}
 \label{Directions_table}
\end{table}

Here, $\widehat{\bm{h}}_{_{SP,i}}$ is the observed satellite plane spin vector of galaxy $i$ (Table \ref{Directions_table}). This table also shows $\sigma_{_{SP,i}}$, how well we expect our model to reproduce the observed satellite plane orbital pole of galaxy $i$. We include an allowance for both observational and modelling uncertainties. Compared to the case of the MW, the M31 satellite plane is observed to be thinner \citep[][tables 2 and 4]{Pawlowski_2014} and also has a smaller angular spread in most of our simulations. Thus, we allow a smaller angular uncertainty for the M31 satellite plane (i.e. $\sigma_{_{SP,M31}} < \sigma_{_{SP,MW}}$).

\subsection{Fraction of tidal debris that counter-rotate}
\label{Counter_rotating_fraction}

The method described in Section \ref{Satellite_plane_angular_momenta} yields the amount of material in the co-rotating and counter-rotating peaks, which we call $W_+$ and $W_-$, respectively. This can be used to obtain a counter-rotating fraction
\begin{eqnarray}
	f_{counter} ~\equiv ~ \frac{W_-}{W_+ ~+~ W_-}
\end{eqnarray}

The values of $f_{counter}$ can be compared with the proportion of material in each satellite plane on a counter-rotating orbit. To penalise models that have a different $f_{counter}$ to what we expect, we add an appropriate contribution to the total $\chi^2$ of the model.
\begin{eqnarray}
	\Delta \chi^2 ~&=&~ \sum_{i = {MW,M31}} \chi^2_{_{CR,i}} \\
	\chi^2_{_{CR,i}} ~&=&~ \left( \frac{f_{counter,i} - f_{counter,obs,i}}{\sigma_{f_{counter,i}}} \right)^2
\end{eqnarray}

Around M31, only ${\ssim \frac{1}{2}}$ of its satellites lie in its satellite plane \citep{Ibata_2013}. 13 of these 15 planar satellites share the same radial velocity trend (i.e. those on different sides of M31 have a different sign of radial velocity, once the systemic motion of M31 is subtracted). The remaining 2 satellites could well be interlopers or genuine counter-rotators. As there is no clear evidence for the latter, we assume that $f_{counter,obs,M31} = 0$ but allow an uncertainty of $\sigma_{f_{counter,M31}} = 0.05$. This could be refined with proper motion measurements of M31 satellites, especially those which break the radial velocity trend (And XIII and And XXVII).

For the MW, the observational picture is clearer due to the availability of proper motions for most of its classical satellites \citep[][and references therein]{Pawlowski_2013}. Sculptor is counter-rotating within the MW satellite plane \citep{Sohn_2017}. As it is one object out of ${\approx 10}$ Galactic satellites with known proper motions, we require our models to yield $f_{counter,obs,MW} = 0.10 \pm 0.05$.

\subsection{Milky Way thick disk}
\label{MW_thick_disk_z_rms}

\subsubsection{Age}

Our simulations of a past MW-M31 flyby must involve significant tidal torques on their disks in order to explain their significant misalignments with their satellite planes ($\ssim 50^\circ$ for M31 and $\ssim 75^\circ$ for the MW). As a result, the MW and M31 disks would have been dynamically heated. It is difficult to test this for M31 because it is an external galaxy viewed nearly face-on \citep{Chemin_2009}. M31 has also experienced more recent interactions with other galaxies \citep{Sadoun_2014}, significantly complicating any analysis.

The merger history of the MW has been more quiescent \citep{Wyse_2009}. It is also much easier to get detailed ages, chemistry and kinematics for MW stars due to their proximity. Thus, one might expect our Galaxy to retain some evidence of a past interaction with M31. We suggest that its thick disk was formed in this event.

Originally, the Galactic thick disk was identified based on the density of stars having a double-exponential law with distance out of the Galactic disk plane \citep{Gilmore_1983}. The lower mass thick disk component may well have arisen from dynamical heating of the MW thin disk during its close encounter with M31 \citep{Zhao_2013}. Indeed, the thick disk seems to have formed fairly rapidly from its thin disk \citep{Hayden_2015}. The associated burst of star formation \citep[][figure 2]{Snaith_2014} could have further thickened the MW disk \citep{Kroupa_2002}.


If this scenario is correct, then there ought to be a sudden jump in the age-velocity dispersion relation of nearby MW stars. This is in fact observed ${9 \pm 1}$ Gyr ago \citep{Quillen_2001}, so the MW-M31 flyby almost certainly occurred then if at all. We consider this constraint on their trajectory by adding a contribution to the model $\chi^2$ of
\begin{eqnarray}
    \chi^2_{_{TD, age}} ~=~ \left(\frac{t_f - t_{flyby} - 9 Gyr}{1 Gyr} \right)^2
\end{eqnarray}

\subsubsection{Thickness}
\label{MW_thick_disk_height}

In addition to its age, another important property of the thick disk is its thickness or scale height. As our simulation lacks dissipation, it is most analogous to that part of the MW stellar disk which existed before the flyby. The gas disk would likely have settled down after the flyby, leading to a thin component to the MW stellar disk that would not be accurately included in our model. Thus, we expect to roughly match the observed scale height of the old MW thick disk. This was recently measured by \citet{Ferguson_2017}, who fit its vertical density profile using
\begin{eqnarray}
    \rho \left( z \right) ~\propto~ \sech^2 \left( \frac{z}{2H_{_2}}\right)
\end{eqnarray}

Based on their figure 9, we take a value of $H_{_2} = 0.7 \pm 0.1$ kpc, though we inflate the uncertainty to 0.15 kpc for comparison with our simulations as these also have some uncertainty. This is partly because they lack the self-gravity required to obtain a $\sech^2$ profile. To facilitate a comparison, we obtain the rms thickness of the observed and simulated thick disks.
\begin{eqnarray}
	z_{_{rms,obs}} ~=~ \frac{\pi}{\sqrt{3}}H_{_2} 
\end{eqnarray}


The observed thickness $z_{_{rms,obs}}$ must be at least partly due to processes internal to the MW such as interactions with stars, molecular clouds and spiral arms \citep[e.g.][]{Carlberg_1985, Kroupa_2002}. Our simulations lack such processes, so any disk heating arises almost entirely due to the effect of M31. To account for this, we need to estimate how thick the Galactic disk might have been without such secular heating effects. We estimate this using \citet{Quillen_2001}, who showed that very young stars have a rather small vertical velocity dispersion $\sigma_{_z}$. This is presumably because stars form out of dissipational gas which has settled itself into a rather thin disk \citep[][figure 6]{Nakanishi_2016}. Older stars have higher $\sigma_{_z}$, but this quickly saturates at ${\ssim 20}$ km/s. We assume that this is what $\sigma_{_z}$ would have been without the M31 flyby. However, the thick disk has $\sigma_{_z} \approx 40$ km/s \citep[][figure 1]{Minchev_2014}. This suggests that internal processes were responsible for ${\ssim \frac{1}{4}}$ of the disk heating required to explain the Galactic thick disk, with the rest arising from the close passage of M31 or the massive MW star clusters which formed in the associated starburst. Thus, our models should yield a thick disk which is thinner than observed. To account for this, we only require our models to reach
\begin{eqnarray}
    z_{_{TD}} ~=~ \frac{3}{4} z_{_{rms,obs}}
		\label{Secular_heating_adjustment}
\end{eqnarray}

This can be compared with the simulated value of $z_{_{rms}}$. Following \citet[][figure 12]{Ferguson_2017}, we obtain this based on particles having a cylindrical Galactocentric radius within 1 kpc of the Solar Circle, which we take to be 8.2 kpc from the Galactic centre \citep{McMillan_2017}. We neglect effects due to non-axisymmetry, thus using the entire annulus corresponding to Galactocentric cylindrical radii of ${7.2-9.2}$ kpc. This maximises the number of particles used to determine the $z_{_{rms}}$ of our simulated thick disk.

Vertically, we allow particles up to 21.5 kpc either side of the MW disk, ensuring this is properly included. We then find $z_{_{rms}}$ using a plane fitting procedure similar to that described in section 5.1 of \citet{Banik_2017_anisotropy} augmented by $\sigma$-clipping with a threshold of $2.58 \sigma$, corresponding to the 99\% confidence interval of a Gaussian distribution. We use as many $\sigma$-clipping stages as we need for the algorithm to converge, without reintroducing particles excluded at a previous stage. Usually, this process requires ${\ssim 10}$ iterations. Once it converges, we can determine $\chi^2_{_{TD, height}}$ based on the appropriately scaled uncertainty. This allows us to add the thick disk contribution to the total $\chi^2$ of each model.
\begin{eqnarray}
    \chi^2_{_{TD}} ~=~ \chi^2_{_{TD, age}} + \chi^2_{_{TD, height}}
\end{eqnarray}

\subsection{Alignment with the high-velocity galaxy plane}
\label{HVG_alignment}

So far, we have focused on how the MW and M31 may have affected each other. Their high relative velocity around the time of their flyby (668 km/s, Figure \ref{MW_M31_separation_SP}) would have enabled them to gravitationally slingshot any passing LG dwarf galaxy out at a high speed. \citet{Pawlowski_McGaugh_2014} previously identified several LG dwarfs whose radial velocities are indeed much higher than would be expected in $\Lambda$CDM. Similar results were subsequently obtained using more detailed dynamical models of the LG \citep{Banik_Zhao_2016, Banik_Zhao_2017, Peebles_2017}. These HVGs mostly lie very close to a well-defined plane which goes near both the MW and M31 \citep{Banik_2017_anisotropy}. Section 3 of that work shows how this naturally arises in a MOND simulation of the LG because the MW and M31 would be most efficient at scattering LG dwarfs parallel to their velocity. We therefore expect the HVG plane to coincide with the MW-M31 orbital plane.

Unfortunately, there are some serious difficulties with testing if this is indeed the case. The main problem is the very small number of independent HVGs and thus the uncertain orientation of what plane they prefer. Most of the HVGs we identified belong to the NGC 3109 association. In $\Lambda$CDM, it is unlikely to have been gravitationally bound as this would require a rather large and massive dark matter halo \citep{Bellazzini_2013, Kourkchi_2017}. Given the high radial velocities of these galaxies, they need to have been rather close to the MW at some point in the past. Dynamical friction with its dark matter halo probably rules out this scenario, which is why we treated these galaxies as independent HVGs in \citet{Banik_2017_anisotropy}.

However, this consideration would not arise in MOND as dynamical friction would only operate on baryons. Thus, the NGC 3109 association might have been gravitationally bound and still have passed rather close to the MW without merging. Indeed, this seems rather likely given the filamentary nature of the association \citep{Bellazzini_2013}. It may be an unbound structure somewhat similar to a tidal stream, defined by dwarf galaxies instead of stars.

In this scenario, we would have only a very small number of independent HVGs with which to reliably determine the HVG plane normal. The angular dispersion of HVGs with respect to the MW-M31 orbital plane would be $\sim \cos^{-1} 0.9 \approx 25^\circ$ \citep[][figure 7]{Banik_2017_anisotropy}.\footnote{We use $\sim$ for order of magnitude estimates and $\approx$ when making more precise statements.} An uncertainty of this order seems reasonable given the $16^\circ$ angle of the MW-M31 line out of the HVG plane. As there are not very many clearly independent HVGs and our simulations also have some uncertainty, we allow an angular mismatch of $\sigma_{_{HVG}} = 25^\circ$ between the HVG plane normal $\widehat{\bm{n}}_{_{HVG}}$ and the MW-M31 orbital pole $\widehat{\bm{h}}_{_{LG}}$. Also allowing for the unknown sense in which the HVGs rotate within their preferred plane, we thus add an extra contribution to the model $\chi^2$ of
\begin{eqnarray}
	\chi^2_{_{HVG}} ~=~ \left( \frac{\cos^{-1} \left| \widehat{\bm{n}}_{_{HVG}} \cdot \widehat{\bm{h}}_{_{LG}} \right|}{\sigma_{_{HVG}}}\right)^2
	\label{chi_sq_HVG}
\end{eqnarray}

\section{Results and discussion}
\label{Results}

To see if MOND can explain the observed properties of the LG, we run a grid of models in the present values of the EF on the LG, MW-M31 orbital pole and Galactocentric tangential speed of M31. We then obtain a ${\chi^2}$ statistic for each model (Equation \ref{chi_sq}) and thereby select the best-fitting model. We improve its accuracy by using the Newton-Raphson method to adjust the initial orientations of the MW and M31 disks so as to best match their presently observed orientations (Section \ref{Disk_orientations}). We then use this in a new grid of models, focusing on the most promising region of parameter space in the previous grid. After alternating between the Newton-Raphson and grid searches a few times, we improve the resolution in parameter space to several times better than the expected modelling and observational uncertainties. Our best-fitting model is compared with LG observations in Table \ref{Best_model_results} and discussed next.

\begin{table}
 \begin{tabular}{lllll}
\hline
  Quantity & Observed & $\sigma$ & Model & $\chi^2$\\ 
  \hline
  Time since flyby, Gyr & 9 & 1 & 7.65 & 1.82\\ [5pt]
  Milky Way (MW) & \multirow{2}{*}{$\begin{bmatrix} 176.4^\circ \\ -15.0^\circ \end{bmatrix}$} & $25^\circ$ & \multirow{2}{*}{$\begin{bmatrix} 180.6^\circ \\ -42.8^\circ \end{bmatrix}$} & 1.26\\
  satellite plane  & & & &\\ [5pt]
  Andromeda (M31) & \multirow{2}{*}{$\begin{bmatrix} 206.2^\circ \\ 7.8^\circ \end{bmatrix}$} & $15^\circ$ & \multirow{2}{*}{$\begin{bmatrix} 212.4^\circ \\ 0.14^\circ \end{bmatrix}$} & 0.43\\
  satellite plane  & & & &\\ [5pt]
  MW-M31 orbital pole & \multirow{2}{*}{$\begin{bmatrix} 26.6^\circ \\ 31.8^\circ \end{bmatrix}$} & $25^\circ$ & \multirow{2}{*}{$\begin{bmatrix} 24.6^\circ \\ -16.3^\circ \end{bmatrix}$} & 3.70\\
  (expect HVG plane) & & & & \\ [5pt]
  Counter-rotating & 0 & 0.05 & 0 & 0\\
	fraction for M31 & & & &\\
  Counter-rotating & 0.1 & 0.05 & 0.22 & 5.84 \\
	fraction for MW & & & &\\
	MW thick disk $z_{_{rms}}$ & 0.95 & 0.20 & 0.75 & 1.03 \\
	at Solar Circle, kpc & & & &\\
  \hline
 \end{tabular} 
 \caption{Comparison of our best-fitting model with observational constraints on the LG. The uncertainties $\sigma$ are shown in the central column e.g. $\sigma_{_{HVG}} = 25^\circ$. 0.0045 of the mass in the MW ends up in its satellite plane, including both the co-rotating and counter-rotating peaks in the orbital pole distribution of material in its satellite region (Table \ref{Satellite_plane_filter}). For M31, the equivalent quantity is ${6.5 \times 10^{-6}}$. As discussed in Section \ref{HVG_alignment}, we expect the MW-M31 orbital plane to align with the plane of high-velocity galaxies discovered by \citet{Banik_2017_anisotropy}.}
 \label{Best_model_results}
\end{table}

The total $\chi^2$ of our best model is 14.08. There are three model parameters and 7 contributions to $\chi^2$. However, the orientation of a plane is a 2D concept such that the satellite plane and HVG constraints should really count for 6 constraints rather than 3. In this case, there are 10 constraints $-$ 3 model parameters $=$ 7 `degrees of freedom', for which the chance of an even higher $\chi^2$ is $P = 4.98\%$ based on the chi2cdf function of \textsc{matlab}$^\text{\textregistered}$. However, even if we count the satellite and HVG planes as one constraint each, this still leaves 4 degrees of freedom for which $P = 0.7\%$. Given the crudeness of our model, we believe that it matches a wide array of LG observations well enough to form the basis for more detailed modelling \citep[e.g.][]{Bilek_2017}.

To get an idea of how the different constraints affect the parameters of our best-fitting model, we repeat our $\chi^2$ analysis neglecting one contribution to $\chi^2$ at a time. The best-fitting model is unaffected by the constraints from the thick disk age, the thickness of the MW disk, the orientation of the HVG plane found by \citet{Banik_2017_anisotropy} and the fraction of counter-rotating material around M31 (presumably because our models had no such material). Neglecting the constraint from the MW satellite plane orientation, the preferred model has a slightly smaller closest approach distance of 17.90 kpc. Ignoring the counter-rotating fraction of tidal debris around the MW, the best-fit orbital pole tilts $5^\circ$ further south and the model prefers a weaker present $g_{_{ext}}$ of 0.02$a_{_0}$. Perhaps the most important single constraint is the accurately known M31 satellite plane orientation. Without this information, the best-fitting model prefers an orbital pole $15^\circ$ further north and a smaller closest approach distance of 17.90 kpc. Even these changes are very small, suggesting that no individual observational constraint greatly affects our analysis.

\subsection{Implied MW-M31 trajectory}
\label{Discussion_timing_argument}

\subsubsection{The timing argument}

\begin{figure}
	\centering
		\includegraphics [width = 8.5cm] {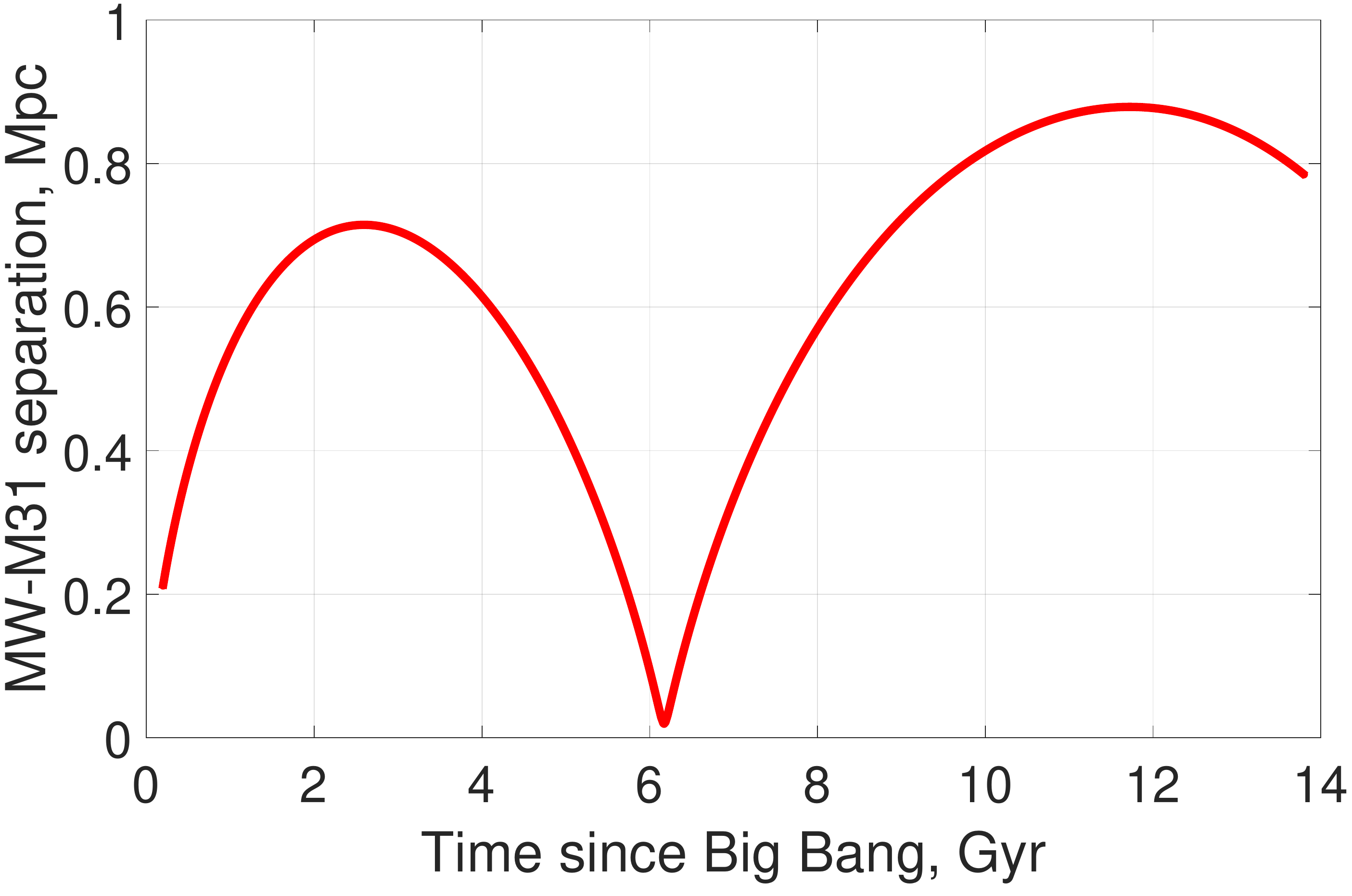}		
		\includegraphics [width = 8.5cm] {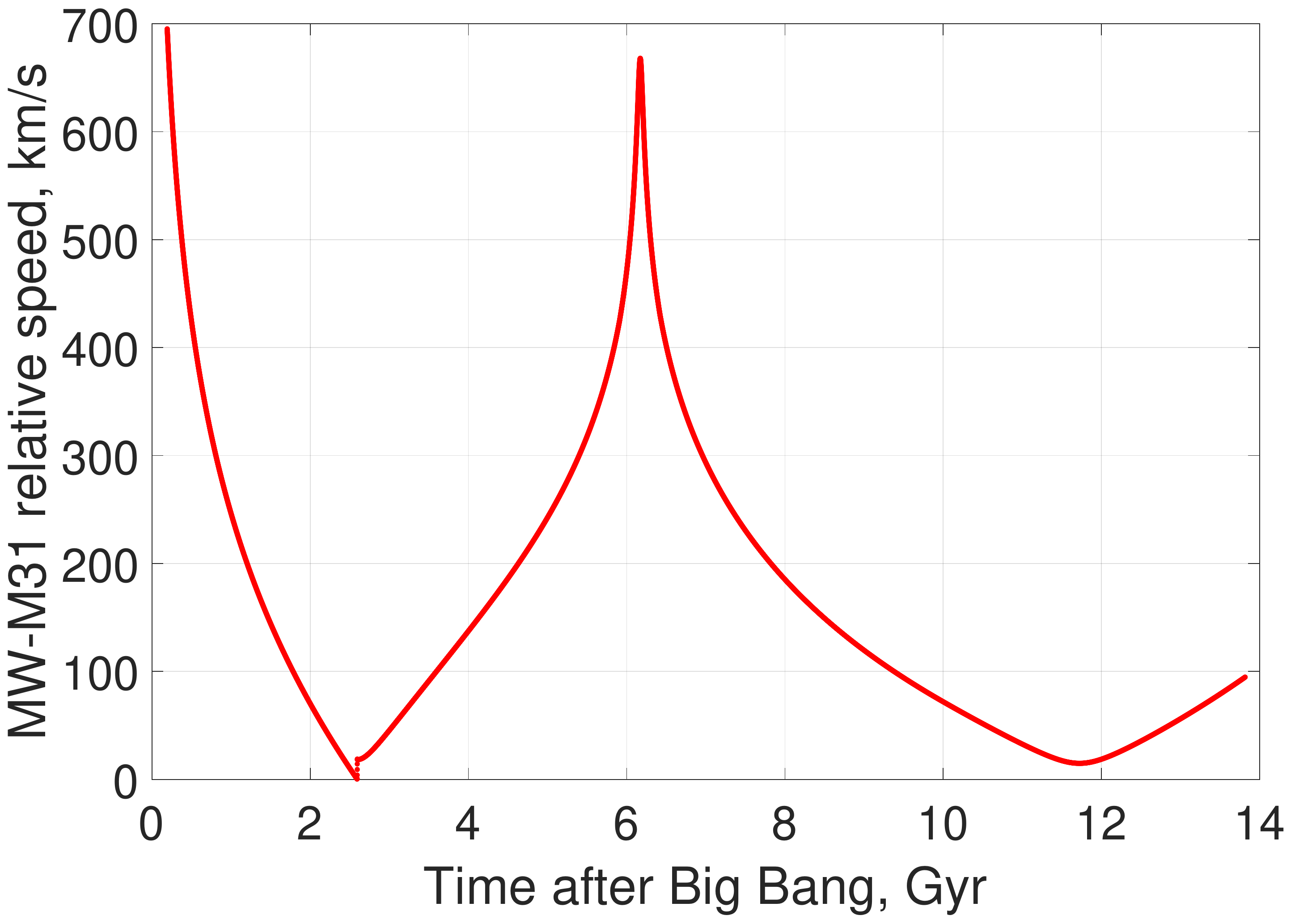}		
		\caption{\emph{Top}: The MW-M31 separation ${d \left(t \right)}$ in our best-fitting model, using the cosmological parameters in Table \ref{Cosmology}. The parameters of this trajectory are given in Table \ref{MW_M31_parameters}. We start integrating test particle trajectories 2 Gyr after the Big Bang (Section \ref{MW_M31_disks}). \emph{Bottom}: The MW-M31 relative velocity, which is very high at early times due to cosmic expansion. The discontinuity at 2.6 Gyr (first turnaround) arises because we assume the orbit is radial prior to this point but has its present angular momentum at later times (Section \ref{Angular_momentum_barrier}). The MW-M31 flyby occurs at 6.17 Gyr with a relative velocity of 668.1 km/s. Notice the very slow relative velocity at the second apocentre $-$ even a tiny perturbation can significantly reorient the MW-M31 orbit. Thus, the present direction of $\widehat{\bm{h}}_{_{LG}}$ is not a good guide to what $\widehat{\bm{h}}_{_{LG}}$ was several Gyr ago.}
	\label{MW_M31_separation_SP}
\end{figure}

Figure \ref{MW_M31_separation_SP} shows the MW-M31 trajectory in our best-fitting model, with the main model parameters summarised in Table \ref{MW_M31_parameters}. The flyby occurs 7.65 Gyr ago, consistent with the thick disk age of ${9 \pm 1}$ Gyr \citep{Quillen_2001}. To satisfy the timing argument (Section \ref{Timing_argument_NR}), our model requires the MW and M31 to have a total mass of $2.81 \times 10^{11} M_\odot$. This is slightly higher than our estimate of the total baryonic mass in their disks, assuming the MW rotation curve flatlines at ${v_{_f} = 180}$ km/s \citep{Kafle_2012} while that of M31 flattens at 225 km/s \citep{Carignan_2006}. Taken at face value, these rotation speeds imply a total mass of $2.3 \times 10^{11} M_\odot$ because the MOND dynamical mass of a galaxy is
\begin{eqnarray}
    M_{dyn} ~~=~~ \frac{{v_{_f}}^4}{Ga_{_0}}
    \label{M_dyn_MOND}
\end{eqnarray}

\begin{table}
 \begin{tabular}{ll}
	\hline
	Quantity & Value \\ \hline
  Orbital pole of MW-M31 orbit & $\left( 24.6^\circ, -16.3^\circ \right)$ \\
	MW-M31 separation & 783 kpc \\
  MW-M31 tangential velocity & 16.70 km/s \\
  MW$+$M31 mass & $2.81 \times 10^{11} M_\odot$ \\
  Flyby time (after Big Bang) & 6.17 Gyr \\
  Closest approach distance & 19.6 kpc \\
  Present external field on LG & $0.022 ~a_{_0}$ \\
  External field direction & $\left(276^\circ, 30^\circ \right)$ \\ [5pt]
  \hline
 \end{tabular}
 \caption{Data concerning the MW-M31 trajectory in our best-fitting model, with the present separation obtained from \citet{McConnachie_2012}. Our adopted cosmology (Table \ref{Cosmology}) implies the Universe is 13.82 Gyr old, so the last MW-M31 pericentre was 7.65 Gyr ago. The direction of the EF on the LG is obtained from \citet{Kogut_1993} and is the same in all models (Section \ref{g_ext_history}). The model has a peak EF strength of ${0.029 a_{_0}}$ when the cosmic scale-factor ${a = 0.75}$. Although the simulated Galactocentric tangential velocity of M31 is currently in almost the opposite direction to that suggested by \citet{Van_der_Marel_2012}, it is consistent within uncertainties as the orbit is nearly radial (see their figure 3). The grid of models we run has a step of $0.001 a_{_0}$ in the present EF on the LG, 1.15 km/s in the tangential velocity of M31 (${\approx 1.7}$ kpc in closest approach distance) and ${5^\circ}$ in the MW-M31 orbital pole.}
 \label{MW_M31_parameters}
\end{table}

When it comes to the MW-M31 timing argument, their large separation requires us to consider any massive satellites as these also contribute to the MW-M31 attraction. The LMC has ${v_{_f} \approx 70}$ km/s \citep{Alves_2000, Van_der_Marel_2014} while ${v_{_f} \approx 120}$ km/s for M33 \citep{Corbelli_2003, Corbelli_2007}. Thus, the most luminous LG satellites only contribute a mass of ${\approx 1.5 \times 10^{10} M_\odot}$ without much affecting the MW:M31 mass ratio. This can only explain ${\approx \frac{1}{3}}$ of the discrepancy between the timing argument mass of the LG and that inferred from rotation curves of LG objects. The remaining discrepancy disappears if actual rotation speeds are 3\% higher than we assume. This is quite possible given expected observational uncertainties for the dominant LG galaxy M31 \citep[e.g.][table 4]{Chemin_2009}.

The high timing argument mass might also be a sign of hot gas halos surrounding the MW and M31. \citet{Nicastro_2016} conducted observations at a range of Galactic latitudes and found that the best-fitting model has a hot gas corona around the MW with a mass of $2 \times 10^{10} M_\odot$ (see their table 2 model A). This is consistent with the estimate of $2.7 \pm 1.4 \times 10^{10} M_\odot$ obtained by \citet{Salem_2015} based on how the LMC gas disc is truncated at a smaller radius than its stellar disk, a phenomenon which they attribute to ram pressure stripping by the MW corona. If this was much more massive, the MOND-predicted MW escape velocity curve would be higher and flatter than inferred from observations \citep[][figure 5]{Banik_2017_escape}. A very massive Galactic hot gas corona would also reduce the apocentre of the Sagittarius tidal stream \citep[][section 4.3]{Thomas_2017}, making it difficult to explain the large measured distances to some parts of the stream \citep{Belokurov_2014}.

A corona has also been detected around M31 based on absorption features in spectra of background quasars \citep{Lehner_2015}. Thus, a small amount of hot gas around the MW and M31 could easily explain why their total MOND timing argument mass slightly exceeds what is implied by their rotation curves and those of their most luminous satellites.

\subsubsection{The apocentre asymmetry}
\label{Apocentre_asymmetry}

An interesting feature of our MW-M31 trajectory is that their second apocentre is larger than their first. This apocentre asymmetry is very important because it is closely related to the timing of the MW-M31 flyby. Naively, one might expect that the MW-M31 system is currently near apocentre so the flyby must have been relatively recent. For example, if we take the most recent apocentre to be 12 Gyr after the Big Bang and consider this to be $\frac{3}{2}$ orbital periods, then the flyby occurred after 1 orbital period i.e. at a time of 8 Gyr, only ${\ssim 6}$ Gyr ago. This is much more recent than the observed ${9 \pm 1}$ Gyr age of the Galactic thick disk which we expect formed due to the flyby.

To match this constraint, our model will try to find some way of getting a small first apocentre so the MW and M31 can quickly turn around and undergo their flyby. Afterwards, some extra effect is needed to push the galaxies further out, thus matching their present separation. Part of the necessary repulsion must come from the cosmological acceleration term in Equation \ref{Combined_MW_M31_attraction}. However, this would cause only a small effect because the MW-M31 dynamics is largely decoupled from cosmic expansion. Indeed, a quick look at the EF-included trajectory in figure 1 of \citet{Zhao_2013} shows that the two apocentres are expected to be very similar despite those authors including the cosmological acceleration.

Although the apocentre asymmetry could be due to our assumed EF history (Section \ref{g_ext_history}), this is not the whole story because the asymmetry is much weaker if $\widehat{\bm{h}}_{_{LG}}$ is changed while not adjusting $\bm{g}_{_{ext}}$. Instead, our investigations reveal that the apocentre asymmetry is mainly due to tides raised by perturbers on the LG (Section \ref{Tides}). To investigate this possibility in more detail, we perform some analytic calculations of how much tides might be expected to influence the MW-M31 trajectory. We treat the MW and M31 as test particles in the EF-dominated gravitational field due to each perturber, whose tidal stress tensor is found in Appendix \ref{Tides_g_ext_domination}. The EF-dominated assumption is likely to hold true at the first MW-M31 apocentre and should be very accurate at their second apocentre.

\begin{table}
 \begin{tabular}{lll}
  \hline
	Perturber & Estimated $g_{_{tide}}$ at & Estimated \\
	& first apocentre, m/s$^2$ & $g_{_{tide}}$ now, m/s$^2$ \\ \hline
    Cen A & $-1.85 \times 10^{-13}$ & $+6.22 \times 10^{-14}$ \\
    M81 & $-6.80 \times 10^{-14}$ & $-4.11 \times 10^{-15}$ \\
    IC 342 & $-5.33 \times 10^{-13}$ & $+9.75 \times 10^{-15}$ \\
    GA & $-8.34 \times 10^{-15}$ & $+1.64 \times 10^{-14}$ \\
		$\frac{\ddot{a}}{a} d_{_{M31}}$ & $-6.47 \times 10^{-13}$ & $+6.06 \times 10^{-14}$ \\ [2pt]
  \hline
 \end{tabular}
\caption{Our estimates for how much each perturber affects the MW-M31 relative acceleration at the times shown. We obtain these estimates by applying the analytic method described in Appendix \ref{Tides_g_ext_domination}. For the Great Attractor (GA), we use Equation \ref{Tide_GA} instead. The last row shows the contribution of the cosmological acceleration term. The MW and M31 are expected to have a relative acceleration of ${\approx 2 \times 10^{-12}}$ m/s$^2$ once the external field is considered (Section \ref{Apocentre_asymmetry}).}
 \label{Tide_estimates}
\end{table}

Table \ref{Tide_estimates} gives our analytic estimates for how much tides from each perturber would have affected the MW-M31 relative acceleration at the time of their first apocentre and at the present time. For reference, the present MW-M31 attraction may be estimated as $\frac{\sqrt{GM_{_{LG}} a_{_0}}}{d_{_{M31}}} = 2.77 \times 10^{-12}$ m/s$^2$, though this would be weakened ${\approx 30\%}$ due to the EF (Figure \ref{QUMOND_residuals}). It is evident that the perturbers we consider would likely have created extra attraction between the MW and M31 in the period before their flyby, thus limiting the distance at which they turned around and causing them to have an earlier flyby. However, the perturbers would not have continued `compressing' the MW-M31 orbit after the flyby. This is no doubt partly because of cosmic expansion, which affects the LG-perturber distances far more than it affects the much more bound MW-M31 system.

However, cosmic expansion can't be the entire explanation because the tidal forcing changes sign for Cen A and IC 342, something that expansion alone can't do. The sign change is an indication that the geometry of the problem is different after the flyby than before it. This is unsurprising given that the MW-M31 line must rotate substantially around the time of their close encounter. In fact, our model has $\widehat{\bm{d}}_{_{M31}}$ rotate ${233^\circ}$ between the first MW-M31 apocentre and today, significantly changing the relative orientations of the MW, M31 and external perturbers. Combined with other effects, this explains the apocentre asymmetry in our best model and thus how the MW-M31 flyby could be as long ago as the formation of the Galactic thick disk.

\subsubsection{MW-M31 orbital pole}

Our best model has a very different MW-M31 orbital pole to that indicated by the toy model of \citet[][section 2.2]{Banik_2017_anisotropy}. That model relied on many assumptions, almost certainly over-simplifying the problem. Of particular concern is the impulse approximation, which is not likely to work given the long duration of the flyby compared to the dynamical time of MW stars. For example, the MW-M31 separation is below twice its minimum value for ${\ssim 120}$ Myr (Figure \ref{MW_M31_separation_SP}), a significant fraction of the time taken for the Sun to orbit around the MW \citep{McMillan_2017}. It is also unrealistic to assume that M31 only significantly affects the MW at the time of closest approach as M31 gets within 20 kpc of the MW (Table \ref{MW_M31_parameters}). In any case, M31 would come closest to different parts of the MW at different times, an effect not included in the toy model. Moreover, the model does not contain any rigorous dynamical arguments and is a purely kinematic analysis. It is therefore quite possible that the toy model did not adequately capture how the MW-M31 flyby would have worked. The main goal of this contribution is to construct a more rigorous dynamical model of the flyby as a stepping stone to $N$-body simulations.

An important constraint on our model is the present relative velocity between the MW and M31. We determine this by adjusting the observed value \citep{Van_der_Marel_2012} for the motion of the LMC \citep{Kallivayalil_2013} and M33 \citep{Brunthaler_2005} as well as the motion of the Sun within the MW \citep{McMillan_2011, Francis_2014}. Applying Equation \ref{M_dyn_MOND} suggests that the LMC has 2.25\% as much mass as the MW \citep{Alves_2000, Kafle_2012} while M33 has 7.5\% as much mass as M31 \citep{Corbelli_2003, Carignan_2006}. Combining this information yields the relative velocity between the MW-LMC barycentre and the M31-M33 barycentre. We fix the radial component of this velocity at -93.4 km/s but let the tangential component vary so as to alter the MW-M31 orbital pole $\widehat{\bm{h}}_{_{LG}}$ and their closest approach distance. In our best-fitting model, the tangential velocity of M31 is similar in magnitude to the observational estimate of \citet{Van_der_Marel_2012} but points in almost the opposite direction. Their figure 3 shows that this is consistent within uncertainties because the MW-M31 orbit is nearly radial.

Future high-precision proper motion measurements of M31 could provide some constraints on $\widehat{\bm{h}}_{_{LG}}$ and thus whether it might be the same as in our best-fitting model. However, this requires a good idea of the LMC and M33 masses as one would need to consider the relative velocity between the MW-LMC barycentre and the M31-M33 barycentre. Both M33 and the LMC have velocities of ${\approx 200}$ km/s orthogonal to the MW-M31 line, affecting their tangential velocity by $\approx \left( \frac{120}{225} \right)^4\times$ as much i.e. by 16 km/s, even if we only consider the more massive M33. If we assume uncertainties of 10\% on its rotation curve, this implies a 6 km/s uncertainty on the present tangential velocity of M31.

In addition to this, M33 also affects the position of M31 as the two orbit around their barycentre. Given a M31-M33 distance of $\approx 200$ kpc \citep[][table 1]{Patel_2017}, the reflex motion of M31 must be ${\approx 16}$ kpc. This is a significant fraction of our estimated 20 kpc distance between the MW and M31 at the time of their closest approach (Table \ref{MW_M31_parameters}). To avoid this becoming a major source of uncertainty, we would need to know how the M31-M33 orbit behaved 8 Gyr into the past.

Given the nearly radial MW-M31 orbit, determining $\widehat{\bm{h}}_{_{LG}}$ also requires accurate knowledge of the motion of the Sun with respect to the MW, in particular the velocity of the Local Standard of Rest \citep{McMillan_2017}. Although these considerations have only a small effect on the MW-M31 orbit over the last few Gyr, it is so nearly radial that such effects could determine whether M31 flew past the MW on one side or on the other. This makes it very difficult to directly determine the geometry of the MW-M31 flyby through backwards integration of present conditions.

Fortunately, some insights can be gained from our MOND timing argument analysis. This indicates that the LG must be unrealistically massive if the MW-M31 orbital pole lies within ${\ssim 75^\circ}$ of Galactic co-ordinates $\left( 227^\circ, -34^\circ \right)$. Other orbital poles can satisfy the timing argument with a LG mass more consistent with the observed baryonic masses of its constituents. As this still leaves a wide range of possibilities, we focus on other ways of testing our model.

\subsection{The Local Group satellite planes}
\label{Discussion_satellite_plane}

\subsubsection{Orientation}
\label{SP_orientation}

A strong test of our models is provided by the observed orientations of the MW and M31 satellite planes (Table \ref{Directions_table}). Although these may have precessed slightly due to the non-spherical nature of the MW and M31, significant precession is unlikely as all satellites are not at the same radial distance \citep[][figure 2]{Kroupa_2013}. Thus, different satellites would precess at different rates, thickening the satellite planes \citep{Fernando_2016}. Their small observed thickness therefore strongly suggests that their orientations have not substantially changed since they were formed.

\begin{figure}
	\centering
		\includegraphics [width = 8.5cm] {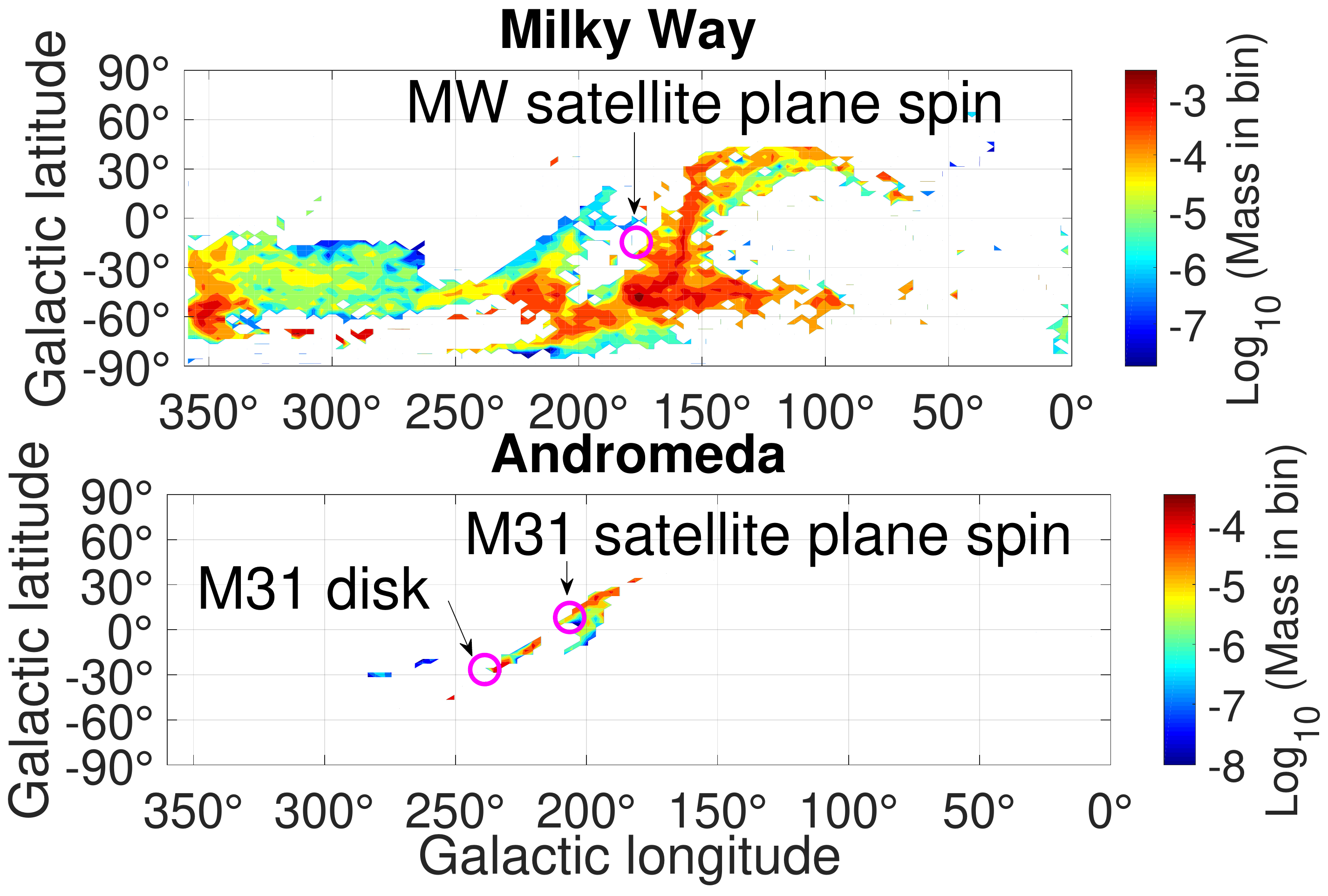}		
\caption{The distribution of orbital angular momentum directions (spin vectors) for tidal debris around the MW and M31 disks at the end of our best-fitting simulation (parameters in Table \ref{MW_M31_parameters}). Satellite particles are selected according to the criteria in Table \ref{Satellite_plane_filter}, thus rejecting the MW and M31 disks themselves. The mass scale used is arbitrary (satellite plane masses are discussed in Section \ref{SP_mass}). For model comparison, the method described in Section \ref{Satellite_plane_angular_momenta} is applied to a low resolution version of these histograms prior to further refinement. \emph{Top}: Results for the MW. Its disk spin vector points at the South Galactic Pole while the open pink circle shows the spin vector of its satellite plane (Table \ref{Directions_table}). \emph{Bottom}: Results for M31. We use open pink circles to show the observed spin vectors of its disk (lower left) and satellite plane (upper right).}
	\label{Orbital_pole_histogram}
\end{figure}

To see how well we can fit these constraints, we show the orbital pole distribution of all satellite particles around the MW and M31 (Figure \ref{Orbital_pole_histogram}). Both orbital pole distributions show a lack of material on a similar orbit to the disk of the host galaxy, a consequence of the filters we applied to our simulation in order to focus on the MW and M31 satellite regions (Table \ref{Satellite_plane_filter}).

For the MW, there is a broad peak in the orbital pole distribution near a Galactic longitude of ${180^\circ}$, similar to the observed spin vector of the MW satellite plane. The Galactic latitude of the simulated satellite plane is ${\ssim 25^\circ}$ lower than observed, indicating that there is not quite enough tidal torque on the MW in our model. However, a discrepancy at this level is reasonable giving expected observational and modelling errors. For example, treating M31 as a disk rather than a point mass could allow parts of it to get much closer to the MW, thus exerting a larger pull on it. Moreover, the orbital poles of MW satellites have an angular dispersion of ${\approx 25^\circ}$ \citep[][section 4]{Kroupa_2013}.

The simulated M31 satellite particles prefer a much narrower range of orbital poles $-$ our orbital pole analysis (Section \ref{Satellite_plane_angular_momenta}) yields an angular dispersion of ${41^\circ}$ for the MW satellite plane but only ${23^\circ}$ for M31. This may explain why the MW satellite plane indeed has nearly twice the thickness of the M31 satellite plane \citep[][tables 2 and 4]{Pawlowski_2014}. However, both structures are observed to be much thinner than in our best-fitting model. This is to be expected given its lack of dissipative processes.

Due to reduced observational and modelling uncertainties in its orientation, the M31 satellite plane provides a strong test of our models. Thus, it is interesting that our best-fitting model matches the observed spin vector of this structure to within ${10^\circ}$ (Table \ref{Best_model_results}).

\subsubsection{Counter-rotating fraction}
\label{SP_f_counter_rotating}

Around the MW, a secondary orbital pole peak is also evident near a Galactic longitude of ${350^\circ}$ (Figure \ref{Orbital_pole_histogram}). This is not quite where a true counter-rotating peak would be, though it is in roughly the right place. Due to dissipation, we expect that the actual tidal debris around the MW would settle into a plane with the counter-rotating material also forced to orbit within this plane but in the opposite sense. Thus, our model strongly suggests that some MW satellites ought to be counter-rotating. Indeed, Sculptor is just such an object \citep{Sohn_2017}.

Taken at face value, our model predicts that 22\% of the MW satellite plane mass ought to be in similar objects, much more than observed. If there is a significant amount of dissipation before the tidal debris coalesce into a few satellites, then the counter-rotating material could fall to much lower radii and perhaps get re-accreted onto the MW. Thus, the counter-rotating fraction in our model only provides an upper limit on how much counter-rotating material might actually be present in the MW satellite plane.

Unlike the MW, our model yields no counter-rotating material around M31. Although it lacks dissipation, it is very difficult to see how this could explain future observations that reveal several counter-rotating satellites within the M31 satellite plane. Thus, our flyby model may be tested with proper motions of the two M31 satellites in its satellite plane that do not share the coherent radial velocity trend of the remaining 13 M31 satellites in this structure \citep{Ibata_2013}. If both of these anomalous satellites (And XIII and And XXVII) have proper motions indicating orbits within the M31 satellite plane but opposite to the majority, then this implies a significant counter-rotating fraction, thus challenging our model. However, if these satellites have orbits that are not well-aligned with the M31 satellite plane, then by definition they are just interlopers. In this situation, the planar M31 satellites would all be co-rotating, much more consistent with our model.

To be sure of this conclusion, one would need to obtain proper motions for any subsequently discovered members of the M31 satellite plane in addition to its already known members. For example, Cassiopeia III \citep{Martin_2013} is another example of a M31 satellite spatially within the M31 satellite plane but whose radial velocity indicates that it is counter-rotating with respect to this structure \citep{Martin_2014}.

\subsubsection{Mass}
\label{SP_mass}

As well as explaining their orientations, our simulations should broadly match the observed masses of the LG satellite planes. In principle, masses can be estimated from luminosities as galaxies lack dark matter halos in MOND. However, the MW and M31 satellite planes consist of rather faint galaxies, many of which were discovered rather recently. Due to the large associated uncertainties, we do not use satellite plane masses to select our best-fitting model. We can therefore use this as an independent check on whichever model best fits the other constraints (Section \ref{Comparison_with_observations}).

In this model, 0.0045 of the MW mass ends up in its satellite plane, including both the co-rotating and counter-rotating peaks in the orbital pole distribution. We can estimate the actual mass of the MW satellite plane by considering its most massive satellite, the LMC. If we assume that it has ${v_{_f} = 70}$ km/s \citep{Alves_2000} while the MW has ${v_{_f} = 180}$ km/s \citep{Kafle_2012}, then we expect the satellite plane mass fraction to be $\ssim \left( \frac{70}{180} \right)^4 = 0.02$ (Equation \ref{M_dyn_MOND}).

For M31, this quantity is only ${6.5 \times 10^{-6}}$ in our model. Observationally, M31 satellite plane galaxies are each ${\approx 4-5}$ orders of magnitude fainter than M31 itself \citep[][table 2]{Martin_2016}. Although M33 is much brighter, it must have an independent origin as it is not part of the M31 satellite plane \citep[][figure 1]{Ibata_2013}. If this structure also contains M32, this would make the M31 satellite plane much more massive \citep[][table 4]{McConnachie_2012}. However, it is unclear whether M32 is really part of this structure due to the very small distance between M32 and M31 \citep[][table 4]{Weisz_2014} $-$ \emph{any} M31 satellite this close in would lie within the M31 satellite plane. Moreover, M32 has likely been accreted recently by M31 \citep{Loeb_2014}. Although that work considered $\Lambda$CDM, this is also possible in MOND because M32 would experience some dynamical friction against the baryonic disk of M31 if the galaxies had a sufficiently close encounter. As this process would eventually cause M32 to merge with M31, the current detectability of M32 and M31 as separate objects makes it unlikely that M32 has maintained its present tight orbit around M31 over the ${\approx 8}$ Gyr since the MW-M31 flyby. Thus, we expect that M32 is unrelated to the M31 satellite plane.

Our model correctly predicts that the M31 satellite plane should be ${\approx 3}$ orders of magnitude fainter than the MW satellite plane, though both satellite plane masses are underestimated by about an order of magnitude. The lower mass of the M31 satellite plane is almost certainly related to the MW having a lower mass than M31. This makes it more difficult for the MW to tidally disrupt M31 than vice versa.

\subsubsection{Initial radial distribution}
\label{SP_r_initial}

\begin{figure}
	\centering
		\includegraphics [width = 8.5cm] {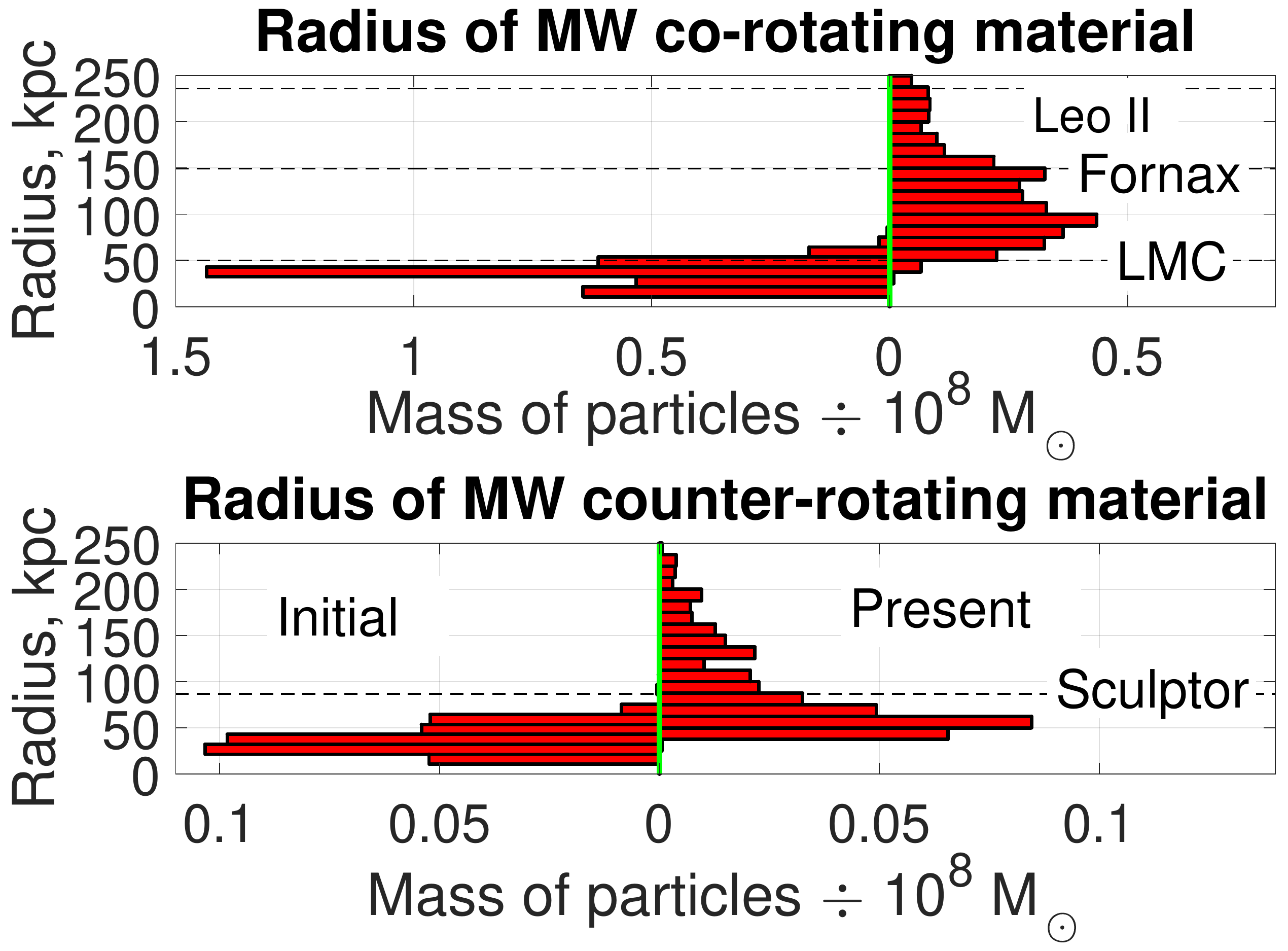}		
		\caption{Histogram of the initial and final Galactocentric distances of MW particles in its satellite region (defined in Section \ref{Satellite_plane_angular_momenta}). Each particle contributes the mass it represents within the MW disk (Section \ref{MW_M31_disks}). The particles are divided according to whether they are currently in the co-rotating peak (\emph{top}) or the counter-rotating peak (\emph{bottom}). The initial radial distribution of the particles is shown to the left of the vertical green lines, while the parts of the figures to the right of these lines show the present-day distribution of the same material. The dashed horizontal lines show distances to the indicated MW satellite. The proper motion of Leo II indicates it is in the MW satellite plane \citep{Piatek_2016}, but most of this structure is defined by less distant satellites \citep[][figure 2]{Kroupa_2013}. The only known counter-rotating MW satellite is Sculptor \citep{Sohn_2017}. The observed radial extents of both satellite planes are nicely illustrated in \citet[][figure 1]{Pawlowski_2018}.}
	\label{MW_radial_distribution}
\end{figure}

\begin{figure}
	\centering
		\includegraphics [width = 8.5cm] {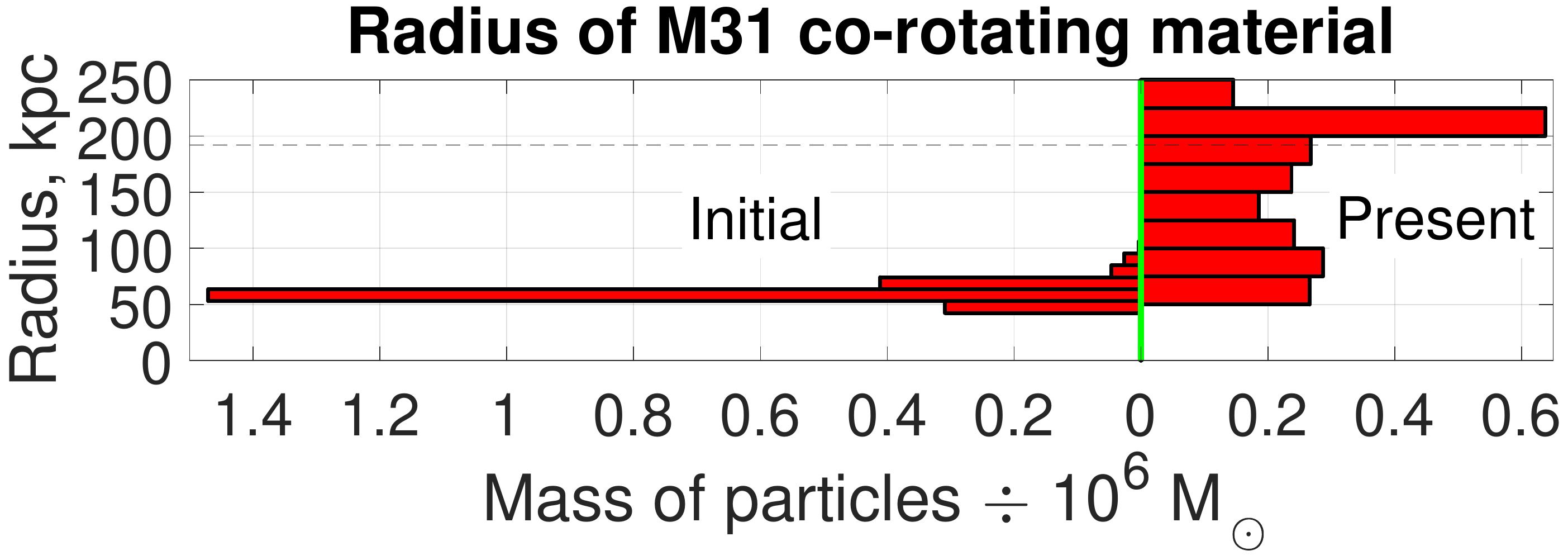}
		\caption{Similar to Figure \ref{MW_radial_distribution}, but for M31. The dashed horizontal line at 192 kpc is the rms radial extent of the observed M31 satellite plane \citep[][section 3]{Ibata_2014}. Our model lacks counter-rotating particles in the M31 satellite region (Figure \ref{Orbital_pole_histogram}).}
	\label{M31_radial_distribution}
\end{figure}

Our low predicted satellite plane masses may be related to our assumed exponential disk law, which could well be inaccurate in the outer parts of each galaxy that go on to form their satellite planes. The importance of this can be seen in Figures \ref{MW_radial_distribution} and \ref{M31_radial_distribution}, whose left sides (left of vertical green lines) show the initial radial distribution of material ending up in the MW and M31 satellite planes, respectively. It is evident that a significant fraction of the MW satellite particles were initially ${\approx 50}$ kpc from the MW, even though its stellar disk has a scale length of only 2.15 kpc. Thus, our simulated satellite plane masses are sensitive to assumptions concerning the MW disk surface density at large radii. Partly to deal with this, we also include a gas disk with a longer scale length of 7 kpc (Table \ref{Disk_parameters}). Thus, we have to consider test particles starting at quite large radii. Our adopted outer limit of 107.5 kpc (Table \ref{Disk_resolution}) seems to be sufficient for our investigation as the initial radial distribution of MW satellite plane particles tails off by this point.

For M31, we use a similar initial radial range $-$ our simulations start with M31 particles out to 106 kpc from M31 \citep[20 disk scale lengths,][]{Courteau_2011}. This seems to be enough for our purposes (Figure \ref{M31_radial_distribution}). However, this is such a large distance that our estimated M31 satellite plane mass is rather sensitive to its disk surface density at large radii. Even if M31 can be treated as an exponential disk, there is a ${10\%}$ uncertainty in its scale length \citep{Courteau_2011}. To estimate how this might affect our results, we consider the M31 mass fraction exterior to 60 kpc, the approximate initial radius of most M31 satellite plane particles (Figure \ref{M31_radial_distribution}). This fraction would be ${2.4 \times}$ higher if we assume the M31 disk scale length is 5.8 kpc instead of 5.3 kpc. Thus, it is unlikely that our model will provide a very good match to the masses of the LG satellite planes, though it should yield the correct order of magnitude.

Although these uncertainties clearly affect the mass of the LG satellite planes, their orientations should be less sensitive $-$ we weighted all MW and M31 particles equally and found very little change to the appearance of Figure \ref{Orbital_pole_histogram}. This is because there is a clear peak in the orbital pole distribution for each galaxy such that its location is not very sensitive to details.

For the MW, there is a clear secondary orbital pole peak which is approximately counter-rotating relative to the primary peak (Figure \ref{Orbital_pole_histogram}). The particles in the secondary peak start at similar radii to the co-rotating particles (Figure \ref{MW_radial_distribution}). This result is different to that obtained by \citet{Pawlowski_2011}, whose figure 8 indicates that the counter-rotating particles start further out. There are many possible reasons for this difference e.g. using a different mass ratio, geometry and gravity law.

\subsubsection{Present radial distribution}
\label{SP_r_final}

The right sides of the panels in Figure \ref{MW_radial_distribution} (right of vertical green lines) show the present radial distribution of MW satellite particles in our best-fitting model. This is similar to the range of Galactocentric distances at which we observe MW satellites in its satellite plane \citep[][figure 2]{Kroupa_2013}. The most luminous MW satellite is the LMC, whose Galactocentric distance is 50 kpc \citep{Pietrzynski_2013}. This is certainly a distance at which our simulations indicate there should be a significant amount of tidal debris from the MW-M31 flyby.

Our model indicates that the counter-rotating particles do not go as far out as the co-rotating particles. This is similar to the result obtained by \citet{Pawlowski_2011}, as can be seen from their figure 6. The observed counter-rotator Sculptor \citep{Sohn_2017} has a Galactocentric distance of 86 kpc \citep{Martinez_2015}, well within the simulated counter-rotating material's final radial distribution. Thus, both the co-rotating and counter-rotating MW satellite particles in our simulation have a radial distribution broadly similar to what we observe for actual MW satellites orbiting within its satellite plane.

Figure \ref{M31_radial_distribution} shows the initial and final radial distribution of M31 satellite particles. We do not show results for counter-rotating M31 satellite particles as our simulation lacks such particles (bottom panel of Figure \ref{Orbital_pole_histogram}). At the present time, our best-fitting model indicates that M31 satellite particles should have a broad range of distances from M31. This is similar to the range of observed distances between M31 and its satellites within its satellite plane \citep[][figure 1]{Ibata_2013}.\footnote{At the 783 kpc distance of M31 \citep{McConnachie_2012}, each square in this figure has a side length of 55 kpc.} Indeed, our simulated M31 satellite particle radial distribution peaks at ${\approx 200}$ kpc, very similar to the observed 192 kpc rms radial extent of the M31 satellite plane \citep[][section 3]{Ibata_2014}. Their results suggest that this structure does not extend much further out, so we impose a maximum distance limit of 250 kpc from M31 when selecting particles belonging to its simulated satellite plane.

\subsubsection{Orbital eccentricity}
\label{SP_eccentricity}

\begin{figure}
	\centering
		\includegraphics [width = 4.1cm] {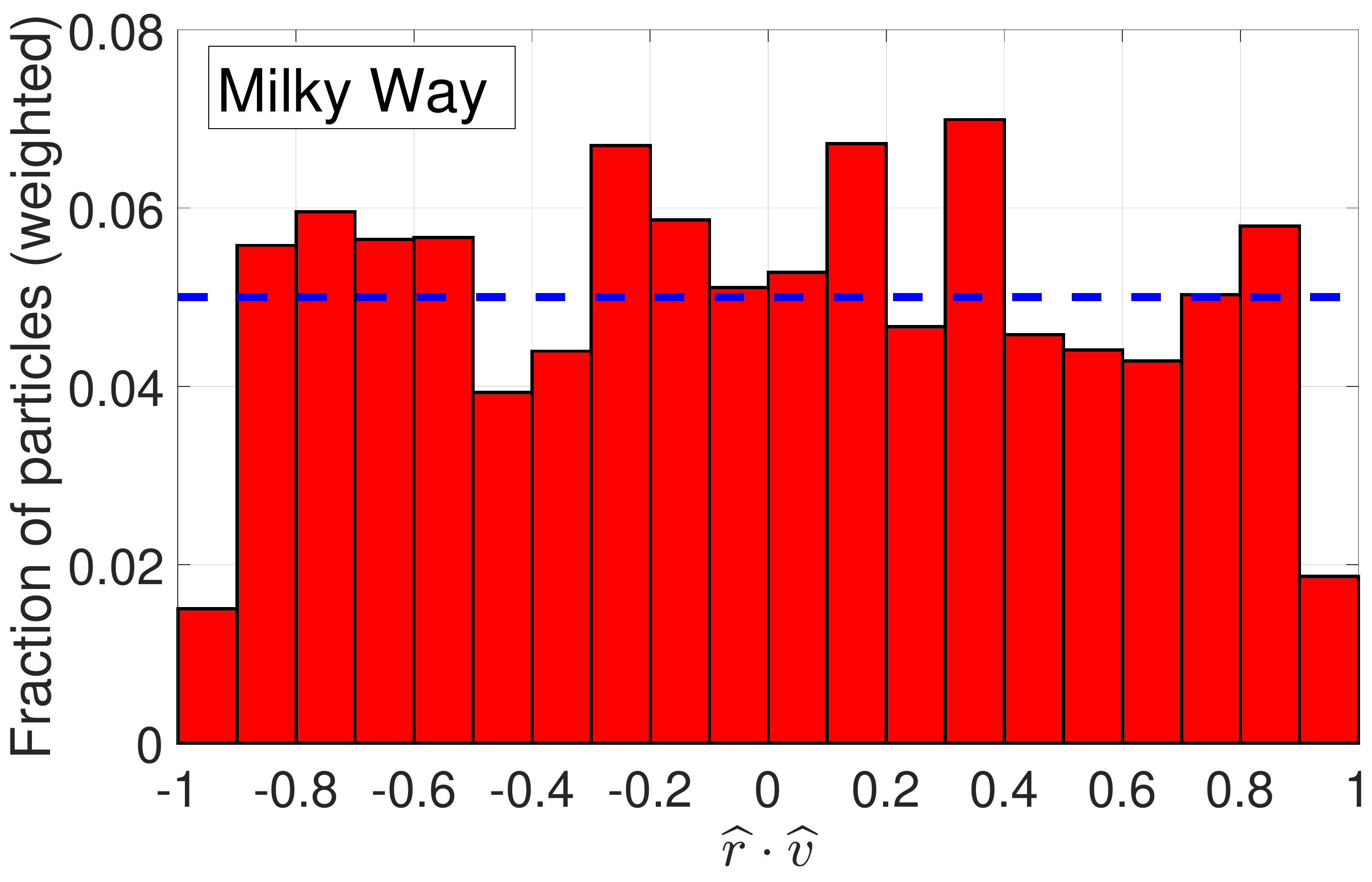}
		\includegraphics [width = 4.1cm] {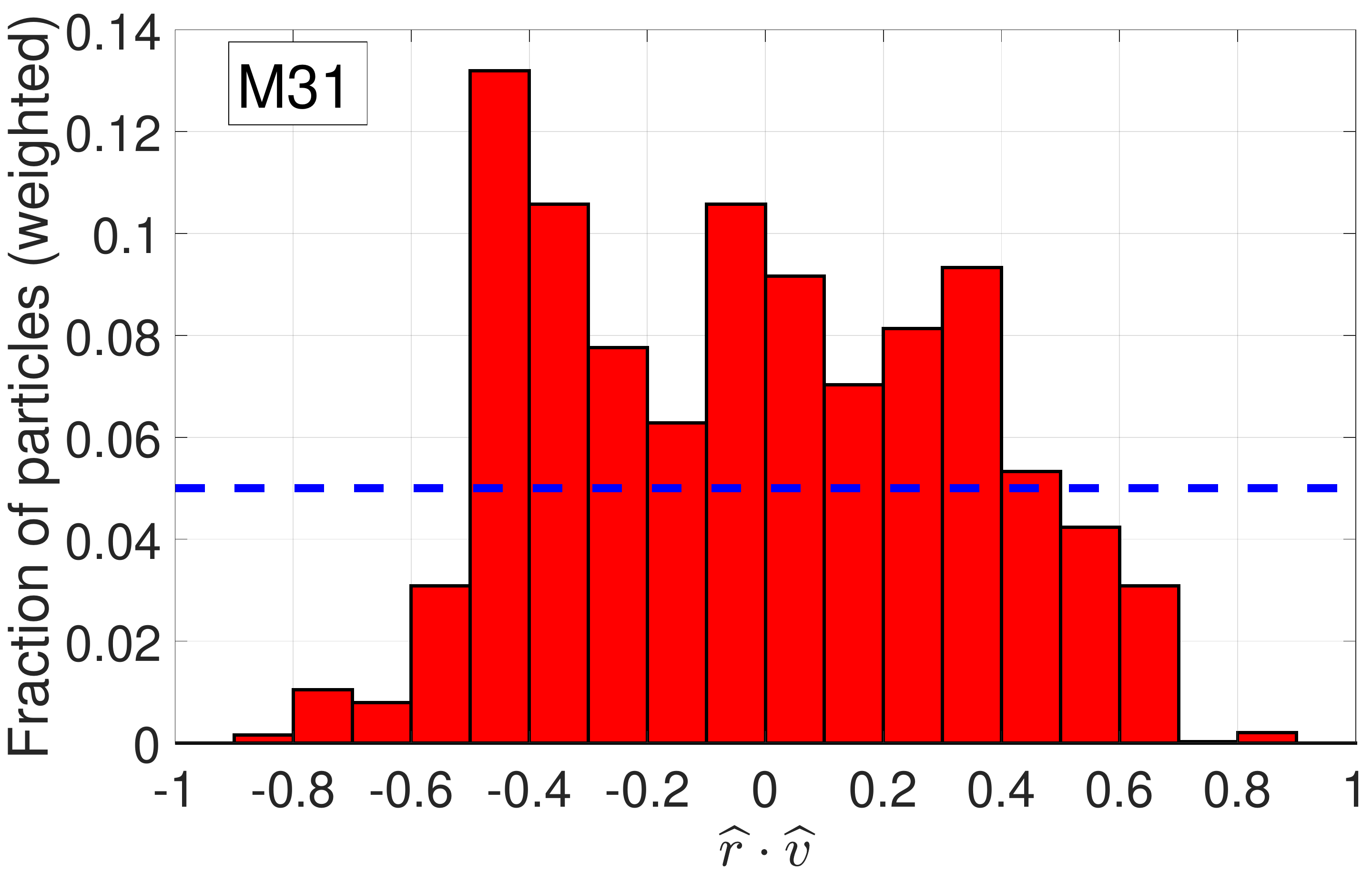}
		\caption{\emph{Left}: Histogram of $\widehat{\bm{r}} \cdot \widehat{\bm{v}}$, where $\bm{r}$ and $\bm{v}$ are, respectively, the position and velocity of each particle relative to the MW. This is a basic measure of orbital eccentricity. We only show particles corresponding to the MW satellite plane according to the criteria in Table \ref{Satellite_plane_filter}. The dashed blue line shows the expected distribution for particles on random orbits around the MW. \emph{Right}: Analogous results for M31. All particles shown are in the co-rotating peak around M31 as there are no counter-rotating particles (Figure \ref{Orbital_pole_histogram}).}
	\label{MW_eccentricity_distribution}
\end{figure}

To better understand the orbits of satellite plane particles in our simulations, we consider how eccentric their orbits are. This is difficult to quantify rigorously due to the non-Keplerian potential (Equation \ref{Potential}). As a proxy, we show the cosine distribution of angles between the position and velocity of each particle with respect to its host galaxy (Figure \ref{MW_eccentricity_distribution}), with weights assigned as in Section \ref{MW_M31_disks}.

For the MW, our results are similar to what they would be if its satellite particles had randomly distributed velocities (dashed blue line). The only major difference is a deficit of particles on extremely radial orbits i.e. whose velocity is within $\cos^{-1}0.9 = 26^\circ$ of the radial direction. This is unlike the actual MW satellite plane as that is co-rotating and mostly rotation-supported \citep{Kroupa_2013}. The discrepancy may be due to the lack of dissipative processes in our simulation. Even so, the nearly symmetric distribution of $\widehat{\bm{r}} \cdot \widehat{\bm{v}}$ suggests that the MW satellite system has reached some sort of equilibrium in the ${\approx 8}$ Gyr since its interaction with M31.

We obtain rather different results for the satellite particles around M31 (right panel of Figure \ref{MW_eccentricity_distribution}). Despite the lack of dissipation, we get much fewer particles on radial orbits than would be expected for randomly distributed velocities. The clear peak near 0 (tangential motion) implies near-circular orbits. This provides a natural explanation for the clear radial velocity gradient across the M31 satellite plane, whereby satellites on different sides of M31 almost always have a different sign of radial velocity relative to that of M31 \citep{Ibata_2013}.

Our models suggest that the satellite plane members of M31 ought to be on more nearly circular orbits than those of the MW. Observationally, even the MW satellite plane is rotationally supported \citep{Kroupa_2013}. Thus, we would certainly expect the M31 satellite plane to be predominantly rotation-supported. It will be interesting to test this prediction once proper motions of M31 satellites become available.

\subsection{The Milky Way thick disk}
\label{Discussion_thick_disk}

\begin{figure}
	\centering
		\includegraphics [width = 8.5cm] {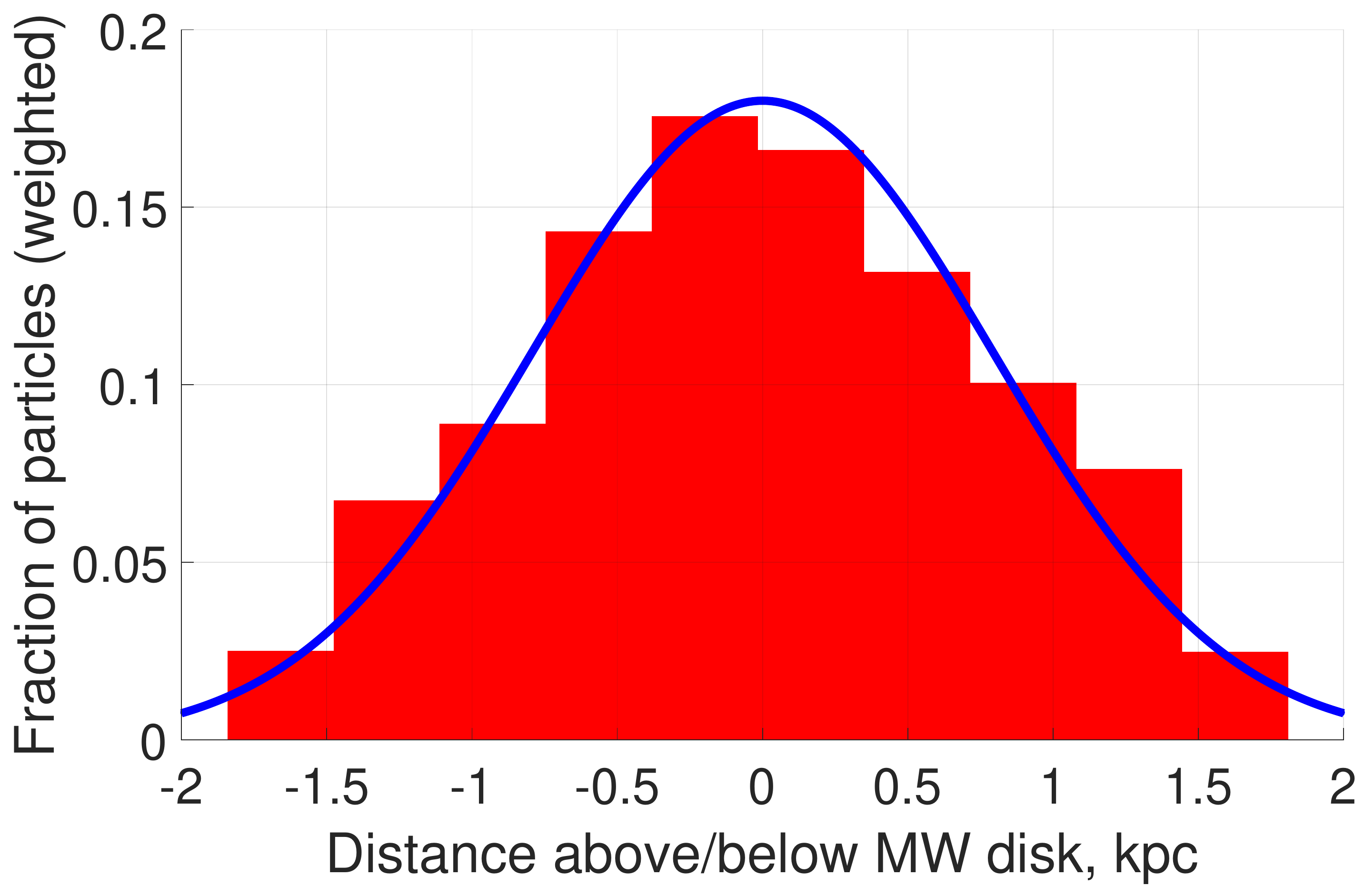}		
		\caption{Weighted histogram of the vertical distances of MW particles from its disk. Here, we only show particles with cylindrical radius between 7.2 kpc and 9.2 kpc, the region constrained by the observations of \citet{Ferguson_2017}. The solid blue line shows a Gaussian whose width is the thickness of the plane best fitting these particles (Table \ref{Best_model_results}).}
	\label{z_histogram_MW}
\end{figure}

An important constraint on our models is the extent to which they disrupt the well-observed MW disk in which we live (Section \ref{MW_thick_disk_z_rms}). For comparison with observations, we extract the rms thickness of the plane best fitting MW particles at Galactocentric cylindrical radii $r_{plane}$ within 1 kpc of the Solar value \citep[][figure 12]{Ferguson_2017}. In our best model, this agrees fairly well with observations (Table \ref{Best_model_results}). To check this result, we show the vertical distribution of these particles (Figure \ref{z_histogram_MW}). Their distance from the MW disk follows a nearly Gaussian distribution whose width is very similar to the rms thickness inferred by our plane-fitting algorithm \citep[][section 5.1]{Banik_2017_anisotropy}.

The lower scale height than observed may be due to the Galactic disk having been thickened by processes other than tides raised during the M31 flyby (Section \ref{MW_thick_disk_height}). Although we allow for this to some extent (Equation \ref{Secular_heating_adjustment}), the MW disk may have experienced a burst of star formation due to the close passage of M31 \citep{{Snaith_2014}}. This may have heated the Galactic disk via `popping' star clusters \citep{Kroupa_2002, Assmann_2011}. As our model does not include this process, it should find a dynamically colder MW than observed, even if secular heating is accounted for.

\begin{figure}
	\centering
		\includegraphics [width = 8.5cm] {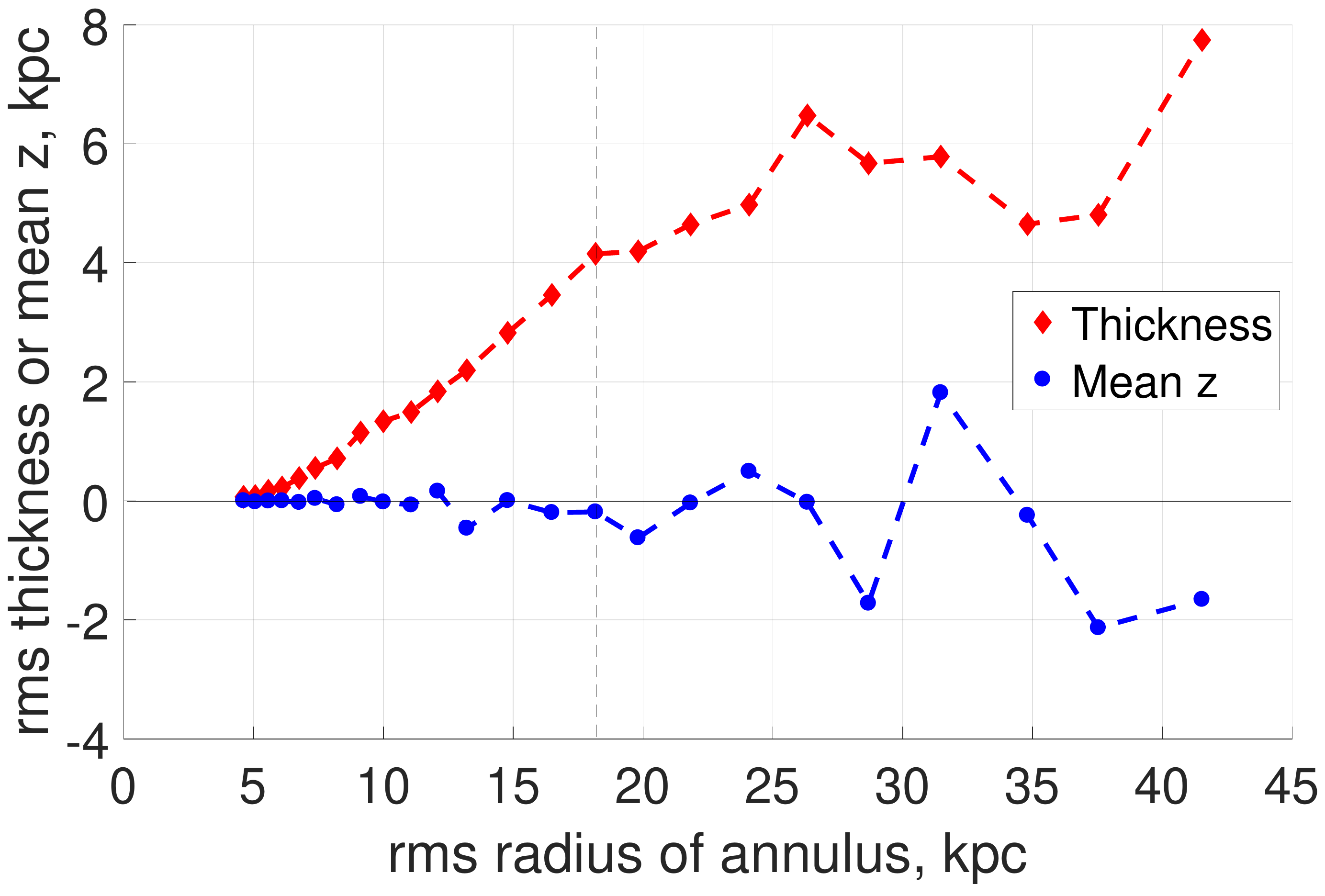}		
		\caption{We use red diamonds to show the rms thickness of MW disk material in various annuli (bins in Galactocentric radii), found using the plane-fitting method described in section 5.1 of \citet{Banik_2017_anisotropy} augmented by $\sigma$-clipping. The blue dots show the mean height (displacement towards the North Galactic Pole) of particles in each annulus relative to the position of the MW particle, our proxy for the position of the Galactic centre. The vertical dashed line at 17.7 kpc is the radius which encloses 95\% of the MW mass in our best model. We have likely underestimated the thickness of the innermost annulus shown here as we terminate the trajectories of all particles that get within 2.15 kpc of the MW.}
	\label{MW_disk_z_rms}
\end{figure}

In the future, it will become possible to obtain detailed observations of a larger region of the thick disk. We thus use Figure \ref{MW_disk_z_rms} to show the rms height of our simulated MW disk for several bins in $r_{plane}$, using a similar method to that used in Section \ref{MW_thick_disk_height}. This reveals that the thick disk ought to be even thicker further from the MW, a natural consequence of its tidal origin. Thus, the inner regions of the MW should suffer less disruption than its more weakly bound outer parts.

However, the trend may well be opposite this if the MW disk formed through internal processes e.g. interactions between stars, whose spatial density is higher closer to the Galactic centre. Thus, it may be possible to determine if the thick disk originated in a tidal heating event. If it did not, this would cast doubt on our scenario.

Interestingly, the thick disk does appear to have a higher velocity dispersion further from the MW \citep{Banik_2014}. The apparent lack of accreted stars suggests that a flyby may be a better explanation than a minor merger of the sort typical in $\Lambda$CDM \citep{Ruchti_2015}.

The expected 7.65 Gyr age of the thick disk in our model is lower than the ${9 \pm 1}$ Gyr estimated by \citep{Quillen_2001}. Although this discrepancy is not significant, more recent observations suggest that the thick disk has an age closer to 7 Gyr \citep{Yu_2018}, more in line with our model.

\subsection{Analogues to the Large Magellanic Cloud}
\label{Discussion_LMC_analogues}

If our model yields a realistic satellite plane around the MW, it ought to yield particles on orbits similar to its most massive member, the LMC. To look for analogues of the LMC, we search for particles whose orbit around the MW takes them to the LMC's Galactocentric distance $r_{_{LMC}}$ with at least as much radial kinetic energy and specific angular momentum $h$ as the LMC (measuring distances $r$ and velocities $v$ with respect to the MW). This allows us to consider particles on orbits similar to that of the LMC even if they are currently at a different point on their orbit e.g. near apocentre rather than pericentre. We thus make use of the adjusted energy
\begin{eqnarray}
	\tilde{E} ~\equiv~ \overbrace{\Phi \left( r \right) + \frac{1}{2}v^2}^E ~-~ \overbrace{\frac{h^2}{2{r_{_{LMC}}}^2}}^{	\begin{array}{cc}
		\text{Tangential kinetic} \\
		\text{energy at $r = r_{_{LMC}}$}	\end{array}}
	\label{E_LMC}
\end{eqnarray}

For particles close to the MW, their motion is mainly governed by their total energy $E$ and angular momentum $h$ with respect to the MW as M31 is very distant. Up to an additive constant, $\tilde{E}$ is thus the specific radial kinetic energy of a particle when it has a Galactocentric distance of $r_{_{LMC}}$ (if $\tilde{E}$ is too low, then the particle is not capable of reaching this distance). We define an analogue of the LMC as a particle whose $\tilde{E}$ and $h$ both exceed the LMC value. The constraint on $\tilde{E}$ ensures that analogues can reach $r = r_{_{LMC}}$ with its observed radial velocity while the constraint on $h$ ensures analogues would do so at the correct tangential velocity. A LMC analogue selected in this way can have a very different present position and velocity to the LMC, as long as the particle belongs to the simulated MW satellite plane.

If we neglect M31 and the EF, the form of the Galactic potential $\Phi \left( r \right)$ follows from integrating the simple interpolating function \citep{Famaey_Binney_2005} for a single point mass using a hyperbolic substitution.
\begin{eqnarray}
	\Phi \left( r \right) &=& \sqrt{GMa_{_0}} \left[ Ln \left(1 + \sqrt{1 + \tilde{r}^2} \right) - \frac{1}{\tilde{r}} - \sqrt{\frac{1}{\tilde{r}^2} + 1} \right] \nonumber \\
	\tilde{r} &=& \frac{2r}{r_{_M}} ~~\text{ where  } r_{_M} = \frac{\sqrt{GM}}{a_{_0}}
  \label{Potential}
\end{eqnarray}

Here, $M$ is the mass of the MW while $r_{_M}$ is its `MOND radius', beyond which the effects of MOND become significant. For the MW, ${r_{_M} \approx 9}$ kpc, not much smaller than the LMC distance of ${r_{_{LMC}} = 50}$ kpc \citep{Pietrzynski_2013}. This makes it inaccurate to assume that the LMC is in the deep-MOND limit.

\begin{figure}
	\centering
		\includegraphics [width = 8.5cm] {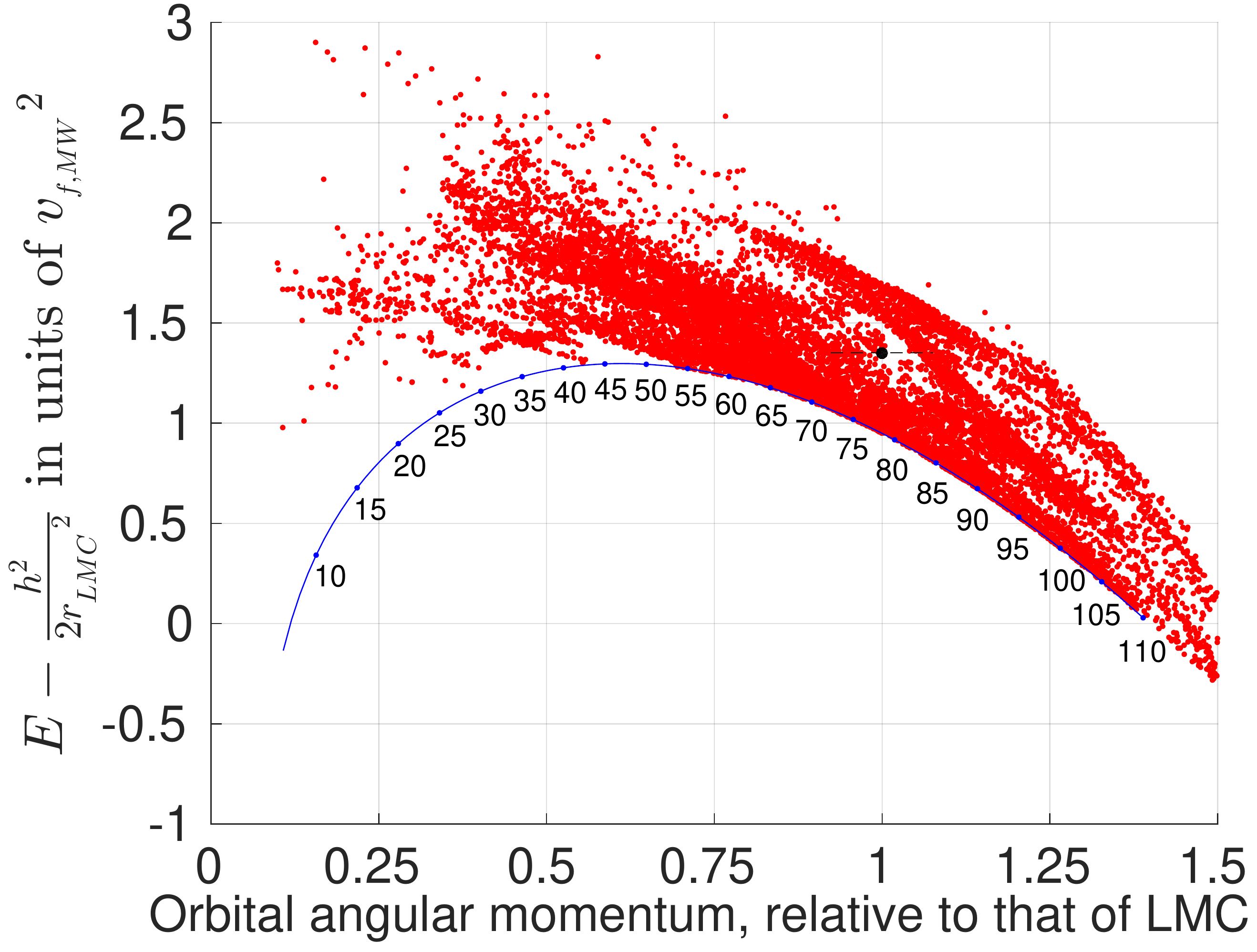}		
		\caption{Values of $\tilde{E}$ (Equation \ref{E_LMC}) against $h$ (Equation \ref{h_rel}) relative to the LMC value for MW satellite particles (red dots, only part of graph shown). The blue curve shows the relation for particles on circular orbits, with text labels along this curve (next to blue dots) giving the corresponding Galactocentric distance in kpc. We consider particles analogous to the LMC if their $\tilde{E}$ and $h$ both exceed the LMC value (black dot, uncertainty indicated with dashed line). We do not show the uncertainty in $\tilde{E}$ for the LMC because we neglect the rather small uncertainty in its radial velocity, whereas its tangential velocity is mainly a function of its less accurately known proper motion \citep[][table 5]{Kallivayalil_2013}. Our adopted distance to the LMC is 49.39 kpc based on a heliocentric distance of 49.97 kpc \citep{Pietrzynski_2013} and a Sun-Galactic centre distance of 8.20 kpc \citep{McMillan_2017}.}
	\label{LMC_analogues}
\end{figure}

In Figure \ref{LMC_analogues}, we show the values of $\tilde{E}$ and $h$ for particles in the satellite region of the MW. It is evident that a large number of particles have orbits analogous to that of the LMC using our definition (above and to the right of the black dot representing the LMC). In fact, the LMC is nearly in the middle of the sea of red dots representing simulated particles. Thus, our best-fitting model suggests that the LMC may well have condensed out of material in the outer part of the MW disk that was raised onto a higher and more inclined orbit by tides from M31. If our simulation had included self-gravity, it is reasonable to suppose that these particles on LMC-like orbits would have formed into a LMC-like object. This is similar to the scenario proposed by \citet{Yang_2014} in which the LMC is a tidal dwarf galaxy.

In this case, the internal dynamics of the LMC need to be explained using its baryons alone as tidal dwarf galaxies are expected to be purely baryonic \citep{Wetzstein_2007}. This is very difficult to do in Newtonian gravity \citep{Van_der_Marel_2002}, a result that was confirmed recently with proper motion-based measurements of the LMC rotation curve \citep{Van_der_Marel_2014, Van_der_Marel_2016}. On the other hand, figure 8 of \citet{Van_der_Marel_2014} shows that the LMC falls on the baryonic Tully-Fisher relation (Equation \ref{M_dyn_MOND}), implying consistency with MOND regardless of how the LMC formed.

In our model, the LMC is bound to the MW. This is also evident if we look at figure 2 of \citet{Patel_2017} and put in a MW mass of ${1.5 \times 10^{12} M_\odot}$ corresponding to what Newtonian gravity requires in order to sustain a flat rotation curve of 180 km/s out to 200 kpc. The bound nature of the LMC is unsurprising given that the observed MW escape velocity at the LMC position \citep[${380 \pm 30}$ km/s,][]{Williams_2017} exceeds its Galactocentric velocity \citep[${321 \pm 24}$ km/s,][table 5]{Kallivayalil_2013}.

In MOND, a bound LMC must have been orbiting the MW for ${\approx 8}$ Gyr as there is almost no dynamical friction 40 kpc from a purely baryonic MW. Assuming the Small Magellanic Cloud (SMC) formed on a similar orbit as part of the MW satellite plane, it would be quite feasible for these two objects to closely interact, helping to explain the observed properties of the Magellanic Stream \citep{Hammer_2015}.

A common origin during a past MW-M31 flyby provides a natural explanation for the alignment between the Galactic orbit of the LMC, the LMC-SMC orbit and the MW satellite plane \citep[][figure 1]{Pawlowski_McGaugh_2015}. It has long been assumed that the LMC and SMC have been bound to the MW for several Gyr \citep{Diaz_Bekki_2011}, consistent with our expectation. If instead they were recently accreted by the MW due to dynamical friction with its dark matter halo \citep{Besla_2007}, then these geometrical alignments would be fortuitous. After all, the MW satellite plane orientation would be a priori unrelated to the LMC-SMC binary orbit which could itself be oriented rather differently to the motion of their barycentre about the MW \citep[][section 5.3.1]{Kroupa_2015}. The competing scenarios might be distinguished using more detailed observations of the Magellanic Stream \citep{Fox_2016}.

\section{Conclusions}
\label{Conclusions}

Using a MOND model of the LG, we showed that the MW-M31 trajectory could be made consistent with cosmological initial conditions, a technique known as the timing argument. Such trajectories always involve a past close MW-M31 flyby (Figure \ref{MW_M31_separation_SP}). Assuming a standard $\Lambda$CDM cosmology (Equation \ref{Expansion_history}), the MW-M31 trajectory satisfies the timing argument with a total implied mass close to the sum of their observed baryonic masses and a close pericentre around the time when the Galactic thick disk formed (Table \ref{MW_M31_parameters}). This conclusion would be affected somewhat if the actual cosmology is non-standard, as seems likely if the MW and M31 indeed had a past close flyby. However, any viable model should yield an expansion history similar to the standard one given that this works well observationally \citep[e.g.][]{Joudaki_2017}.

Using a restricted $N$-body model of the MW and M31 disks advanced according to our MW-M31 trajectory, we simulated how their close flyby would have affected their internal structure. By performing a grid of models covering the parameter space, we obtained a reasonably good match to the observed properties of the MW and M31 (Table \ref{Best_model_results}). In particular, we found clear peaks in their tidal debris orbital pole distributions at orientations similar to those observed for their satellite planes (Figure \ref{Orbital_pole_histogram}).

Our model correctly predicts a much smaller angular dispersion for the M31 satellite plane compared to the MW \citep[][tables 2 and 4]{Pawlowski_2014}. It also matches the observation that the M31 satellite plane is ${\approx 3}$ orders of magnitude less massive than that of the MW. However, it underestimates the masses of both satellite planes by about an order of magnitude. This may be related to the rather strong sensitivity of our models to the disk surface density of the MW and M31 at radii of ${\ssim 50}$ kpc (Section \ref{SP_r_initial}).

Our model naturally yields counter-rotating material in the satellite plane around the MW, though not around M31. Thus, we can naturally explain why Sculptor counter-rotates within the MW satellite plane \citep{Pawlowski_2011, Sohn_2017}. However, our simulated MW satellite plane has a rather high fraction of counter-rotators. Interestingly, the observed MW satellite plane has a smaller radial extent than in our simulations, as can be seen by comparing our Figure \ref{MW_radial_distribution} with figure 2 of \citet{Kroupa_2013}. Both these issues may be related to the lack of dissipative processes in our simulations. We hope to conduct more detailed investigations like those of \citet{Bilek_2017} using the Phantom of RAMSES algorithm \citep{PoR} which adapts the gravity solver of the RAMSES code \citep{Teyssier_2002} widely used by astronomers to conduct $N$-body simulations that include hydrodynamic effects.

Assuming these yield similar results to our flyby model, it may be tested based on proper motions of M31 satellites in its satellite plane. In particular, two of these (And XIII \& And XXVII) do not share the coherent radial velocity trend of the remaining 13 M31 satellites in this structure \citep{Ibata_2013}. If they turn out to orbit within the M31 satellite plane, then this implies a significant value of $f_{counter}$ for M31, thus challenging our model (Table \ref{Best_model_results}). We expect instead that these two dwarfs are just interlopers whose orbits take them far from the M31 satellite plane, as suggested by \citet{Ibata_2013} in their supplementary information. This seems quite feasible given that half the M31 satellites do not belong to its satellite plane and have a more isotropic distribution.

It has proven difficult to reconcile the $\Lambda$CDM paradigm with the highly flattened satellite galaxy planes of the Milky Way and Andromeda while also explaining the high internal velocity dispersions of their constituent satellites. Using the alternative paradigm of MOND, we obtained a reasonable match to several important properties of these structures, in particular their 3D orientations. However, our restricted $N$-body model misses dissipative and self-gravitating effects which are likely important for accurate models of galactic interactions. Thus, more work will be required to test this scenario.

\section{Acknowledgements}
\label{Acknowledgements}

The authors are grateful to the referee for her/his constructive comments. IB was supported by Science and Technology Facilities Council studentship 1506672. The work of DOR was enabled by Royal Astronomical Society undergraduate research bursary 6390/302/001. The algorithms were set up using \textsc{matlab}$^\text{\textregistered}$.

\bibliographystyle{mnras}
\bibliography{SPO_bbl}

\begin{appendix}

\section{The tangential force between two masses}
\label{Tangential_force}

In Section \ref{Perturber_effect}, we used an approximation for the relative tangential gravity $\bm{g}_{tan}$ between two masses. This is the difference in the gravitational field experienced by the masses, less the component of this parallel to the line connecting them. In Newtonian gravity, $\bm{g}_{tan} = 0$. This is also true in MOND if the bodies are isolated. However, $\bm{g}_{tan} \neq 0$ in the presence of an external field, even if one of the objects can be considered a test particle \citep[][equation 39]{Banik_2015}. Our estimate for $\bm{g}_{tan}$ is
\begin{samepage}
\begin{eqnarray}
 	\bm{g}_{tan} = \left( \widehat{\bm{r}} \cos \theta - \bm{\widehat{g}}_{ext} \right)k ~~\text{ where } k = 
\end{eqnarray}
\begin{eqnarray}
    \label{F_tan}
	\left[ \cos \theta - \frac{4}{5} \sin^2 \theta \cos \left( \pi q_{LG} \right) \frac{\tilde{r}}{1 + \tilde{r}^2}\right] \frac{GM_pa_0r}{2g_{ext}\left( r^3 + {r_t}^3\right)}
\end{eqnarray}
\end{samepage}
\begin{eqnarray}
	\tilde{r} &\equiv & \frac{r}{r_t} \\
	r_t &=& \frac{\sqrt{GMa_0}}{Qg_{ext}} \\
	q_{LG} &=& \frac{M_{_{LG}}}{\underbrace{M_{_{LG}} + M_p}_M} \\
	\cos \theta &=& \widehat{\bm{r}} \cdot \bm{\widehat{g}}_{ext}~,~~ \sin^2 \theta = 1 - \cos^2 \theta \\
	\bm{r} &=& \bm{r}_p - \bm{r}_i~,~~\text{ where $i = $ MW or M31}
\end{eqnarray}
 
Here, $M$ refers to the combination of the perturber mass $M_p$ and the LG mass $M_{_{LG}}$. For the case $q_{_{LG}} = 0$ or 1, Equation \ref{F_tan} is a fit to numerical results obtained for a point mass embedded in a constant EF. This is a problem we solved previously using our ring library procedure \citep[][section 4.3]{Banik_2017_escape}. The factor of $\cos \left( \pi q_{_{LG}} \right)$ is our guess for how the tangential force behaves at other values of $q_{_{LG}}$, which we can't solve directly due to the much higher computational cost associated with non-axisymmetric problems. Other functions could also work if they are normalised to be 1 for $q_{_{LG}} = 0$ and are antisymmetric with respect to $q_{_{LG}} \to 1 -q_{_{LG}}$.

\section{The tidal stress tensor in a dominant external field}
\label{Tides_g_ext_domination}

In Section \ref{Apocentre_asymmetry}, we discussed how tides raised by perturbers like Cen A could have influenced the MW-M31 trajectory. To understand this analytically, we consider the tidal stress tensor due to a point mass embedded in a dominant external field $\bm{g}_{_{ext}}$. To facilitate this analysis, we define Cartesian co-ordinates $x,y,z$ centred on the perturber of mass $M$ and based on axes better suited to the geometry of the problem.
\begin{eqnarray}
	\widehat{\bm{z}} ~&=&~ \widehat{\bm{g}}_{_{ext}} \\
	\widehat{\bm{y}} ~&\propto&~ \bm{r} \times \bm{g}_{_{ext}} \\
	\widehat{\bm{x}} ~&=&~ \widehat{\bm{y}} \times \widehat{\bm{z}}
\end{eqnarray}

$\bm{r}$ is the position of some point of interest relative to the perturber, whose potential is $\Phi$. Due to axisymmetry and the curl-free nature of the gravitational field ($\nabla \times \nabla \Phi = 0$) required to prevent energy gain for a particle going around a closed loop, we have that
\begin{eqnarray}
	\Phi_{xy} ~&=&~ \Phi_{yx} ~=~ 0 \\
	\Phi_{yz} ~&=&~ \Phi_{zy} ~=~ 0 \\
	\Phi_{yy} ~&=&~ \frac{\Phi_{x}}{x}
	\label{Easy_tensor_components}
\end{eqnarray}

Here, we use a common shorthand for derivatives according to which e.g. $\Phi_{yz} \equiv \frac{\partial^2 \Phi}{\partial x \partial y}$. The second spatial derivatives of $\Phi$ govern tides raised by a perturber. We approximate that it is sufficiently distant from the LG that the problem is EF-dominated, reducing the potential to that given in equation 37 of \citet{Banik_2015}.
\begin{eqnarray}
	\Phi ~&=&~ -\frac{GM \nu_{ext}}{r} \left( 1 + \frac{K_0}{2} \sin^2 \theta \right) \\
	K_0 ~&\equiv &~ \frac{\partial Ln ~ \nu}{\partial Ln ~ g_{_{N,ext}}}
\end{eqnarray}

This leads to the components of the tidal stress tensor given below, which are not obvious from symmetry arguments like Equation \ref{Easy_tensor_components}.
\begin{eqnarray}
	\label{Hard_tensor_components}
	\frac{\Phi_{xx}}{\alpha} ~&=&~ \left(K_0 - 1 \right)\left(1 - 3\sin^2 \theta \right) + \frac{3}{2}K_0 \sin^2 \theta \left( 5 \sin^2 \theta - 3\right) \nonumber \\
	\frac{\Phi_{zz}}{\alpha} ~&=&~ 3\cos^2 \theta - 1 + \frac{3}{2}K_0 \sin^2 \theta \left( 5 \cos^2 \theta - 1\right) \nonumber \\
	\frac{\Phi_{xz}}{\alpha} ~&=&~ \Phi_{zx} ~=~ \frac{15}{2}K_0 \sin^3 \theta \cos \theta - 3 \left( K_0 - 1 \right) \sin \theta \cos \theta \nonumber \\
	\alpha &\equiv & -\frac{GM \nu_{ext}}{r^3}
\end{eqnarray}

In general, diagonalising the tidal stress tensor requires us to find the eigenvalues of a symmetric ${3 \times 3}$ matrix. However, the alignment of our co-ordinate system with the axisymmetry of the problem means that the stress tensor is already in block diagonal form. Thus, we only need to solve a ${2 \times 2}$ matrix and use Equation \ref{Easy_tensor_components} to obtain the third eigenvalue $\lambda_\phi$ corresponding to the eigenvector $\widehat{\bm{y}}_{eff}$.
\begin{eqnarray}
	\lambda_\phi ~&=&~ \frac{GM \nu_{ext}}{r^3} \left( 1 - K_0 + \frac{3}{2} K_0 \sin^2 \theta \right) \\
	2 \lambda_r ~&=&~ \overbrace{\Phi_{xx} + \Phi_{zz}}^{\alpha \left(K_0 + 1 - \frac{3}{2}K_0 \sin^2 \theta \right)} - \sqrt{4{\Phi_{xz}}^2 + \left(\Phi_{xx} - \Phi_{zz} \right)^2} \nonumber \\
	2 \lambda_\theta ~&=&~ \Phi_{xx} + \Phi_{zz} + \sqrt{4{\Phi_{xz}}^2 + \left(\Phi_{xx} - \Phi_{zz} \right)^2} \nonumber
\end{eqnarray}

The eigenvalues $\lambda_i$ have been named according to the direction of the corresponding eigenvector $\hat{\bm{e}}_i$ in the Newtonian limit (${K_0 = 0}$). Note that $\lambda_r < 0$ while the other eigenvalues are positive.
\begin{eqnarray}
	\hat{\bm{e}}_\phi &=& \widehat{\bm{y}}_{eff} \\
	\hat{\bm{e}}_r &\propto& \widehat{\bm{x}}_{eff} + \frac{\lambda_r - \Phi_{xx}}{\Phi_{xz}} \widehat{\bm{z}}_{eff} \\
	\hat{\bm{e}}_\theta &\propto& \widehat{\bm{x}}_{eff} + \frac{\lambda_\theta - \Phi_{xx}}{\Phi_{xz}} \widehat{\bm{z}}_{eff}
\end{eqnarray}

\begin{figure}
	\centering
		\includegraphics [width = 8.5cm] {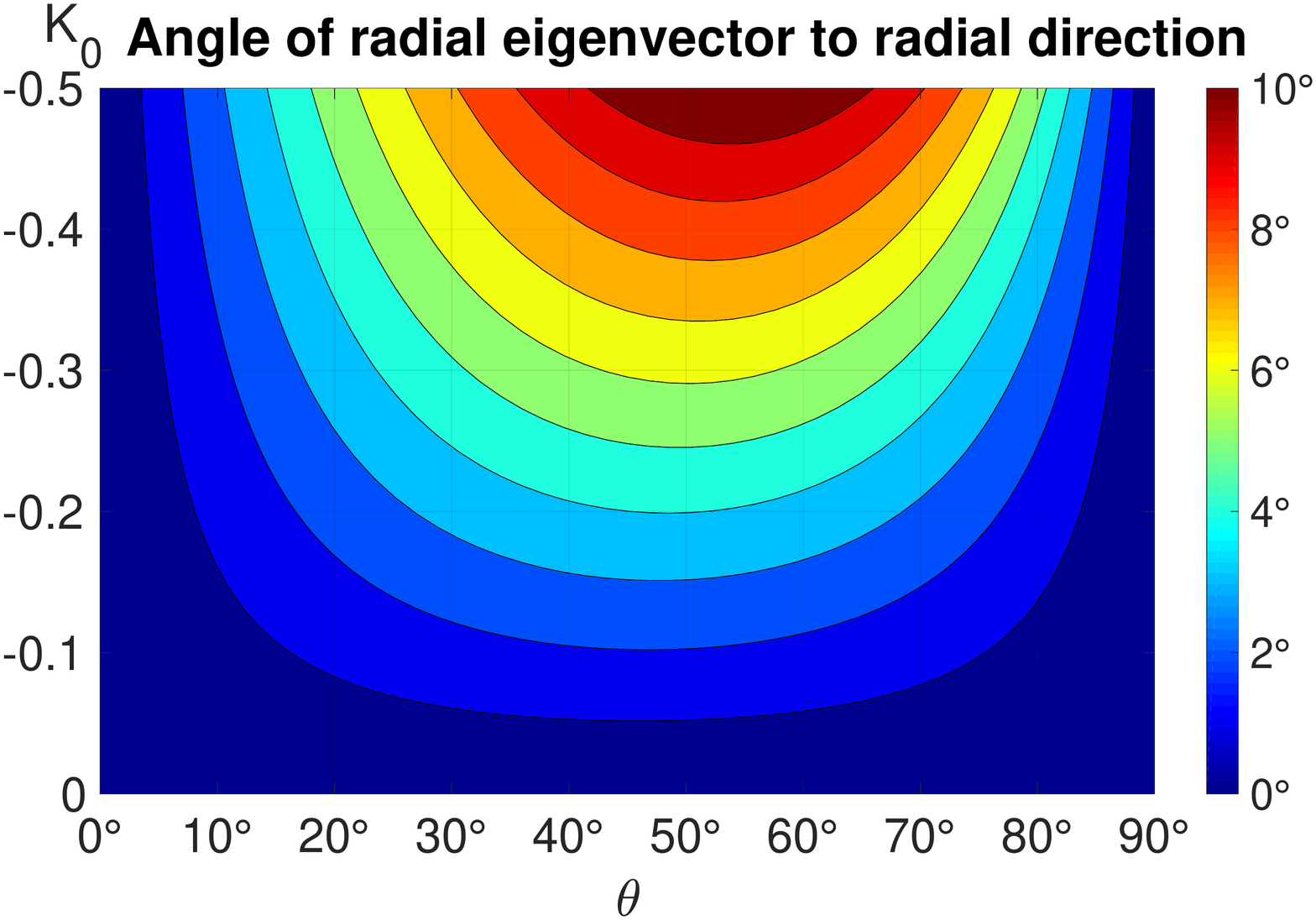}		
		\includegraphics [width = 8.5cm] {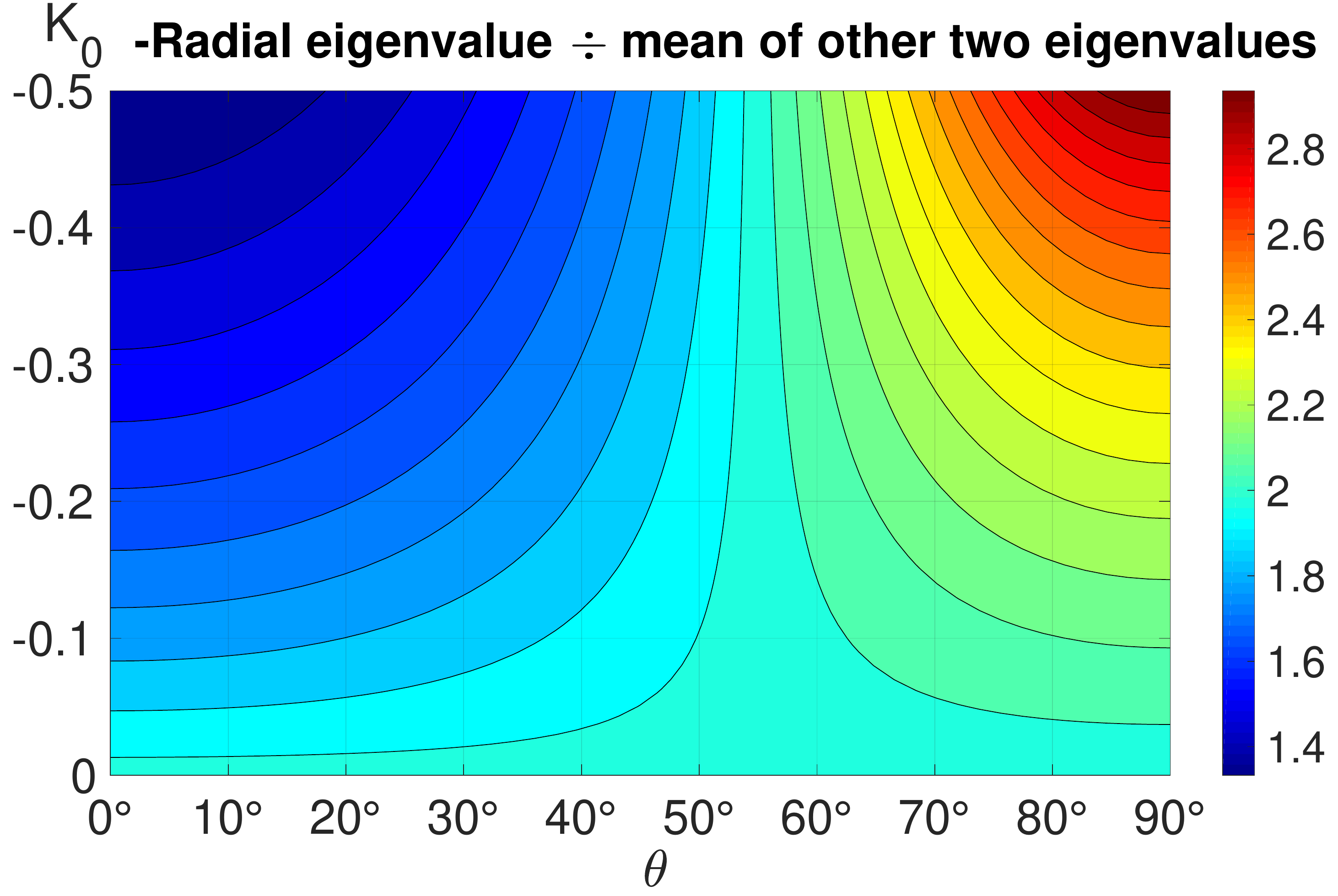}
		\caption{\emph{Top}: Angle between the radial direction and the most radial eigenvector of the tidal stress tensor due to a point mass embedded in a dominant external field. Positive values indicate that $\hat{\bm{e}}_r \cdot \widehat{\bm{g}}_{_{ext}} < \hat{\bm{r}} \cdot \widehat{\bm{g}}_{_{ext}}$ i.e. the radial eigenvector is tilted further away from $\widehat{\bm{g}}_{_{ext}}$ than the radial direction. $K_{_0}$ is the logarithmic derivative of the $\nu$ function, with value 0 in Newtonian gravity and $-\frac{1}{2}$ in the deep-MOND limit. \emph{Bottom}: Values of $-\frac{2\lambda_r}{\lambda_\theta + \lambda_\phi}$, giving an idea about the tidal stress tensor's shape.}
	\label{Eigenvector_angle}
\end{figure}

To help visualise our results, we show the angle between the radial eigenvector $\hat{\bm{e}}_r$ and the radial direction $\hat{\bm{r}}$, with positive values indicating a larger angle to $\widehat{\bm{g}}_{_{ext}}$ (top panel of Figure \ref{Eigenvector_angle}). The bottom panel of this figure shows $\beta \equiv -\frac{2\lambda_r}{\lambda_\theta + \lambda_\phi}$to help give an idea about the tidal stress tensor's shape. As the potential is symmetric with respect to $\bm{g}_{_{ext}} \to -\bm{g}_{_{ext}}$, we only show results for $\theta \leq \frac{\pi}{2}$ i.e. $\bm{r} \cdot \bm{g}_{_{ext}} > 0$.

In Newtonian gravity, $\beta = 2$ and the non-radial directions are equivalent ($\lambda_\theta = \lambda_\phi$). In general, this is not true as there is an additional preferred direction due to the EF. Moreover, $\beta$ can also differ from 2 as the gravitational field need not be divergence-free outside the matter distribution (Equation \ref{QUMOND_equation}).

For two particles separated by a small amount $\bm{s}$ compared to their distance $r$ from the perturber, we can now find the tidal contribution to the acceleration of their separation $\ddot{s}$. This can be decomposed as a sum over contributions from the three eigenvalues $\lambda_i$ which depend on how well $\bm{s}$ aligns with the corresponding eigenvector $\hat{\bm{e}}_{_i}$.
\begin{eqnarray}
    \left. \frac{\ddot{s}}{s} \right|_{tides} ~=~ -\sum_{i = 1}^3 \lambda_i \left( \hat{\bm{s}} \cdot \hat{\bm{e}}_{_i} \right)^2
\end{eqnarray}

The actual value of $\ddot{s}$ will include contributions from other sources like the cosmological acceleration.

\end{appendix}

\bsp
\label{lastpage}
\end{document}